# New Generalized Expressions for the Electromagnetic Fields of the Hybrid Modes of an Optical Fiber Using Complex and Bi-Complex Mathematics

W. Astar

St. Petersburg College (SPC), P.O. Box 13489, St. Petersburg, Florida 33733-3489, USA

**Abstract -** The electromagnetic fields of the hybrid modes of an optical fiber are reformulated using bi-complex mathematics, which leads to simpler expressions relative to those found with the widely used, conventional complex formulation. Generalized expressions for the electromagnetic fields are also found using both formulations for comparative purposes. Furthermore, the feasibility of bi-complex mathematics is investigated for the derivation of the modal powers, orthogonality relations, and the weakly-guided fiber approximation

## 1. Introduction

A practical optical fiber is typically analyzed by beginning first with the simpler, ideal optical fiber (or the fiber). The fiber is characterized by conditions related to its geometry, and its constitutive materials. Geometrically, the fiber is circular in cross-section, with an eccentricity of zero, and is spatially and temporally invariant along its entire cylindrical length. The cross-section of the fiber considered for this report, is comprised of 2 circularly contiguous, concentric regions: a core of refractive index $n_1$ and a radius $a$, and an annular cladding of refractive index $n_2 < n_1$, and a radial width of $(b - a) \gg a$. The fiber is thus a single-step, step-index fiber, as shown in **fig. 1**. The cross-section of the fiber is assumed to be co-incidental with the $xy$-plane of a right-handed coordinate system, resulting in electromagnetic (EM-) field propagation along the positive $z$-direction.

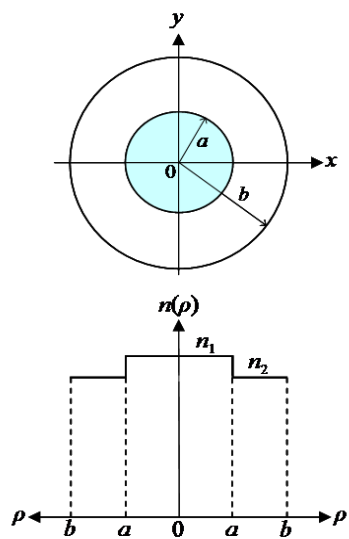

**Fig. 1.** Cross-section of the ideal step-index fiber showing the radii and the refractive indices of the core and the cladding. **The schematic is not to scale, since it is typical that the cladding thickness $(b-a) \gg a$**.

The ideal fiber is also considered to be comprised of homogeneous, isotropic, and non-magnetic, lossless materials. Since the fiber is assumed to be longitudinally and temporally invariant, then the spatial profile of its EM-field is functionally independent of both time and the longitudinal direction (*z*). *Most generally*, the EM-field vector of a hybrid mode has 4 transverse components and 2 longitudinal components, for each cross-sectional region of the fiber, resulting in a total of 12 distinct components. Furthermore, a hybrid mode also exhibits polarization dependence in 2 states, which are degenerate in the propagation constant $\beta$. Lastly, and due to the cylindrical symmetry of the fiber, the spatial profile of the EM-field is derived in cylindrical coordinates, which in this report is described by the triplet ($\rho$, $\varphi$, $z$).

**The EM-field is assumed to be a sinusoidal function of time in this report, and devoid of any data modulation**. In **the complex formulation**, and based on the stated assumptions, the EM-field may be expressed as a **spatiotemporal vector** that oscillates at an angular frequency of $\omega_0$ [1]:

$$\tilde{\mathbf{V}}(\mathbf{r},t) = \vec{\mathbf{V}}(\zeta) e^{j(\omega_0 t - \beta z)}, \quad \mathbf{V} \in \{\mathbf{E}, \mathbf{H}\}. \tag{1.1}$$

It is also a **phasor** with respect to **the imaginary number j,** and is the product of a spatial vector with a phasor, represented by a spatiotemporal, complex exponential factor. Consequently, (1.1) is actually a vector as well as a phasor, or a "**vecsor**". Furthermore, *r* is short-form for the triplet $(\zeta, z)$, whereas $\zeta$ (Greek zeta) is short-form for the polar coordinate couple $(\rho, \varphi)$. The phasor approach is traditionally used as a mathematical convenience in order to simplify EM vector calculus, which would otherwise be complicated by trigonometric functions. The spatial vector used in (1.1) is synthesized as

$$\vec{\mathbf{V}}(\zeta) = \vec{\mathbf{V}}_\rho(\zeta) + \vec{\mathbf{V}}_\varphi(\zeta) + \vec{\mathbf{V}}_z(\zeta) = \sum_{\xi = \rho, \varphi, z} \vec{\xi} V_\xi(\zeta). \tag{1.2}$$

It is constructed from its constituent scalar components as

$$V_\xi(\zeta) = \vec{\xi} \bullet \vec{\mathbf{V}}(\zeta) = |V_\xi(\zeta)| \exp(j\theta_\xi); \quad \xi \in \{\rho, \varphi, z\}, \; \theta_\xi \in \mathbb{R} \backslash \mathbb{Q} \tag{1.3}$$

The spatial vector (1.2) may still be complex with respect to j, but is not a phasor, since the exponent in (1.3) is just a complex constant. The physical EM-field vector is a real quantity, and is simply obtained from its corresponding phasor (1.1) by a real-operation:

$$\vec{\mathbf{v}}(\mathbf{r},t) = \text{Re}\big[\tilde{\mathbf{V}}(\mathbf{r},t)\big] = \sum_{\xi = \rho, \varphi, z} \vec{\xi} |V_\xi(\zeta)| \cos(\omega_0 t - \beta z + \theta_\xi), \quad \vec{\mathbf{v}} \in \{\vec{\mathbf{e}}, \vec{\mathbf{h}}\}. \tag{1.4}$$

**Among the objectives of this report is to derive a new, closed-form, compact expression for the complex spatial vector $\vec{\mathbf{V}}(\zeta)$ (1.2) and its constituent components $V_\xi(\zeta)$, that would incorporate all 12 components of the EM-field of a hybrid mode, without the introduction of new variables**.

In **the bi-complex formulation**, the **spatial vector** itself (1.2), is derived from a vecsor, but this time with respect to **another imaginary number, i**. Then the physical, spatial vector (1.2, 3) is obtained by another real-operation, with respect to i, as follows

$$\vec{V}(\zeta) = \text{Re}_i\left[\tilde{U}(\zeta)\right], \quad U \in \{E, H\} \tag{1.5}$$

which is in terms of the **spatial vecsor** given by

$$\tilde{U}(\zeta) = \vec{U}(\rho)e^{-i\phi_m} = \sum_{\xi=\rho,\varphi,z} \xi \tilde{U}_\xi(\zeta). \tag{1.6}$$

**The composite angle $\phi_m$ incorporates the cylindrical azimuth $\varphi$, as well as a non-arbitrary phase-factor related to *m*odal polarization**, which is discussed later. The radial vector found in (1.6) is generally complex with respect to *both* i and j, but is not a phasor in either. When a phasor is solely a function of the polar couple $\zeta$ or $(\rho, \varphi)$ as is (1.6), it is implicitly understood that it is a phasor with respect to i *only*. The vecsor (1.6) can be constructed from its constituent phasor components, each generally given by

$$\tilde{U}_\xi(\zeta) = \xi \cdot \tilde{U}(\zeta) = |\tilde{U}_\xi(\zeta)|\exp(i\gamma_\xi + j\theta_\xi)e^{-i\phi_m}, \quad \xi \in \{\rho, \varphi, z\}, \gamma_\xi \in \mathbb{R}\backslash\mathbb{Q} \tag{1.7}$$

Substituting (1.5) into (1.1), and the result into (1.4), yields an alternative, more explicit expression for the real, EM-field vector, in terms of its 3 components:

$$\vec{v}(r,t) = \text{Re}_j \text{Re}_i\left[\hat{U}(r,t)\right] = \sum_{\xi=\rho,\varphi,z} \xi |\tilde{U}_\xi(\zeta)|\cos(\phi_m + \gamma_\xi)\cos(\omega_0 t - \beta z + \theta_\xi). \tag{1.8}$$

The use of different complex operators in the same expression is addressed in the next section. Furthermore, the real-operator used in (1.4) has been subscripted with a j, in (1.8), to distinguish it from that of (1.5). This leads to a new, **bi-complex vecsor**,

$$\hat{U}(r,t) = \tilde{U}(\zeta)e^{j(\omega_0 t - \beta z)} = \vec{U}(\rho)e^{-i\phi_m}e^{j(\omega_0 t - \beta z)} \tag{1.9}$$

which is completely separable over $(\rho, \varphi, z)$, unlike its counterpart (1.1) in the complex formulation. It will be found that this leads to simpler and more compact expressions for the EM-field of a hybrid mode.

**Analytical expressions for the spatial vecsor $\tilde{U}(\zeta)$ and its components $\tilde{U}_\xi(\zeta)$ will be found in this report, and compared with their counterparts in the complex formulation.** These comparisons are carried out in various applications, such as modal power flow, orthogonality, and the weakly-guided fiber approximation.

Lastly, this report is largely based on Okamoto's nomenclature [1]. Using the earlier work of Snitzer [2], Okamoto [1] significantly simplified the EM-field component expressions, upon the recognition of an *s*-parameter (which is Snitzer's *P*-parameter) that is recurrent in these expressions. They will also be found to be further simplified upon the identification of a certain, generating function.

## 2. Complex and Bi-complex mathematics

The conventional complex ($\mathbb{C}$) number $z$ (which is not used as such beyond this section, to minimize confusion with **the $z$-coordinate**), is defined in set-builder notation as

$$\mathbb{C} = \{z : z = x + iy \,|\, x, y \in \mathbb{R}\}, \tag{2.1}$$

and is expressed in terms of the real ($\mathbb{R}$) coefficients $x$ and $y$. In mathematics and physics, i is traditionally identified with the square-root of -1. In Okamoto's nomenclature [1], as is the case in electrical engineering, j is used instead of i, and is preferred to i, in order to minimize confusion with the time-dependent electric current, which is traditionally assigned the letter i or *i*. **In fact, j may be used *in lieu* of i to define $z$ in (2.1)**.

   **In this report however, i and j are 2 distinct imaginary numbers used in accordance with a convention due to Corrado Segre**, who introduced it in 1892 [3, 4],

$$i^2 = -1; \quad j^2 = -1; \quad ij = ji \neq -1;$$
$$i^\circ = -i; \quad j^* = -j, \tag{2.2}$$

**which uses different conjugation superscripts in this report**. The conjugation in i is expressed with a superscript of '°', to distinguish it from a conjugation in j, which is expressed with a superscript of '*'. No other rule is used or required in this report beyond those listed in (2.2). This convention is used to define the set of bi-complex numbers [4],

$$\mathbb{BC} = \{w : w = z_1 + j z_2 \,|\, z_1, z_2 \in \mathbb{C}\}, \tag{2.3}$$

in which $\mathbb{C}$ is the set of complex numbers with respect to i as defined by (2.1), whereas bi-complex numbers $w$ may be regarded as numbers $z_n$ that are complex with respect to a distinct, 2nd imaginary number j, in accordance with (2.2, 3), or as complex numbers in j whose coefficients $z_n$ are themselves complex, but in i. For this reason, bi-complex numbers are termed "hypercomplex numbers". It was later found that Segre's bi-complex numbers are algebraically *isomorphic* to tessarines, conventionally defined as [5]

$$\mathbb{T} = \{t : t = t_1 + i t_2 + j t_3 + k t_4 \,|\, t_1, t_2, t_3, t_4 \in \mathbb{R}\} \tag{2.4}$$

and which were first proffered by James Cockle in 1848 [5]. The tessarines are also based on 2 distinct imaginary numbers i and j, but according to the following convention

$$i^2 = -1; \quad j^2 = +1; \quad k^2 = (ij)^2 = (ji)^2 = -1, \tag{2.5}$$

which indeed yield Segre's convention (2.2), if k is identified with Segre's j, whereas j is assigned to Segre's ij (or ji), since the square of the latter yields +1. It appears that Segre was either oblivious of Cockle's earlier work, or perhaps just chose to ignore it [6].

   Bi-complex numbers may also be viewed as Hamilton quaternions [7], which are four-dimensional numbers. However, they are not quaternions in the strictest sense, since the product of the imaginary numbers according to Segre (2.2) (or to Cockle (2.5)) is commutative, which is not the case for quaternions [7].

It may seem perplexing to have 2 imaginary numbers with identical properties in the same mathematical expressions. However, they are being proffered here in a similar manner as the basis vectors **x** and **y** in the two-dimensional Cartesian coordinates for instance, which also have mostly identical properties, such as

$$|\mathbf{x}|=|\mathbf{y}|=1; \quad \mathbf{x}\cdot\mathbf{x}=\mathbf{y}\cdot\mathbf{y}=1; \quad \mathbf{x}\times\mathbf{x}=\mathbf{y}\times\mathbf{y}=0. \tag{2.6}$$

The most frequently encountered bi-complex quantity in this report, is in a product form, and not in the sum given by (2.3). In polar form, this product is expressed as:

$$z_1 z_2 = \exp(i\theta_1)\exp(j\theta_2), \quad \{\theta_1,\theta_2\}\in\mathbb{R}\backslash\mathbb{Q} \tag{2.7}$$

It may be re-expressible as (2.3) if desired, **although with different $z_n$**. Since the RHS is commutative, the LHS would also be expected to be so. If the exponents are both variable, they impart different, phasor-like behavior whenever (2.7) is used in a vector. For this reason, such a vector would become a bi-complex phasor, separable in i and j.

In order to find the real part of the product (2.7), consecutive real-operations must be carried out over i and j, but in no particular order, which renders the real operations commutative for i and j. If the product (2.7) is defined such that

$$z_1 z_2 = (x_1 + iy_1)(x_2 + jy_2), \quad \{x_1,y_1,x_2,y_2\}\in\mathbb{R} \tag{2.8}$$

then

$$\text{Re}_i\,\text{Re}_j[z_1 z_2] = \text{Re}_i\,\text{Re}_j[x_1 x_2 + jx_1 y_2 + ix_2 y_1 + ij\,y_1 y_2] = \text{Re}_i[x_1 x_2 + ix_2 y_1] = x_1 x_2, \tag{2.9}$$

with the argument of the RHS of the 1st equality clearly in the set (2.3), whereas

$$\text{Re}_j\,\text{Re}_i[z_1 z_2] = \text{Re}_j\,\text{Re}_i[x_1 x_2 + jx_1 y_2 + ix_2 y_1 + ij\,y_1 y_2] = \text{Re}_j[x_1 x_2 + jx_1 y_2] = x_1 x_2, \tag{2.10}$$

and confirms the commutative property. **It should be clear that j (i) is to be treated as a real constant under a $\text{Re}_i$ -operation ($\text{Re}_j$ -operation).** Consequently, it is also true that

$$\text{Re}_i\,\text{Re}_j[z_1 z_2] = \text{Re}_i[z_1]\text{Re}_j[z_2] = \text{Re}_j[z_2]\text{Re}_i[z_1] = \text{Re}_j\,\text{Re}_i[z_1 z_2]. \tag{2.11}$$

As for complex conjugation, **a conjugation in i (j) treats j (i) as a real constant**,

$$(z_1 z_2)^{\circ *} = (x_1 + iy_1)^{\circ *}(x_2 + jy_2)^{\circ *} = (x_1 - iy_1)(x_2 - jy_2) = (x_1 + iy_1)^{\circ}(x_2 + jy_2)^{*} = z_1^{\circ} z_2^{*}. \tag{2.12}$$

The following general relation is applicable to the complex numbers in (2.8),

$$\text{Re}_\mu [z_p] = \text{Re}_\mu [z_p^\diamond], \quad \{\mu,p\} \in \{\{i,1\},\{j,2\}\} \tag{2.13}$$

for which the superscript '$\diamond$' denotes either conjugation('$\circ$') with respect to i (if $p = 1$), or conjugation ('*') with respect to j (if $p = 2$). Then it is also true that

$$\text{Re}_i \text{Re}_j [z_1^{\circ*} z_2^{\circ*}] = \text{Re}_i [z_1^\circ] \text{Re}_j [z_2^*] = \text{Re}_i [z_1] \text{Re}_j [z_2] = \text{Re}_i \text{Re}_j [z_1 z_2] \tag{2.14}$$

since **the real-part of a complex number is always identical with the real-part of its complex-conjugate.**

Furthermore, although the following relation is true for either $z_1$ or $z_2$,

$$\text{Re}_\mu [z_p] = \frac{1}{2}(z_p^\diamond + z_p), \quad \{\mu,p\} \in \{\{i,1\},\{j,2\}\} \tag{2.15}$$

it is not generally true that

$$\text{Re}_i \text{Re}_j [z_1 z_2] = \frac{1}{2}\left((z_1 z_2)^{\circ*} + z_1 z_2\right). \tag{2.16}$$

Instead, each Re-operation must be carried out in turn on the product, yielding

$$\text{Re}_i \text{Re}_j [z_1 z_2] = \frac{1}{4}\left((z_1 z_2)^{\circ*} + (z_1 z_2)^* + (z_1 z_2)^\circ + z_1 z_2\right) \tag{2.17}$$

which simplifies to

$$\text{Re}_i \text{Re}_j [z_1 z_2] = \frac{1}{4}\left(z_1^\circ z_2^* + z_1 z_2\right) + \frac{1}{4}\left(z_1^\circ z_2 + z_1 z_2^*\right). \tag{2.18}$$

The general vector cross-product of the real-parts, with respect to a general imaginary number $\mu$, of 2 (bi-)complex vecsors is often encountered in this report during the derivation of power, and is found as

$$\text{Re}_\mu \hat{\mathbf{X}}(\mathbf{r},t) \times \text{Re}_\mu \hat{\mathbf{Y}}(\mathbf{r},t) = \frac{1}{2}\left[\hat{\mathbf{X}}^\diamond(\mathbf{r},t) + \hat{\mathbf{X}}(\mathbf{r},t)\right] \times \frac{1}{2}\left[\hat{\mathbf{Y}}^\diamond(\mathbf{r},t) + \hat{\mathbf{Y}}(\mathbf{r},t)\right]. \tag{2.19}$$

After carrying out the vector cross-product, there results

$$\text{Re}_\mu \hat{\mathbf{X}}(\mathbf{r},t) \times \text{Re}_\mu \hat{\mathbf{Y}}(\mathbf{r},t) = \frac{1}{2}\begin{bmatrix} \hat{\mathbf{X}}^\diamond(\mathbf{r},t) \times \hat{\mathbf{Y}}^\diamond(\mathbf{r},t) + \hat{\mathbf{X}}^\diamond(\mathbf{r},t) \times \hat{\mathbf{Y}}(\mathbf{r},t) \\ \hat{\mathbf{X}}(\mathbf{r},t) \times \hat{\mathbf{Y}}(\mathbf{r},t) + \hat{\mathbf{X}}(\mathbf{r},t) \times \hat{\mathbf{Y}}^\diamond(\mathbf{r},t) \end{bmatrix} \tag{2.20}$$

which finally yields, after a simplification, and in one possible form,

$$\text{Re}_\mu \hat{\mathbf{X}}(\mathbf{r},t) \times \text{Re}_\mu \hat{\mathbf{Y}}(\mathbf{r},t) = \frac{1}{2}\text{Re}_\mu \left[\hat{\mathbf{X}}^\diamond(\mathbf{r},t) \times \hat{\mathbf{Y}}(\mathbf{r},t) + \hat{\mathbf{X}}(\mathbf{r},t) \times \hat{\mathbf{Y}}(\mathbf{r},t)\right]. \tag{2.21}$$

# 3. The electromagnetic field of a hybrid mode of an ideal fiber
## 3.1 From the original complex expressions to bi-complex expressions

As explained in §**1**, the ideal fiber is a step-index fiber of the single-step type, and is comprised of the core, and the lower index cladding. Due to its assumedly *perfect* cylindrical symmetry, the EM-field components of the fiber's modes are derived in cylindrical coordinates, as the solutions of the EM-field wave equation.

In Okamoto's nomenclature [1], the 6 complex EM-field components of a hybrid mode **in the core of the fiber**, geometrically described by $\rho \in [0, a]$ as in **fig. 1** of §**1**, are expressed in terms of **Bessel functions of the first kind (*J*)**, and trigonometric functions,

$$E_\rho(\rho,\varphi) = -j\beta \frac{aA}{2u}\left[(1-s)J_{n-1}\left(\frac{u}{a}\rho\right) - (1+s)J_{n+1}\left(\frac{u}{a}\rho\right)\right]\cos(n\varphi+\psi_m) \quad (3.1.1)$$

$$E_\varphi(\rho,\varphi) = j\beta \frac{aA}{2u}\left[(1-s)J_{n-1}\left(\frac{u}{a}\rho\right) + (1+s)J_{n+1}\left(\frac{u}{a}\rho\right)\right]\sin(n\varphi+\psi_m) \quad (3.1.2)$$

$$E_z(\rho,\varphi) = AJ_n\left(\frac{u}{a}\rho\right)\cos(n\varphi+\psi_m) \quad (3.1.3)$$

$$H_\rho(\rho,\varphi) = -j\omega_0\varepsilon_1 \frac{aA}{2u}\left[(1-s_1)J_{n-1}\left(\frac{u}{a}\rho\right) + (1+s_1)J_{n+1}\left(\frac{u}{a}\rho\right)\right]\sin(n\varphi+\psi_m) \quad (3.1.4)$$

$$H_\varphi(\rho,\varphi) = -j\omega_0\varepsilon_1 \frac{aA}{2u}\left[(1-s_1)J_{n-1}\left(\frac{u}{a}\rho\right) - (1+s_1)J_{n+1}\left(\frac{u}{a}\rho\right)\right]\cos(n\varphi+\psi_m) \quad (3.1.5)$$

$$H_z(\rho,\varphi) = -\frac{A\beta s}{\omega_0\mu_0}J_n\left(\frac{u}{a}\rho\right)\sin(n\varphi+\psi_m) \quad (3.1.6)$$

In the above equations, *A* is a generally complex amplitude constant which is found from the power of the EM-field of the mode; *n* is the azimuthal eigenvalue and is unity for the fundamental $HE_{11}$-mode, but is otherwise larger than unity for higher-order hybrid modes; *β* is the propagation constant, which is obtained from the dispersion relation for hybrid modes [1]; $\varepsilon_1$, the permittivity of the core; and $\mu_0$, the magnetic permeability of a vacuum. Moreover, the permittivity instead of the refractive index is being preferentially used in the component expressions wherever it occurs, to render the expressions more compact. **In this report, the couple (*ρ*, *φ*) is being used to represent the cross-section in cylindrical coordinates, which differs from the mixed Greek/Roman $(r,\theta)$ used in Okamoto's nomenclature [1]**. Otherwise, the nomenclature is identical to that of Okamoto, and a full **Nomenclature** section is found at the end of this report.

Generally, the phase-factor $\psi_m$ determines the 2 polarization states of a hybrid mode, which are indistinguishable in their radial dependence. For *n* = 1, for instance, it determines whether the EM-field represents the $HE_{11}^x$-mode for which *m* = 1, or the $HE_{11}^y$-mode, for which *m* = 2, which are the 2 eigenmodes of a single-mode fiber.

The 6 complex EM-field components of a hybrid mode **in the cladding of the fiber**, geometrically described by $\rho \in (a, b]$ as in **fig. 1** of §**1**, are expressed in terms of **modified Bessel functions of the second kind ($K$)**, and trigonometric functions,

$$E_\rho(\rho,\varphi) = -j\beta \frac{aA}{2w} \left( \frac{J_n(u)}{K_n(w)} \right) \left[ (1-s) K_{n-1}\left(\frac{w}{a}\rho\right) + (1+s) K_{n+1}\left(\frac{w}{a}\rho\right) \right] \cos(n\varphi + \psi_m) \quad (3.1.7)$$

$$E_\varphi(\rho,\varphi) = j\beta \frac{aA}{2w} \left( \frac{J_n(u)}{K_n(w)} \right) \left[ (1-s) K_{n-1}\left(\frac{w}{a}\rho\right) - (1+s) K_{n+1}\left(\frac{w}{a}\rho\right) \right] \sin(n\varphi + \psi_m) \quad (3.1.8)$$

$$E_z(\rho,\varphi) = A \frac{J_n(u)}{K_n(w)} K_n\left(\frac{w}{a}\rho\right) \cos(n\varphi + \psi_m) \quad (3.1.9)$$

$$H_\rho(\rho,\varphi) = -j\omega_0\varepsilon_2 \frac{aA}{2w} \left( \frac{J_n(u)}{K_n(w)} \right) \left[ (1-s_2) K_{n-1}\left(\frac{w}{a}\rho\right) - (1+s_2) K_{n+1}\left(\frac{w}{a}\rho\right) \right] \sin(n\varphi + \psi_m) \quad (3.1.10)$$

$$H_\varphi(\rho,\varphi) = -j\omega_0\varepsilon_2 \frac{aA}{2w} \left( \frac{J_n(u)}{K_n(w)} \right) \left[ (1-s_2) K_{n-1}\left(\frac{w}{a}\rho\right) + (1+s_2) K_{n+1}\left(\frac{w}{a}\rho\right) \right] \cos(n\varphi + \psi_m) \quad (3.1.11)$$

$$H_z(\rho,\varphi) = -\frac{A\beta s}{\omega_0\mu_0} \frac{J_n(u)}{K_n(w)} K_n\left(\frac{w}{a}\rho\right) \sin(n\varphi + \psi_m) \quad (3.1.12)$$

where $\varepsilon_2$ is the permittivity of the cladding. Apart from the free-space wave-number $k_0$, (3.1.1-12) are being expressed in terms of Okamoto's **non-dimensional parameters**,

$$k_0 = \omega_0/c \quad (3.1.13)$$

$$u = a\left(k_0^2 n_1^2 - \beta^2\right)^{1/2} \quad (3.1.14)$$

$$w = a\left(\beta^2 - k_0^2 n_2^2\right)^{1/2} \quad (3.1.15)$$

$$v^2 = u^2 + w^2 \quad (3.1.16)$$

$$s = \frac{n v^2 J_n(u) K_n(w)}{u w^2 J_n'(u) K_n(w) + u^2 w K_n'(w) J_n(u)}, \quad n \geq 1, \; v > 0 \quad (3.1.17)$$

$$s_r = s\left(\beta/k_0 n_r\right)^2; \quad r \in \{1, 2\} \quad (3.1.18)$$

$$\psi_m = (m-1)\pi/2, \quad m \in \{1, 2\} \quad (3.1.19)$$

with c being the speed of light in vacuum. It can be clearly seen that the *s*-parameter is indeed recurrent throughout (3.1.1-12), either directly as (3.1.17), or indirectly as (3.1.18). It is expressed in terms of Bessel functions and their derivatives. The parameter *v*, which is restricted to being less than 2.405 for a single-mode fiber, is found from a Pythagorean relation (3.1.16) with the normalized, transverse wave-numbers *u* (3.1.14) and *w* (3.1.15). Any EM-field vector for a given coordinate $\xi$, is obtainable using the relation

$$\vec{V}_\xi(\boldsymbol{r},t) = \xi \, \text{Re}\left[V_\xi(\rho,\varphi) e^{j(\omega_0 t - \beta)}\right]; \quad V \in \{E, H\}, \; \xi \in \{\rho, \varphi, z\}, \; \rho \in [a,b] \quad (3.1.20)$$

with $\beta$, its propagation constant, being independent of the modal index *m* for the case of the ideal fiber. Each vector component $V_\xi(\rho,\varphi)$ is found from the set of components (3.1.1-12). The three-dimensional vector of the EM-field is constructed using (3.1.20).

**The corresponding bi-complex EM-field phasors ($\tilde{U}_\xi(\rho,\varphi)$), are found from the complex components ($V_\xi(\rho,\varphi)$) given by (3.1.1-12), using the following transform:**

$$\tilde{U}_\xi(\rho,\varphi) = \left[\frac{(\delta_{\xi\rho} + \delta_{\xi z})\delta_{VE} + \delta_{\xi\varphi}\delta_{VH}}{\cos(n\varphi + \psi_m)} + i\frac{\delta_{\xi\varphi}\delta_{VE} + (\delta_{\xi\rho} + \delta_{\xi z})\delta_{VH}}{\sin(n\varphi + \psi_m)}\right] V_\xi(\rho,\varphi) e^{-i(n\varphi + \psi_m)} \quad (3.1.21)$$

which makes extensive use of the well-known, Kronecker delta function [8], defined as

$$\delta_{pq} = \delta[p - q] = \begin{cases} 1, & p = q \\ 0, & p \neq q \end{cases} \quad (3.1.22)$$

which is **unity only when its argument is zero**, and vanishes otherwise (see **APPENDIX A**). The transform can be described as a division by a cos ($n\varphi + \psi_m$) if the component carries this function (which is true for either the $\rho$- or the *z*-components of the *E*-field, or the $\varphi$-component of the *H*-field), OR a division by -i sin($n\varphi + \psi_m$) if the component carries a sin($n\varphi + \psi_m$)-function (which is true for either the $\varphi$-component of the *E*-field, or the $\rho$- and *z*-components of the *H*-field). Lastly, the result is multiplied by a spatial phasor regardless of the trigonometric dependence of a given component. Another possibility, also using Kronecker deltas as in (3.1.21), is the following transform:

$$\tilde{U}_\xi(\rho,\varphi) = \left\{\frac{\delta[V_\xi(\rho,(\pi/2 - \psi_m)/n)]}{\cos(n\varphi + \psi_m)} + i\frac{\delta[V_\xi(\rho,(\pi - \psi_m)/n)]}{\sin(n\varphi + \psi_m)}\right\} V_\xi(\rho,\varphi) e^{-i(n\varphi + \psi_m)}.$$

(3.1.23)

**The sign of a complex component is preserved under either transformation**. These expressions are termed the Complex-To-Bi-complex (CTB) transforms. The reverse transform from the bi-complex phasor to the complex scalar, can be found by solving (3.1.21) or (3.1.23) for $V_\xi(\rho,\varphi)$ since the bracketed term in each, is never zero. However, it is easier to take the real-part with respect to i instead, as

$$V_\xi(\rho,\varphi) = \text{Re}_i \tilde{U}_\xi(\rho,\varphi). \quad (3.1.24)$$

**The bi-complex phasor equivalents** of the complex scalars (3.1-12) are thus given by

$$\tilde{E}_\rho(\rho,\varphi) = -\mathrm{j}\beta \frac{aA}{2u}\left[(1-s)J_{n-1}\left(\frac{u}{a}\rho\right) - (1+s)J_{n+1}\left(\frac{u}{a}\rho\right)\right]e^{-\mathrm{i}(n\varphi+\psi_m)} \qquad (3.1.25)$$

$$\tilde{E}_\varphi(\rho,\varphi) = \mathrm{ij}\beta \frac{aA}{2u}\left[(1-s)J_{n-1}\left(\frac{u}{a}\rho\right) + (1+s)J_{n+1}\left(\frac{u}{a}\rho\right)\right]e^{-\mathrm{i}(n\varphi+\psi_m)} \qquad (3.1.26)$$

$$\tilde{E}_z(\rho,\varphi) = AJ_n\left(\frac{u}{a}\rho\right)e^{-\mathrm{i}(n\varphi+\psi_m)} \qquad (3.1.27)$$

$$\tilde{H}_\rho(\rho,\varphi) = -\mathrm{ij}\omega_0\varepsilon_1 \frac{aA}{2u}\left[(1-s_1)J_{n-1}\left(\frac{u}{a}\rho\right) + (1+s_1)J_{n+1}\left(\frac{u}{a}\rho\right)\right]e^{-\mathrm{i}(n\varphi+\psi_m)} \qquad (3.1.28)$$

$$\tilde{H}_\varphi(\rho,\varphi) = -\mathrm{j}\omega_0\varepsilon_1 \frac{aA}{2u}\left[(1-s_1)J_{n-1}\left(\frac{u}{a}\rho\right) - (1+s_1)J_{n+1}\left(\frac{u}{a}\rho\right)\right]e^{-\mathrm{i}(n\varphi+\psi_m)} \qquad (3.1.29)$$

$$\tilde{H}_z(\rho,\varphi) = -\mathrm{i}\frac{A\beta s}{\omega_0\mu_0}J_n\left(\frac{u}{a}\rho\right)e^{-\mathrm{i}(n\varphi+\psi_m)} \qquad (3.1.30)$$

$$\tilde{E}_\rho(\rho,\varphi) = -\mathrm{j}\beta \frac{aA}{2w}\left(\frac{J_n(u)}{K_n(w)}\right)\left[(1-s)K_{n-1}\left(\frac{w}{a}\rho\right) + (1+s)K_{n+1}\left(\frac{w}{a}\rho\right)\right]e^{-\mathrm{i}(n\varphi+\psi_m)} \qquad (3.1.31)$$

$$\tilde{E}_\varphi(\rho,\varphi) = \mathrm{ij}\beta \frac{aA}{2w}\left(\frac{J_n(u)}{K_n(w)}\right)\left[(1-s)K_{n-1}\left(\frac{w}{a}\rho\right) - (1+s)K_{n+1}\left(\frac{w}{a}\rho\right)\right]e^{-\mathrm{i}(n\varphi+\psi_m)} \qquad (3.1.32)$$

$$\tilde{E}_z(\rho,\varphi) = A\frac{J_n(u)}{K_n(w)}K_n\left(\frac{w}{a}\rho\right)e^{-\mathrm{i}(n\varphi+\psi_m)} \qquad (3.1.33)$$

$$\tilde{H}_\rho(\rho,\varphi) = -\mathrm{ij}\omega_0\varepsilon_2 \frac{aA}{2w}\left(\frac{J_n(u)}{K_n(w)}\right)\left[(1-s_2)K_{n-1}\left(\frac{w}{a}\rho\right) - (1+s_2)K_{n+1}\left(\frac{w}{a}\rho\right)\right]e^{-\mathrm{i}(n\varphi+\psi_m)} \qquad (3.1.34)$$

$$\tilde{H}_\varphi(\rho,\varphi) = -\mathrm{j}\omega_0\varepsilon_2 \frac{aA}{2w}\left(\frac{J_n(u)}{K_n(w)}\right)\left[(1-s_2)K_{n-1}\left(\frac{w}{a}\rho\right) + (1+s_2)K_{n+1}\left(\frac{w}{a}\rho\right)\right]e^{-\mathrm{i}(n\varphi+\psi_m)} \qquad (3.1.35)$$

$$\tilde{H}_z(\rho,\varphi) = -\frac{\mathrm{i}A\beta s}{\omega_0\mu_0}\frac{J_n(u)}{K_n(w)}K_n\left(\frac{w}{a}\rho\right)e^{-\mathrm{i}(n\varphi+\psi_m)} \qquad (3.1.36)$$

## 3.2 Generalization of the EM-field component expressions

It is possible to re-express the EM-field over *both* cross-sectional regions of the fiber **using just 6 equations, instead of the twelve (3.1.1-12) of Okamoto's nomenclature**:

$$E_{r\rho}(\zeta) = -j\beta \frac{aA}{2u}\left(\frac{uJ_n(u)}{wK_n(w)}\right)^{r-1}\left[(1-s)\frac{J_{n-1}^{2-r}(u\rho/a)}{K_{n-1}^{1-r}(w\rho/a)} + \lambda_r(1+s)\frac{J_{n+1}^{2-r}(u\rho/a)}{K_{n+1}^{1-r}(w\rho/a)}\right]\cos\phi_m \tag{3.2.1}$$

$$E_{r\varphi}(\zeta) = j\beta \frac{aA}{2u}\left(\frac{uJ_n(u)}{wK_n(w)}\right)^{r-1}\left[(1-s)\frac{J_{n-1}^{2-r}(u\rho/a)}{K_{n-1}^{1-r}(w\rho/a)} - \lambda_r(1+s)\frac{J_{n+1}^{2-r}(u\rho/a)}{K_{n+1}^{1-r}(w\rho/a)}\right]\sin\phi_m \tag{3.2.2}$$

$$E_{rz}(\zeta) = A\frac{J_n^{r-1}(u)}{K_n^{r-1}(w)}\frac{J_n^{2-r}(u\rho/a)}{K_n^{1-r}(w\rho/a)}\cos\phi_m \tag{3.2.3}$$

$$H_{r\rho}(\zeta) = -j\omega_0\varepsilon_r \frac{aA}{2u}\left(\frac{uJ_n(u)}{wK_n(w)}\right)^{r-1}\left[(1-s_r)\frac{J_{n-1}^{2-r}(u\rho/a)}{K_{n-1}^{1-r}(w\rho/a)} - \lambda_r(1+s_r)\frac{J_{n+1}^{2-r}(u\rho/a)}{K_{n+1}^{1-r}(w\rho/a)}\right]\sin\phi_m \tag{3.2.4}$$

$$H_{r\varphi}(\zeta) = -j\omega_0\varepsilon_r \frac{aA}{2u}\left(\frac{uJ_n(u)}{wK_n(w)}\right)^{r-1}\left[(1-s_r)\frac{J_{n-1}^{2-r}(u\rho/a)}{K_{n-1}^{1-r}(w\rho/a)} + \lambda_r(1+s_r)\frac{J_{n+1}^{2-r}(u\rho/a)}{K_{n+1}^{1-r}(w\rho/a)}\right]\cos\phi_m \tag{3.2.5}$$

$$H_{rz}(\zeta) = -\frac{A\beta s}{\omega_0\mu_0}\frac{J_n^{r-1}(u)}{K_n^{r-1}(w)}\frac{J_n^{2-r}(u\rho/a)}{K_n^{1-r}(w\rho/a)}\sin\phi_m \tag{3.2.6}$$

Since factors that appear in the core and the cladding expressions must now both be represented in this generalization, 2 new parameters have been introduced in (3.2.1-6) in order to maintain the compactness of these new expressions. They are given by

$$\lambda_\alpha = e^{j\pi\alpha}, \ \alpha \in \mathbb{Z}^+, \tag{3.2.7}$$

$$\phi_m = n\varphi + \psi_m, \ m \in \{1,2\}, n \in \mathbb{Z}^+. \tag{3.2.8}$$

In Okamoto's original nomenclature [1], the core is designated as the 1st region, whereas the cladding, the zero-th region. The latter unfortunately leads to a cladding permittivity of $\varepsilon_0$, which is traditionally reserved for the vacuum permittivity. **For this reason, a different regional assignment is used here, and Okamoto's expressions (3.1-12) have been recast to reflect this new assignment. Thus, Okamoto's expressions for the core (3.1-6) are obtained simply by setting in (3.2.1-6), $r = 1$, whereas those for the cladding (3.7-12) are recovered by setting $r = 2$.** Operationally, recovery of any component (3.1-12) from (3.2.1-6) can be achieved by specifying the coordinate $\xi$, and the region $r'$,

$$V_\xi(\zeta) = \delta_{rr'}V_{r\xi}(\zeta); \ \xi \in \{\rho,\varphi,z\}, \ r' \in \{1,2\}, \ \rho \in [a,b]. \tag{3.2.9}$$

**The Greek variable $\xi$ (xi) should not be confused with the Greek spatial argument $\zeta$ (zeta), which has already been defined as short-form for the polar coordinate couple ($\rho$, $\varphi$). It should also be emphasized that $\varepsilon_r$ is the *regional* permittivity, and *not* the relative permittivity. Moreover, $\lambda$ is only a non-dimensional parameter frequently used throughout this report, and is *not* being used as the wavelength of the EM-field. The *wavelength of the EM-field* is actually *not used* anywhere in this report, which consistently uses the angular frequency $\omega_0$ of the EM-field, instead.**

Bi-complex versions of the component expressions are also found using the same generalization, or by applying the CTB transform (3.1.21 or 23) to (3.2.1-6), yielding

$$\tilde{E}_{r\rho}(\rho,\varphi) = -j\beta \frac{aA}{2u}\left(\frac{uJ_n(u)}{wK_n(w)}\right)^{r-1}\left[(1-s)\frac{J_{n-1}^{2-r}(u\rho/a)}{K_{n-1}^{1-r}(w\rho/a)} + (1+s)e^{j\pi r}\frac{J_{n+1}^{2-r}(u\rho/a)}{K_{n+1}^{1-r}(w\rho/a)}\right]e^{-i(n\varphi+\psi_m)}$$
(3.2.10)

$$\tilde{E}_{r\varphi}(\rho,\varphi) = ij\beta \frac{aA}{2u}\left(\frac{uJ_n(u)}{wK_n(w)}\right)^{r-1}\left[(1-s)\frac{J_{n-1}^{2-r}(u\rho/a)}{K_{n-1}^{1-r}(w\rho/a)} - (1+s)e^{j\pi r}\frac{J_{n+1}^{2-r}(u\rho/a)}{K_{n+1}^{1-r}(w\rho/a)}\right]e^{-i(n\varphi+\psi_m)}$$
(3.2.11)

$$\tilde{E}_{rz}(\rho,\varphi) = A\frac{J_n^{r-1}(u)}{K_n^{r-1}(w)}\frac{J_n^{2-r}(u\rho/a)}{K_n^{1-r}(w\rho/a)}e^{-i(n\varphi+\psi_m)}$$
(3.2.12)

$$\tilde{H}_{r\rho}(\rho,\varphi) = -ij\omega_0\varepsilon_r \frac{aA}{2u}\left(\frac{uJ_n(u)}{wK_n(w)}\right)^{r-1}\left[(1-s_r)\frac{J_{n-1}^{2-r}(u\rho/a)}{K_{n-1}^{1-r}(w\rho/a)} - (1+s_r)e^{j\pi r}\frac{J_{n+1}^{2-r}(u\rho/a)}{K_{n+1}^{1-r}(w\rho/a)}\right]e^{-i(n\varphi+\psi_m)}$$
(3.2.13)

$$\tilde{H}_{r\varphi}(\rho,\varphi) = -j\omega_0\varepsilon_r \frac{aA}{2u}\left(\frac{uJ_n(u)}{wK_n(w)}\right)^{r-1}\left[(1-s_r)\frac{J_{n-1}^{2-r}(u\rho/a)}{K_{n-1}^{1-r}(w\rho/a)} + (1+s_r)e^{j\pi r}\frac{J_{n+1}^{2-r}(u\rho/a)}{K_{n+1}^{1-r}(w\rho/a)}\right]e^{-i(n\varphi+\psi_m)}$$
(3.2.14)

$$\tilde{H}_{rz}(\rho,\varphi) = -i\frac{A\beta s}{\omega_0\mu_0}\frac{J_n^{r-1}(u)}{K_n^{r-1}(w)}\frac{J_n^{2-r}(u\rho/a)}{K_n^{1-r}(w\rho/a)}e^{-i(n\varphi+\psi_m)}$$
(3.2.15)

Comparing the complex expressions (3.2.1-6) to the bi-complex (3.2.10-15) ones, it is seen that the latter are more explicit, and just as compact without the use of new variables (3.2.7, 8), or sacrificing font-size. They can clearly be rendered even more compact with the use of the new variables, and the short-form coordinate $\zeta$. Okamoto's original expressions for any $\xi$-component (3.1-12), in either region $r'$, may be reproduced using the following operations

$$V_\xi(\zeta) = \delta_{rr'}\,\text{Re}_i\,\tilde{U}_{r\xi}(\zeta); \quad \xi \in \{\rho,\varphi,z\},\ r' \in \{1,2\},\ \rho \in [a,b].$$
(3.2.16)

The Re-operation allows the recovery of the generalized complex expressions (3.2.1-6).

A generalization of the component expressions in terms of **a single generating function** is also possible. The longitudinal components are first re-expressed, with the help of the well-known Bessel function recurrence relations, valid for $n \geq 1$,

$$\frac{2n}{\sigma} J_n(\sigma) = J_{n+1}(\sigma) + J_{n-1}(\sigma),$$
$$\frac{2n}{\tau} K_n(\tau) = K_{n+1}(\tau) - K_{n-1}(\tau), \tag{3.2.17}$$

yielding for the radial multiplier common to both *z*-components of the EM-field

$$\frac{J_n^{2-r}(u\rho/a)}{K_n^{1-r}(w\rho/a)} = -\lambda_r \frac{(u\rho/2an)^{2-r}}{(w\rho/2an)^{1-r}} \left[ \frac{J_{n-1}^{2-r}(u\rho/a)}{K_{n-1}^{1-r}(w\rho/a)} - \lambda_r \frac{J_{n+1}^{2-r}(u\rho/a)}{K_{n+1}^{1-r}(w\rho/a)} \right] \tag{3.2.18}$$

and after a simplification and a re-arrangement of the multiplicative factor,

$$\frac{J_n^{2-r}(u\rho/a)}{K_n^{1-r}(w\rho/a)} = -\frac{\rho \lambda_r}{a^2 n} \frac{w^{2r-2}}{u^{2r-4}} \frac{a}{2u} \frac{u^{r-1}}{w^{r-1}} \left[ \frac{J_{n-1}^{2-r}(u\rho/a)}{K_{n-1}^{1-r}(w\rho/a)} - \lambda_r \frac{J_{n+1}^{2-r}(u\rho/a)}{K_{n+1}^{1-r}(w\rho/a)} \right]. \tag{3.2.19}$$

Moreover, using (3.1.13, 18), it is found that

$$s = s_r \frac{\omega_0^2 \mu_0 \varepsilon_r}{\beta^2}. \tag{3.2.20}$$

Substituting (3.2.19, 20) into (3.2.3, 6) where appropriate, yields the *z*-components of the electric and magnetic fields in the alternative forms of

$$E_{rz}(\zeta) = -\frac{\rho \lambda_r w^{2r-2}}{a^2 n u^{2r-4}} \frac{aA}{2u} \left( \frac{uJ_n(u)}{wK_n(w)} \right)^{r-1} \left[ \frac{J_{n-1}^{2-r}(u\rho/a)}{K_{n-1}^{1-r}(w\rho/a)} - \lambda_r \frac{J_{n+1}^{2-r}(u\rho/a)}{K_{n+1}^{1-r}(w\rho/a)} \right] \cos\phi_m$$

$$H_{rz}(\zeta) = \frac{\omega_0 \varepsilon_r}{\beta} \frac{\rho \lambda_r s_r w^{2r-2}}{a^2 n u^{2r-4}} \frac{aA}{2u} \left( \frac{uJ_n(u)}{wK_n(w)} \right)^{r-1} \left[ \frac{J_{n-1}^{2-r}(u\rho/a)}{K_{n-1}^{1-r}(w\rho/a)} - \lambda_r \frac{J_{n+1}^{2-r}(u\rho/a)}{K_{n+1}^{1-r}(w\rho/a)} \right] \sin\phi_m$$

(3.2.21)

In this form, they are now more amenable to generalizations. Examining the transverse components (3.2.1, 2, 4, 5) together with the above *z*-components, it is found that the following expression, termed **the generating function**, is common to **all** components,

$$\Lambda_{nr}(\rho; \eta, \lambda_r) = \frac{aA}{2u} \left( \frac{uJ_n(u)}{wK_n(w)} \right)^{r-1} \left[ (1-\eta) \frac{J_{n-1}^{2-r}(u\rho/a)}{K_{n-1}^{1-r}(w\rho/a)} + \lambda_r (1+\eta) \frac{J_{n+1}^{2-r}(u\rho/a)}{K_{n+1}^{1-r}(w\rho/a)} \right],$$
$$\lambda_r = e^{j\pi r}, \quad \eta \in \{s, s_r, 0\}; \quad r \in \{1, 2\}.$$

(3.2.22)

**The case for $\eta = s$ corresponds to the transverse E-field, that for $\eta = s_r$, to the transverse H-field, and that for $\eta = 0$, to the z-components.** Combining the transverse (3.2.1, 2, 4, 5) and the longitudinal (3.2.21) components together, with (3.2.22) substituted where appropriate, the EM-field components in the **complex formulation** simplify to the following expressions

$$E_{r\rho}(\rho,\varphi) = -j\beta \Lambda_{nr}(\rho;s,\lambda_r)\cos(n\varphi+\psi_m) \tag{3.2.23}$$

$$E_{r\varphi}(\rho,\varphi) = j\beta \Lambda_{nr}(\rho;s,-\lambda_r)\sin(n\varphi+\psi_m) \tag{3.2.24}$$

$$E_{rz}(\rho,\varphi) = -\beta \frac{\rho \lambda_r w^{2r-2}}{a^2 n\beta u^{2r-4}} \Lambda_{nr}(\rho;0,-\lambda_r)\cos(n\varphi+\psi_m) \tag{3.2.25}$$

$$H_{r\rho}(\rho,\varphi) = -j\omega_0\varepsilon_r \Lambda_{nr}(\rho;s_r,-\lambda_r)\sin(n\varphi+\psi_m) \tag{3.2.26}$$

$$H_{r\varphi}(\rho,\varphi) = -j\omega_0\varepsilon_r \Lambda_{nr}(\rho;s_r,\lambda_r)\cos(n\varphi+\psi_m) \tag{3.2.27}$$

$$H_{rz}(\rho,\varphi) = \omega_0\varepsilon_r \frac{\rho \lambda_r s_r w^{2r-2}}{a^2 n\beta u^{2r-4}} \Lambda_{nr}(\rho;0,-\lambda_r)\sin(n\varphi+\psi_m) \tag{3.2.28}$$

The component expressions are now in their most compact form. However, they are not in the same elegant form presented by Okamoto, as they are no longer transparently in terms of Bessel functions.

The **bi-complex versions** are obtained by applying either CTB transform (3.1.21) or (3.1.23), to (3.2.23-28), with the result

$$\tilde{E}_{r\rho}(\rho,\varphi) = -j\beta \Lambda_{nr}(\rho;s,\lambda_r)e^{-i(n\varphi+\psi_m)} \tag{3.2.29}$$

$$\tilde{E}_{r\varphi}(\rho,\varphi) = ij\beta \Lambda_{nr}(\rho;s,-\lambda_r)e^{-i(n\varphi+\psi_m)} \tag{3.2.30}$$

$$\tilde{E}_{rz}(\rho,\varphi) = -\beta \frac{\rho \lambda_r w^{2r-2}}{a^2 n\beta u^{2r-4}} \Lambda_{nr}(\rho;0,-\lambda_r)e^{-i(n\varphi+\psi_m)} \tag{3.2.31}$$

$$\tilde{H}_{r\rho}(\rho,\varphi) = -ij\omega_0\varepsilon_r \Lambda_{nr}(\rho;s_r,-\lambda_r)e^{-i(n\varphi+\psi_m)} \tag{3.2.32}$$

$$\tilde{H}_{r\varphi}(\rho,\varphi) = -j\omega_0\varepsilon_r \Lambda_{nr}(\rho;s_r,\lambda_r)e^{-i(n\varphi+\psi_m)} \tag{3.2.33}$$

$$\tilde{H}_{rz}(\rho,\varphi) = i\omega_0\varepsilon_r \frac{\rho \lambda_r s_r w^{2r-2}}{a^2 n\beta u^{2r-4}} \Lambda_{nr}(\rho;0,-\lambda_r)e^{-i(n\varphi+\psi_m)} \tag{3.2.34}$$

All these expressions are still in the format of a list or a look-up table, like Okamoto's original nomenclature (3.1-12), although there are now 6 compact expressions, instead of 12 elaborate ones. This may be the preferred approach, but alternatives are now explored.

Upon examination of the transverse (3.2.1, 2, 4, 5) and the longitudinal (3.2.21) components together again, which are in **the complex formulation**, it is deduced that a general *explicit* expression for *any* component of the EM-field of a hybrid mode, for either region of the fiber's cross-section, can be efficiently given by just **a pair of scalar equations**, one for the electric field, and another, for the magnetic field, and **presented here in compact form by adopting a quotient configuration:**

$$E_{r\xi}(\rho,\varphi) = \frac{\left[(1-s+s\delta_{\xi z})\frac{J_{n-1}^{2-r}(u\rho/a)}{K_{n-1}^{1-r}(w\rho/a)} + \frac{(1+s-s\delta_{\xi z})e^{j\pi r}}{\delta_{\xi\rho}-\delta_{\xi\varphi}-\delta_{\xi z}}\frac{J_{n+1}^{2-r}(u\rho/a)}{K_{n+1}^{1-r}(w\rho/a)}\right]\cos(n\varphi+\psi_m)}{\frac{2ju}{aA\beta}\left(\frac{uJ_n(u)}{wK_n(w)}\right)^{1-r}\left[\delta_{\xi\rho}-\delta_{\xi\varphi}\cot(n\varphi+\psi_m)+jna^2\beta\frac{u^{2r-4}e^{j\pi r}}{\rho w^{2r-2}}\delta_{\xi z}\right]}$$

$$H_{r\xi}(\rho,\varphi) = \frac{\left[(1-s_r+s_r\delta_{\xi z})\frac{J_{n-1}^{2-r}(u\rho/a)}{K_{n-1}^{1-r}(w\rho/a)} - \frac{(1+s_r-s_r\delta_{\xi z})e^{j\pi r}}{\delta_{\xi\rho}-\delta_{\xi\varphi}+\delta_{\xi z}}\frac{J_{n+1}^{2-r}(u\rho/a)}{K_{n+1}^{1-r}(w\rho/a)}\right]\sin(n\varphi+\psi_m)}{\frac{2ju}{aA\omega_0}\left(\frac{uJ_n(u)}{wK_n(w)}\right)^{1-r}\left[\delta_{\xi\rho}+\delta_{\xi\varphi}\tan(n\varphi+\psi_m)-jna^2\beta\frac{u^{2r-4}e^{j\pi r}}{\rho s_r w^{2r-2}}\delta_{\xi z}\right]}\varepsilon_r$$

(3.2.35)

with $\xi \in \{\rho,\varphi,z\}$ as before. **The 2 scalar equations may replace all 12 equations of Okamoto's original nomenclature (3.1.1-12), or the 6 regionally dependent equations (3.2.1-6) of the new nomenclature, dependent on whether *r* = 1 or 2**. In order to recover any of (3.2.1-6) for instance, the relevant $\xi$ is substituted into (3.2.35), which are then evaluated *term-wise* beginning with their numerators, followed by a similar evaluation of their denominators. For instance, setting $\xi = \varphi$ results in the immediate resolution to unity of $\delta_{\xi\varphi}$, and the simultaneous extinction of any term carrying either $\delta_{\xi\rho}$ or $\delta_{\xi z}$. Either of the 2 expressions may also be decomposed into a sum of smaller quotients, but the result would be far more cumbersome than (3.2.35). These generalized expressions are the largest yet, as they must each represent 6 of the 12 components of the EM-field, before any approximation. However, alternative, although less-explicit, but more *vertically compact* versions of (3.2.35) are presented in **Appendix B**.

Simpler, and more compact forms of the above scalar equations are also possible, using (3.2.35) and the generating function (3.2.22), although the resultant equations would no longer transparently retain Snitzer's original Bessel function dependence [2],

$$E_{r\xi}(\rho,\varphi) = -j\beta \frac{\Lambda_{nr}\left[\rho; s(1-\delta_{\xi z}), \lambda_r(\delta_{\xi\rho}-\delta_{\xi\varphi}-\delta_{\xi z})\right]\cos(n\varphi+\psi_m)}{\delta_{\xi\rho}-\delta_{\xi\varphi}\cot(n\varphi+\psi_m)+jna^2\beta\frac{u^{2r-4}\lambda_r}{\rho w^{2r-2}}\delta_{\xi z}}$$

$$H_{r\xi}(\rho,\varphi) = -j\omega_0\varepsilon_r \frac{\Lambda_{nr}\left[\rho; s_r(1-\delta_{\xi z}), -\lambda_r(\delta_{\xi\rho}-\delta_{\xi\varphi}+\delta_{\xi z})\right]\sin(n\varphi+\psi_m)}{\delta_{\xi\rho}+\delta_{\xi\varphi}\tan(n\varphi+\psi_m)-jna^2\beta\frac{u^{2r-4}\lambda_r}{\rho s_r w^{2r-2}}\delta_{\xi z}}$$

(3.2.36)

In the **bi-complex formulation**, the generalized EM-field phasors are given by

$$\tilde{E}_{r\xi}(\rho,\varphi) = \frac{\left[(1-s+s\delta_{\xi z})\dfrac{J_{n-1}^{2-r}(u\rho/a)}{K_{n-1}^{1-r}(w\rho/a)} + \dfrac{(1+s-s\delta_{\xi z})e^{j\pi r}}{\delta_{\xi\rho}-\delta_{\xi\varphi}-\delta_{\xi z}}\dfrac{J_{n+1}^{2-r}(u\rho/a)}{K_{n+1}^{1-r}(w\rho/a)}\right]e^{-i(n\varphi+\psi_m)}}{\dfrac{2ju}{aA\beta}\left(\dfrac{uJ_n(u)}{wK_n(w)}\right)^{1-r}\left[\delta_{\xi\rho}+i\delta_{\xi\varphi}+jna^2\beta\dfrac{u^{2r-4}e^{j\pi r}}{\rho w^{2r-2}}\delta_{\xi z}\right]}$$

$$\tilde{H}_{r\xi}(\rho,\varphi) = \frac{\left[(1-s_r+s_r\delta_{\xi z})\dfrac{J_{n-1}^{2-r}(u\rho/a)}{K_{n-1}^{1-r}(w\rho/a)} - \dfrac{(1+s_r-s_r\delta_{\xi z})e^{j\pi r}}{\delta_{\xi\rho}-\delta_{\xi\varphi}+\delta_{\xi z}}\dfrac{J_{n+1}^{2-r}(u\rho/a)}{K_{n+1}^{1-r}(w\rho/a)}\right]\varepsilon_r ie^{-i(n\varphi+\psi_m)}}{\dfrac{2ju}{aA\omega_0}\left(\dfrac{uJ_n(u)}{wK_n(w)}\right)^{1-r}\left[\delta_{\xi\rho}+i\delta_{\xi\varphi}-jna^2\beta\dfrac{u^{2r-4}e^{j\pi r}}{\rho s_r w^{2r-2}}\delta_{\xi z}\right]}$$

(3.2.37)

with $\xi \in \{\rho,\varphi,z\}$ as before. For any cylindrical coordinate component $\xi$, it can be seen that the 2 equations are in quadrature with respect to $\varphi$ due to the presence of the multiplicative imaginary number (i) in the numerator of the magnetic field component, which is absent in that of the electric field[1]. These bi-complex expressions, besides being separable with respect to $\rho$ and $\varphi$, and functionally quite similar, are evidently also more compact than their complex counterparts (3.2.35), due to their extensive use of trigonometric functions.

Using (3.2.22), more compact versions of (3.2.37) are deduced to be

$$\tilde{E}_{r\zeta}(\rho,\varphi) = -j\beta\frac{\Lambda_{nr}\left[\rho;s(1-\delta_{\xi z}),\lambda_r(\delta_{\xi\rho}-\delta_{\xi\varphi}-\delta_{\xi z})\right]}{\delta_{\xi\rho}+i\delta_{\xi\varphi}+jna^2\dfrac{\beta w^2}{u^4}\dfrac{\lambda_r u^{2r}}{\rho w^{2r}}\delta_{\xi z}}e^{-i(n\varphi+\psi_m)}$$

$$\tilde{H}_{r\zeta}(\rho,\varphi) = -j\omega_0\varepsilon_r\frac{\Lambda_{nr}\left[\rho;s_r(1-\delta_{\xi z}),-\lambda_r(\delta_{\xi\rho}-\delta_{\xi\varphi}+\delta_{\xi z})\right]}{\delta_{\xi\rho}+i\delta_{\xi\varphi}-jna^2\dfrac{\beta w^2}{s_r u^4}\dfrac{\lambda_r u^{2r}}{\rho w^{2r}}\delta_{\xi z}}ie^{-i(n\varphi+\psi_m)}$$

(3.2.38)

Any component in either of the 2 regions of the fiber, and in either formulation (3.2.35-36 or 3.2.37-38), is operationally recovered using the relation

$$W_{r'\xi}(\zeta) = \delta_{rr'}W_{r\xi}(\zeta); \quad W_{r\xi} \in \{\tilde{U}_{r\xi},V_{r\xi}\}, \ r' \in \{1,2\}, \ \xi \in \{\rho,\varphi,z\} \quad (3.2.39)$$

with *W* being a generic EM-field component, which could be either *V*, a scalar described in the complex formulation, (3.2.35, 36), or $\tilde{U}$, a phasor in the bi-complex formulation (3.2.37, 38).

---

[1] **A multiplication by i represents a counter-clockwise rotation by π/2 radians in the complex plane**

## 3.3 Bi-regional generalizations using distributions

Knowing the expressions for the EM-field in the core and the cladding, its description over the entire core geometry, is in Okamoto's original nomenclature [1], given by the list

$$W_\xi(\zeta) = \begin{cases} W_{1\xi}(\zeta), & \forall \rho \in [0,a] \\ W_{2\xi}(\zeta), & \forall \rho \in (a,b] \end{cases}, \quad \xi \in \{\rho, \varphi, z\} \tag{3.3.1}$$

where $W_\xi$ is a generic EM-field component, and could be either $V$, a vector component described in the complex formulation, or for this report, $\tilde{U}$, a phasor component in the bi-complex formulation. **The domains of validity in (3.3.1) are mutually exclusive**. The boundary condition for the tangential components, which include the azimuthal and the longitudinal components for the electric and the magnetic fields, in addition to the radial components for the latter (assuming non-magnetic constitutive media), is given by

$$W_{1\xi}(a,\varphi) = W_{2\xi}(a,\varphi). \tag{3.3.2}$$

In this section, a more analytical approach is pursued that can lead to more compact expressions. It will be found possible to obtain an expression for the EM-field over the entire cross-section of the ideal fiber as the sum of its components in the core and in the cladding, as

$$W_\xi(\zeta) = W_{1\xi}(\zeta) f_1(\rho) + W_{2\xi}(\zeta) f_2(\rho) = \sum_{r=1}^{2} W_{r\xi}(\zeta) f_r(\rho). \tag{3.3.3}$$

This is achieved using the well-known Heaviside step-function (which is henceforth termed the step-function). **However, it should be stressed that this approach also requires the generalization of the 2 regional components to a single compact expression $W_{r\xi}(\zeta)$ as in the summand on the RHS of (3.3.3), without which any desired compactness is not attainable.**

On the real number line or $\mathbb{R}$, the original definition of **the step-function** due to O. Heaviside [9], is given by

$$\mathrm{H}(t - t_0) = \begin{cases} 0, & t < t_0 \\ 1, & t \geq t_0 \end{cases} \tag{3.3.4}$$

and **should not be confused with a magnetic field component *H* (which is *italicized*)**. In other definitions [9], the function is only non-zero if its argument is greater than zero, instead of being greater than or equal to zero as in (3.3.4). In the above definition however, the function is being assumed to be rightward-continuous, by regularization. A step-function is locally integrable[2] everywhere, and its integral is possible under either the strict Riemann integration, or in the more general Lebesgue integration. By contrast,

---

[2] A "locally integrable" function is one for which the **integral of its absolute value** is finite over any compact set within its domain

differentiation is not tolerant of discontinuities in a function: since the step-function is discontinuous, it has no general Leibniz derivative valid over its entire domain, which can only exists if a function is locally continuous[3] *everywhere*. However, the Schwartz distribution theory (SDT) [10] makes it possible to differentiate functions such as the step-function. Whereas a *general* function $g$ is defined by the map $g: \mathbb{R}^n \to \mathbb{R}$, a distribution $f$ in the SDT is defined as the map $f: \psi \to \mathbb{R}$, or by its action on a test-function[4] $\psi \in \mathcal{C}_0^\infty(\Omega)$ over an open set $\Omega \subset \mathbb{R}^n$, which is explicitly stated as[5]

$$f: \psi \to \int_\Omega f(\mathbf{r})\psi(\mathbf{r})\mathrm{d}^n\mathbf{r} = \langle f, \psi \rangle \qquad (3.3.5)$$

which also applies to *any* locally integrable function on $\Omega$. In SDT, the map is also expressed as $\mathcal{D}(\Omega) \to \mathcal{D}'(\Omega)$ or from the vector-space of smooth functions, to the vector-space of Schwartz distributions, which is the topological dual of the former [12]. **In $\mathbb{R}$, for instance**, using (3.3.5), for the derivative of the step-function (3.3.4) under SDT,

$$\langle \mathrm{H}', \psi \rangle = \int_{-\infty}^{\infty} \mathrm{H}'(t-t_0)\psi(t)\mathrm{d}t = \left[\mathrm{H}(t-t_0)\psi(t)\right]_{-\infty}^{\infty} - \int_{-\infty}^{\infty} \mathrm{H}(t-t_0)\psi'(t)\mathrm{d}t = -\int_{t_0}^{\infty}\psi'(t)\mathrm{d}t = \psi(t_0)$$
(3.3.5.1)

which is resolved using Taylor's integration by parts (IBP)[6]. **It is thus concluded that the action of the derivative of the step-function on a test-function, results in the evaluation of that test-function at the discontinuity of the step-function, assuming that $t_0 \in \Omega$**. The result leads to the Dirac delta-function.

The Dirac delta-function is defined according to P.A.M. Dirac as [13],

$$\delta(t-t_0) = \begin{cases} 0, & \text{if } t \neq t_0 \\ \infty, & \text{if } t = t_0 \end{cases} \qquad (3.3.6)$$

**The Dirac delta-function should not be confused with the Kronecker delta defined by (3.1.22), whose argument in this report, is consistently expressed as a subscript, instead of with round parentheses as in (3.3.6)**. Under Lebesgue or Lebesgue-Stieltjes (but not Riemann) integration, the delta-function is defined by its action on a test-function in accordance with (3.3.5), yielding

$$\langle \delta, \psi \rangle = \int_{-\infty}^{\infty}\delta(t-t_0)\psi(t)\mathrm{d}t = \psi(t_0)\int_{-\infty}^{\infty}\delta(t-t_0)\mathrm{d}t = \psi(t_0). \qquad (3.3.6.1)$$

It is thus concluded that, **in the distributional sense** (3.3.5), the derivative of the step-function is equivalent to the Dirac delta-function,

$$\mathrm{H}'(t-t_0) = \delta(t-t_0) \qquad (3.3.6.2)$$

---

[3] For a function to be "locally continuous" at a given point within its domain, there must exist an open neighborhood of arbitrary size around that point, where the function is continuous when restricted to that neighborhood
[4] $\mathcal{C}_0^\infty$ is the vector space of infinitely differentiable functions, of non-void compact support within $\Omega$, outside of which, they vanish [11]. Such functions $\psi$ are also termed "smooth functions"
[5] Equation (3.3.5) is also alternatively expressed as $(g, \psi)$ in mathematics, or as $\langle g^* | \psi \rangle$ in physics
[6] IBP requires the derivative of one of the quantities in the integrand. However, in order to avoid a Leibniz derivative of the distribution, this derivative is usually carried out on the test-function, unless a distributional derivative is available

which is valid when the step-function is defined as (3.3.4). This notation should be construed here as short-form for the **distributional derivative** but *not* the Leibniz derivative, since it is traditionally reserved for the latter.

In polar coordinates, the radial step-function can be adapted from the one-dimensional step-function (3.3.4) to describe the $\mathbb{R}^2$-region outside a circle of radius $\rho_0$, but inclusive of its periphery. It is expressed as the outwardly continuous function

$$\mathrm{H}(\rho - \rho_0) = \begin{cases} 0, & \rho < \rho_0 \\ 1, & \rho \geq \rho_0 \end{cases} ; \quad \rho \geq 0, \tag{3.3.7}$$

regardless of the value of the azimuth $\varphi$. **In this definition, the polar variable $\rho$ is implicitly restricted to being greater than or equal to zero as shown above, since it otherwise has no physical meaning in cylindrical coordinates.**

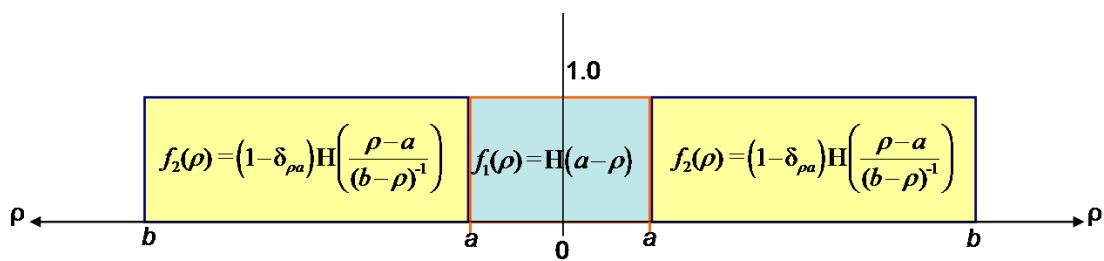

**Fig. 2.** An illustration of the two orthogonal functions (3.3.8, 9) in polar coordinates, with the azimuth-plane taken to be normal to the page. Note that $f_1$ terminates at $\rho = a$, whereas $f_2$ begins at $\rho > a$. Furthermore, $b \gg a$ in practice, which is not reflected in the schematic.

Since the radial domain is restricted *a priori* to be greater than or equal to zero, the core of the fiber can be represented geometrically by a single step-function,

$$f_1(\rho) = \mathrm{H}(a - \rho), \tag{3.3.8}$$

whose argument is a linear function of the radial coordinate. The annular cladding, is defined by an inner radius $a$, which is the radius of the core, and an outer radius $b \gg a$. It can also be represented by a *single* step-function, but in order to account for both its inner and outer radii, the argument of the step-function would have to be non-linear. The most compact representation of the annular cladding is expressed by a step-function, whose argument is a product of two linear functions, **each of which, must be greater than or equal to zero**, in consistence with Heaviside's original definition of the step-function (3.3.4). It is expressible in one of two possible ways, as follows[7],

$$f_2(\rho) = \left(1 - \delta_{\rho a}\right) \mathrm{H}\big((\rho - a)(b - \rho)\big) = \mathrm{H}\big((\rho - a)(b - \rho)\big) - \delta_{\rho a}. \tag{3.3.9}$$

The Kronecker delta is used to ensure that the function excludes the boundary $\rho = a$, in accordance with Okamoto's definition (3.3.1). **The expression effectively represents**

---

[7]In Matlab for instance, (3.3.9) can be easily expressed as ((rho - a).*(b - rho) - (rho == a)) >= 0, where rho is a Matlab vector of values [0: 0.01: 5*b], that represents the radial variable $\rho$ in cylindrical coordinates

a rectangle- or a *gate-function* [9] that limits ρ to (*a*, *b*)**. An equivalence that will prove useful later, is the following decomposition,

$$\mathrm{H}((\rho-a)(b-\rho)) = \mathrm{H}(\rho-a) - \mathrm{H}(\rho-b) + \delta_{\rho b}; \quad a < b. \tag{3.3.10}$$

The Kronecker delta is required to ensure that the RHS resolves to unity for $\rho = b$, in accordance with (3.3.9). A gate-function may also be constructed using 2 step-functions in a multiplicative relation, or by using the absolute value within the argument of a single step-function, both of which, however, lead to less compact expressions. The 2 functions are shown in **fig. 2**, and can be generalized to the regionally dependent function

$$f_r(\rho) = \left(2^{2-r} - \delta_{\rho a}^{r-1}\right) \mathrm{H}\left(\frac{a-\rho}{(\rho-b)^{1-r}}\right); \quad r \in \{1,2\}. \tag{3.3.11}$$

Any component of the EM-field of a hybrid mode may now be analytically expressed over the entire cross-section of the fiber as

$$W_\xi(\zeta) = W_{1\xi}(\zeta)\mathrm{H}(a-\rho) + W_{2\xi}(\zeta)(1-\delta_{\rho a})\mathrm{H}((\rho-a)(b-\rho)). \tag{3.3.12}$$

To verify the domains of validity (3.3.1), (3.3.12) is evaluated at $\rho = a$,

$$W_\xi(a,\varphi) = W_{1\xi}(a,\varphi)\mathrm{H}(0) + W_{2\xi}(a,\varphi)(1-\delta_{aa})\mathrm{H}(0). \tag{3.3.13}$$

Since the step-function resolves to unity for arguments greater than or equal to zero in accordance with its definition (3.3.4), the 1st term is extant, but the 2nd term is extinguished due to the Kronecker delta term, so that

$$W_\xi(a,\varphi) = W_{1\xi}(a,\varphi). \tag{3.3.14}$$

On the other hand, evaluating (3.3.12) for some $\rho' \in (a,b)$,

$$W_\xi(\rho',\varphi) = W_{1\xi}(\rho',\varphi)\mathrm{H}(-|\rho'-a|) + W_{2\xi}(\rho',\varphi)(1-\delta_{\rho'a})\mathrm{H}((\rho'-a)(b-\rho')). \tag{3.3.15}$$

Consequently, the 1st term vanishes, whereas the 2nd term survives, although with the extinction of the Kronecker delta, leading to

$$W_\xi(\rho',\varphi) = W_{2\xi}(\rho',\varphi). \tag{3.3.16}$$

Lastly, it should be noted that without the Kronecker delta bracket used in (3.3.12),

$$W_\xi(a,\varphi) = W_{1\xi}(a,\varphi) + W_{2\xi}(a,\varphi) \tag{3.3.17}$$

and it is seen that its omission has led to a result that violates Okamoto's boundary conditions (3.3.2).

It can be shown by integration, that the area in polar coordinates due to the step-function (3.3.9) is indeed that of the annular cladding, as follows:

$$\int_0^{2\pi}\int_0^{\infty}\left[H\left(\frac{a-\rho}{(\rho-b)^{-1}}\right)-\delta_{\rho a}\right]\rho\,d\rho\,d\varphi = \int_0^{2\pi}\int_a^b \rho\,d\rho\,d\varphi - \int_0^{2\pi}\int_0^{\infty}\delta_{\rho a}\,\rho\,d\rho\,d\varphi. \qquad (3.3.18)$$

After a simplification,

$$\int_0^{2\pi}\int_0^{\infty}\left[H\left(\frac{a-\rho}{(\rho-b)^{-1}}\right)-\delta_{\rho a}\right]\rho\,d\rho\,d\varphi = \pi(b^2-a^2) - 2\pi a\int_0^{\infty}\delta_{\rho a}\,d\rho \qquad (3.3.18.1)$$

whereas the integral of the Kronecker delta, which is possible under Lebesgue integration, resolves to

$$\int_0^{\infty}\delta_{\rho a}\,d\rho = 0, \qquad (3.3.18.2)$$

but is stated here without a proof[8]. The area integral (3.3.18) may also be carried out using IBP, as follows,

$$\int_0^{2\pi}\int_0^{\infty}H\left(\frac{a-\rho}{(\rho-b)^{-1}}\right)\rho\,d\rho\,d\varphi = \left[\pi\rho^2 H\left(\frac{a-\rho}{(\rho-b)^{-1}}\right)\right]_0^{\infty} - \pi\int_0^{\infty}\rho^2\frac{d}{d\rho}H\left(\frac{a-\rho}{(\rho-b)^{-1}}\right)d\rho. \qquad (3.3.19)$$

Now, the first RHS term vanishes, since the step-function evaluates to zero at the bracket bounds. As for the remaining integral, it can be resolved in a distributional approach using the relations (3.3.6.1, 2), and the equivalence (3.3.10), yielding under Lebesgue integration,

$$-\pi\int_0^{\infty}\rho^2\frac{d}{d\rho}H\left(\frac{a-\rho}{(\rho-b)^{-1}}\right)d\rho = -\pi\int_0^{\infty}\rho^2\delta(\rho-a)\,d\rho + \pi\int_0^{\infty}\rho^2\delta(\rho-b)\,d\rho = \pi(b^2-a^2), \qquad (3.3.19.1)$$

since both *a* and *b* are bigger than zero, whereas the derivative of the Kronecker delta vanishes (see **APPENDIX A**). Regardless of which approach is used, (3.3.18) correctly reduces to the surface area of an annular cladding of inner radius *a* and outer radius *b*:

$$\int_0^{2\pi}\int_0^{\infty}\left[H((a-\rho)(\rho-b))-\delta_{\rho a}\right]\rho\,d\rho\,d\varphi = \pi(b^2-a^2). \qquad (3.3.19.2)$$

---

[8] It can be shown at WolframAlpha.com that the text input "integrate Kroneckerdelta[rho,4] rho drho from rho = 0 to inf" **resolves to zero**. The radius *a* is arbitrary, but is set to the value of 4 (which is typical for a SMF) in this instance, to allow for the numerical resolution of the integral. Another approach can also be found in **APPENDIX A**

A step-function is reproduced for integral powers, since it is finite everywhere,

$$f_r^2(\rho) = f_r(\rho), \quad r \in \{1, 2\} \tag{3.3.20}$$

as the point-wise product of 2 *bounded* functions of identical domains is always valid, regardless of discontinuities. However, the LHS can also be construed as a tensor product ($\otimes$) of distributions, and is sometimes termed a hyper or a non-linear distribution. Although the LHS of (3.3.20) is valid as a distribution in the Schwartz space $\mathcal{D}'$ [14], it is not possible to preserve the validity of the above relation within the context of differentiation, whether as Leibniz, or in the distributional sense, in SDT. This may be demonstrated by taking the distributional derivative of the LHS of (3.3.20), using (3.3.5),

$$\langle 2HH', \psi \rangle = 2\int_{-\infty}^{\infty} H'(t-t_0)H(t-t_0)\psi(t)dt = -2\langle H', H\psi \rangle = 2\langle \delta, H\psi \rangle = 2H(0)\psi(t_0) = 2\psi(t_0) \tag{3.3.20.1}$$

which along with a differentiation of the RHS of (3.3.20), using (3.3.5.1), leads to

$$2\psi(t_0) = \psi(t_0), \tag{3.3.20.2}$$

which is a self-contradiction. Therefore, although (3.3.20) is valid in the distributional sense [14 - 16], its utility is not without constraints. In fact, tensor products of distributions such as $\delta^2$, or in functional compositions such as $e^\delta$, are impermissible under SDT, *with very few exceptions*: For instance, **under Lebesgue integration, and thus in SDT, the tensor-product of step-functions as an expression (3.3.20) is still *valid*, since its LHS and RHS differ at a single discontinuity that has a Lebesgue measure of zero** [15]. In SDT however, no associative, commutative *differential* algebra ($\mathcal{A}; \partial, +, \otimes$) that incorporates the space of distributions $\mathcal{D}'$, can also preserve the general product of continuous functions [17 - 19]. This constraint can be alleviated by appealing to the more complex, *differential* Colombeau Algebra $\mathcal{G}$ and its variants [15], which restrict the tensor product to the space of *smooth* functions $C_0^\infty$ instead of to that $C^0$ of *continuous* functions, as in the original SDT [10][9]. However, since functions are distinguished by their microscopic behavior in **such algebra**, the equality in (3.3.20) is only valid in the weak or associative sense, since the LHS and the RHS of (3.3.20) are macroscopically similar, but microscopically disparate about the discontinuity [10] [20]. Thus, **whereas the expression (3.3.20) is strictly valid under SDT but to the exclusion of Leibniz differentiation, it is *associatively* valid under the Colombeau Algebra, while being inclusive of such differentiation**. Fortunately here, the LHS of (3.3.20) only arises during power integral computations, which requires no differentiation along either of the transverse coordinates, nor any IBP that would usually involve at least one differentiation. Thus, the Colombeau Algebra should not be required for this work, although it is important to be cognizant of the limitations of (3.3.20).

---

[9] The former space of functions relates to that of infinitely differentiable functions; whereas that of the latter also includes continuous functions exhibiting abrupt changes such as $|t|$ or any of its functional compositions, none of which has a Leibniz derivative at $t = 0$

[10] This is usually explained by considering the step-function as an idealization of a parameterized, generalized smooth function $G$ (such as erfc($-(t - t_0)/\varepsilon$)/2, as one of many instances) that only approaches the step-function H($t - t_0$) in the limit of the parameter ($\varepsilon$). Therefore $G$ would be expected to behave differently about the location of the discontinuity ($t = t_0$), when raised to an integral power [20]. **By contrast, the SDT gives no consideration to such microscopic behavior**

In general, the square of the *regionally* dependent step-function (3.3.11) is given by

$$f_r^2(\rho) = \left(2^{2-r} - \delta_{\rho a}^{r-1}\right)^2 H^2\left(\frac{a-\rho}{(\rho-b)^{-1}}\right). \tag{3.3.21}$$

Since the point-wise multiplication of any function with itself, including piece-wise continuous functions like the step-function, is valid under SDT,

$$f_r^2(\rho) = \left(2^{2-r} - \delta_{\rho a}^{r-1}\right)^2 H\left(\frac{a-\rho}{(\rho-b)^{-1}}\right). \tag{3.3.22}$$

Furthermore, the following equivalence is also true

$$\left(2^{2-r} - \delta_{\rho a}^{r-1}\right)^2 = 2^{2-r} - \delta_{\rho a}^{r-1}. \tag{3.3.23}$$

Without expanding the square of the quantity, it is observed that since the parenthesized quantity on the LHS is either 0 or 1 for any combination of $r$ and $\rho$, it is then reproducible for any integral power. Thus, the equality (3.3.20) is verified for the regionally dependent step-function (3.3.11), and will be invoked later, for either $r = 1$ or 2.

The 2 step-functions (3.3.8, 9) also obey the multiplicative orthogonality

$$f_{r_1}(\rho) f_{r_2}(\rho) = f_{r_2}(\rho) \delta_{r_1 r_2}, \qquad r_1, r_2 \in \{1, 2\} \tag{3.3.24}$$

that is, given $f_{r_2}$, the product of $f_{r_1}$ with $f_{r_2}$ is only non-zero if $f_{r_1}$ is identical with $f_{r_2}$, or equivalently, when $r_1$ is identical with $r_2$. This is obtained with the help of (3.3.20). This relation will be useful in the derivation of the modal power flow in the fiber, which requires the cross-product of a mode's electric field vector with its magnetic field counterpart. The LHS of (3.3.24) also represents a product of step-functions, like (3.3.20), but its calculus is not problematic because the step-functions of (3.3.24) have disjoint supports [16]. The orthogonality relation (3.3.24) for the 2 step-functions (3.3.8, 9) yields

$$f_1(\rho) f_2(\rho) = H(a-\rho)\left[H\left(\frac{a-\rho}{(\rho-b)^{-1}}\right) - \delta_{\rho a}\right] = H(a-\rho) H\left(\frac{a-\rho}{(\rho-b)^{-1}}\right) - H(0) \delta_{\rho a}. \tag{3.3.25}$$

For the 1st term on the RHS, the 1st step-function in the product is unity over [0, *a*] inclusive of $\rho = a$, but vanishes for $\rho > a$, whereas the 2nd step-function is unity over [*a*, *b*] inclusive of $\rho = a$, but vanishes for $\rho < a$. Consequently, the product of the 2 step-functions is zero everywhere, except at $\rho = a$, which can be summarized by a Kronecker delta at this location. The product (3.3.25) thus simplifies to

$$f_1(\rho) f_2(\rho) = \delta_{\rho a} - H(0) \delta_{\rho a} = \delta_{\rho a} - \delta_{\rho a} = 0 \tag{3.3.26}$$

which validates the multiplicative orthogonality relation (3.3.24) for the 2 functions.

If desired, the regionally dependent components of the EM-field of a hybrid mode can be re-cast as bi-regional forms using (3.3.11). **In the complex formulation**, which is given by (3.2.1-6), application of (3.3.11) yields

$$E_\rho(\zeta) = \beta \frac{aA}{2u} \sum_{r=1}^{2} \frac{f_r(\rho)}{\mathrm{j}\sec\phi_m} \left(\frac{uJ_n(u)}{wK_n(w)}\right)^{r-1} \left[(1-s)\frac{J_{n-1}^{2-r}(u\rho/a)}{K_{n-1}^{1-r}(w\rho/a)} + \lambda_r(1+s)\frac{J_{n+1}^{2-r}(u\rho/a)}{K_{n+1}^{1-r}(w\rho/a)}\right]$$
(3.3.27)

$$E_\varphi(\zeta) = -\beta \frac{aA}{2u} \sum_{r=1}^{2} \frac{f_r(\rho)}{\mathrm{j}\cosec\phi_m} \left(\frac{uJ_n(u)}{wK_n(w)}\right)^{r-1} \left[(1-s)\frac{J_{n-1}^{2-r}(u\rho/a)}{K_{n-1}^{1-r}(w\rho/a)} - \lambda_r(1+s)\frac{J_{n+1}^{2-r}(u\rho/a)}{K_{n+1}^{1-r}(w\rho/a)}\right]$$
(3.3.28)

$$E_z(\zeta) = A \sum_{r=1}^{2} \frac{J_n^{r-1}(u)}{K_n^{r-1}(w)} \frac{J_n^{2-r}(u\rho/a)}{K_n^{1-r}(w\rho/a)} f_r(\rho)\cos\phi_m$$
(3.3.29)

$$H_\rho(\zeta) = \omega_0 \frac{aA}{2u} \sum_{r=1}^{2} \frac{\bar{\varepsilon}_r(\rho)}{\mathrm{j}\cosec\phi_m} \left(\frac{uJ_n(u)}{wK_n(w)}\right)^{r-1} \left[(1-s_r)\frac{J_{n-1}^{2-r}(u\rho/a)}{K_{n-1}^{1-r}(w\rho/a)} - \lambda_r(1+s_r)\frac{J_{n+1}^{2-r}(u\rho/a)}{K_{n+1}^{1-r}(w\rho/a)}\right]$$
(3.3.30)

$$H_\varphi(\zeta) = \omega_0 \frac{aA}{2u} \sum_{r=1}^{2} \frac{\bar{\varepsilon}_r(\rho)}{\mathrm{j}\sec\phi_m} \left(\frac{uJ_n(u)}{wK_n(w)}\right)^{r-1} \left[(1-s_r)\frac{J_{n-1}^{2-r}(u\rho/a)}{K_{n-1}^{1-r}(w\rho/a)} + \lambda_r(1+s_r)\frac{J_{n+1}^{2-r}(u\rho/a)}{K_{n+1}^{1-r}(w\rho/a)}\right]$$
(3.3.31)

$$H_z(\zeta) = -\frac{A\beta s}{\omega_0 \mu_0} \sum_{r=1}^{2} \frac{J_n^{r-1}(u)}{K_n^{r-1}(w)} \frac{J_n^{2-r}(u\rho/a)}{K_n^{1-r}(w\rho/a)} f_r(\rho)\sin\phi_m$$
(3.3.32)

The step function under the summations in the above equations may be generalized as

$$f_r(\rho) = \frac{\mathrm{H}\left((\rho-a)(b-\rho)^{r-1}\right)}{\Gamma\left(2^{2-r} - \delta_{\rho a}^{r-1}\right)} \quad ; \quad r \in \{1,2\},$$
(3.3.33)

which is in its most compact form, with the use of Euler's gamma function (see **APPENDIX B**). Another, useful form is found in §**5.1**. The more explicit, analytical form of the EM-field components would not have been possible with the original set of components (3.1.1-12) [1], which must cover each region of the fiber separately, due to a lack of compactness. In the transverse magnetic field components, the regional permittivity $\varepsilon_r$ has been combined with the general step-function $f_r(\rho)$ to form the *r*egionally dependent permittivity function $\bar{\varepsilon}_r(\rho)$:

$$\bar{\varepsilon}_r(\rho) = \varepsilon_r f_r(\rho).$$
(3.3.34)

A most compact, bi-regional version of the EM-field components (3.3.27-32) is also possible, which are in terms of the generating function (3.2.22):

$$E_\rho(\rho,\varphi) = -\mathrm{j}\beta \sum_{r=1}^{2} \Lambda_{nr}(\rho;s,\lambda_r) f_r(\rho)\cos(\varphi+\psi_m) \tag{3.3.35}$$

$$E_\varphi(\rho,\varphi) = \mathrm{j}\beta \sum_{r=1}^{2} \Lambda_{nr}(\rho;s,-\lambda_r) f_r(\rho)\sin(\varphi+\psi_m) \tag{3.3.36}$$

$$E_z(\rho,\varphi) = -\beta \sum_{r=1}^{2} \frac{\lambda_r w^{2r-2} \rho}{a^2 \beta n u^{2r-4}} \Lambda_{nr}(\rho;0,-\lambda_r) f_r(\rho)\cos(\varphi+\psi_m) \tag{3.3.37}$$

$$H_\rho(\rho,\varphi) = -\mathrm{j}\omega_0 \sum_{r=1}^{2} \Lambda_{nr}(\rho;s_r,-\lambda_r) \bar{\varepsilon}_r(\rho)\sin(\varphi+\psi_m) \tag{3.3.38}$$

$$H_\varphi(\rho,\varphi) = -\mathrm{j}\omega_0 \sum_{r=1}^{2} \Lambda_{nr}(\rho;s_r,\lambda_r) \bar{\varepsilon}_r(\rho)\cos(\varphi+\psi_m) \tag{3.3.39}$$

$$H_z(\rho,\varphi) = \omega_0 \sum_{r=1}^{2} \frac{\lambda_r s_r w^{2r-2} \rho}{a^2 \beta n u^{2r-4}} \Lambda_{nr}(\rho;0,-\lambda_r) f_r(\rho)\sin(\varphi+\psi_m) \tag{3.3.40}$$

The approach is identical for **the bi-complex formulation** of the EM-field of a hybrid mode can be obtained by applying the CTB transform (3.1.21 or 23) to (3.3.27-32),

$$\tilde{E}_\rho(\zeta) = -\mathrm{j}\beta \frac{aA}{2u} \sum_{r=1}^{2} \left(\frac{uJ_n(u)}{wK_n(w)}\right)^{r-1} \left[(1-s)\frac{J_{n-1}^{2-r}(u\rho/a)}{K_{n-1}^{1-r}(w\rho/a)} + \lambda_r(1+s)\frac{J_{n+1}^{2-r}(u\rho/a)}{K_{n+1}^{1-r}(w\rho/a)}\right] f_r(\rho) \mathrm{e}^{-\mathrm{i}\phi_m} \tag{3.3.41}$$

$$\tilde{E}_\varphi(\zeta) = \mathrm{ij}\beta \frac{aA}{2u} \sum_{r=1}^{2} \left(\frac{uJ_n(u)}{wK_n(w)}\right)^{r-1} \left[(1-s)\frac{J_{n-1}^{2-r}(u\rho/a)}{K_{n-1}^{1-r}(w\rho/a)} - \lambda_r(1+s)\frac{J_{n+1}^{2-r}(u\rho/a)}{K_{n+1}^{1-r}(w\rho/a)}\right] f_r(\rho) \mathrm{e}^{-\mathrm{i}\phi_m} \tag{3.3.42}$$

$$\tilde{E}_z(\zeta) = A \sum_{r=1}^{2} \frac{J_n^{r-1}(u)}{K_n^{r-1}(w)} \frac{J_n^{2-r}(u\rho/a)}{K_n^{1-r}(w\rho/a)} f_r(\rho) \mathrm{e}^{-\mathrm{i}\phi_m} \tag{3.3.43}$$

$$\tilde{H}_\rho(\zeta) = -\mathrm{ij}\omega_0 \frac{aA}{2u} \sum_{r=1}^{2} \left(\frac{uJ_n(u)}{wK_n(w)}\right)^{r-1} \left[(1-s_r)\frac{J_{n-1}^{2-r}(u\rho/a)}{K_{n-1}^{1-r}(w\rho/a)} - \lambda_r(1+s_r)\frac{J_{n+1}^{2-r}(u\rho/a)}{K_{n+1}^{1-r}(w\rho/a)}\right] \bar{\varepsilon}_r(\rho) \mathrm{e}^{-\mathrm{i}\phi_m} \tag{3.3.44}$$

$$\tilde{H}_\varphi(\zeta) = -\mathrm{j}\omega_0 \frac{aA}{2u} \sum_{r=1}^{2} \left(\frac{uJ_n(u)}{wK_n(w)}\right)^{r-1} \left[(1-s_r)\frac{J_{n-1}^{2-r}(u\rho/a)}{K_{n-1}^{1-r}(w\rho/a)} + \lambda_r(1+s_r)\frac{J_{n+1}^{2-r}(u\rho/a)}{K_{n+1}^{1-r}(w\rho/a)}\right] \bar{\varepsilon}_r(\rho) \mathrm{e}^{-\mathrm{i}\phi_m} \tag{3.3.45}$$

$$\tilde{H}_z(\zeta) = -\mathrm{i}\frac{A\beta s}{\omega_0\mu_0} \sum_{r=1}^{2} \frac{J_n^{r-1}(u)}{K_n^{r-1}(w)} \frac{J_n^{2-r}(u\rho/a)}{K_n^{1-r}(w\rho/a)} f_r(\rho) \mathrm{e}^{-\mathrm{i}\phi_m} \tag{3.3.46}$$

To help retain the compactness of the transverse components in the complex formulation (3.3.27-32), the trigonometric functions have to be relocated to the denominators of these components. By contrast, (3.3.41-46) are less cumbersome, due to the elimination of the trigonometric functions. In general, the compactness of the bi-regional expressions (3.3.27-32) and (3.3.41-46) would not be possible without the use of one or both of the new variables (3.2.7, 8), which are not used in Okamoto's original expressions (3.1.1-12). The expressions still retain Okamoto's Bessel function dependence, however, even after this generalization.

Applying the CTB transform (3.1.21 or 23) to (3.35-40) yields the more compact versions of the bi-regional components in the bi-complex approach,

$$E_\rho(\rho,\varphi) = -\mathrm{j}\beta \sum_{r=1}^{2} \Lambda_{nr}(\rho;s,\lambda_r) f_r(\rho) \mathrm{e}^{-\mathrm{i}(\varphi+\psi_m)} \tag{3.3.47}$$

$$E_\varphi(\rho,\varphi) = \mathrm{ij}\beta \sum_{r=1}^{2} \Lambda_{nr}(\rho;s,-\lambda_r) f_r(\rho) \mathrm{e}^{-\mathrm{i}(\varphi+\psi_m)} \tag{3.3.48}$$

$$E_z(\rho,\varphi) = -\beta \sum_{r=1}^{2} \frac{\lambda_r w^{2r-2}\rho}{a^2 \beta n u^{2r-4}} \Lambda_{nr}(\rho;0,-\lambda_r) f_r(\rho) \mathrm{e}^{-\mathrm{i}(\varphi+\psi_m)} \tag{3.3.49}$$

$$H_\rho(\rho,\varphi) = -\mathrm{ij}\omega_0 \sum_{r=1}^{2} \Lambda_{nr}(\rho;s_r,-\lambda_r) \bar{\varepsilon}_r(\rho) \mathrm{e}^{-\mathrm{i}(\varphi+\psi_m)} \tag{3.3.50}$$

$$H_\varphi(\rho,\varphi) = -\mathrm{j}\omega_0 \sum_{r=1}^{2} \Lambda_{nr}(\rho;s_r,\lambda_r) \bar{\varepsilon}_r(\rho) \mathrm{e}^{-\mathrm{i}(\varphi+\psi_m)} \tag{3.3.51}$$

$$H_z(\zeta) = \mathrm{i}\omega_0 \sum_{r=1}^{2} \frac{\lambda_r s_r w^{2r-2}\rho}{a^2 \beta n u^{2r-4}} \Lambda_{nr}(\rho;0,-\lambda_r) \bar{\varepsilon}_r(\rho) \mathrm{e}^{-\mathrm{i}(\varphi+\psi_m)} \tag{3.3.52}$$

In both the complex (3.3.35-40) and bi-complex (3.3.47-52) approaches, it is seen that the use of the generating function (3.2.22) has resulted in compact transverse components, but significantly larger expressions for the longitudinal components. By contrast, retaining Okamoto's Bessel function expressions for the longitudinal components in (3.3.29, 32) and in (3.3.43, 46) maintains these components as more compact relative to the transverse components.

The bi-regional expressions are all in the form of (3.3.12). In the next section, compact **vectorial expressions** of the EM-field are constructed using (3.3.12), by taking a vector-sum in cylindrical coordinates of the complex (3.2.35, 36) or the bi-complex (3.2.37, 38) generalized scalars, with the help of the general step-function (3.3.33),

$$\mathbf{W}_\xi(\zeta) = \sum_{r=1}^{2} \sum_{\xi=\rho,\varphi,z} \xi W_{r\xi}(\zeta) f_r(\rho). \tag{3.3.53}$$

## 3.4 Vector expressions of the EM-field of the hybrid modes

**In the complex formulation**, a generalized expression for the vector $\vec{\mathbf{V}}(\zeta)$ of the EM-field of any hybrid mode, for either polarization state, and valid over the entire cross-section of the fiber, can be constructed by applying (3.3.53) to (3.2.35),

$$\vec{\mathbf{E}}(\zeta) = \sum_{r=1}^{2} \sum_{\xi=\rho,\varphi,\mathbf{z}} \xi \frac{\left[(1-s|\xi\times\mathbf{z}|)\dfrac{J_{n-1}^{2-r}(u\rho/a)}{K_{n-1}^{1-r}(w\rho/a)} + \dfrac{\lambda_r(1+s|\xi\times\mathbf{z}|)}{\xi\cdot\boldsymbol{\rho}-\xi\cdot\boldsymbol{\varphi}-\xi\cdot\mathbf{z}}\dfrac{J_{n+1}^{2-r}(u\rho/a)}{K_{n+1}^{1-r}(w\rho/a)}\right] f_r(\rho)\cos\phi_m}{\dfrac{2ju}{aA\beta}\left(\dfrac{uJ_n(u)}{wK_n(w)}\right)^{1-r}\left[\xi\cdot\boldsymbol{\rho}-\xi\cdot\boldsymbol{\varphi}\cot\phi_m + jna^2\beta\dfrac{\lambda_r u^{2r-4}}{\rho w^{2r-2}}\xi\cdot\mathbf{z}\right]}$$

$$\vec{\mathbf{H}}(\zeta) = \sum_{r=1}^{2} \sum_{\xi=\rho,\varphi,\mathbf{z}} \xi \frac{\left[(1-s_r|\xi\times\mathbf{z}|)\dfrac{J_{n-1}^{2-r}(u\rho/a)}{K_{n-1}^{1-r}(w\rho/a)} - \dfrac{\lambda_r(1+s_r|\xi\times\mathbf{z}|)}{\xi\cdot\boldsymbol{\rho}-\xi\cdot\boldsymbol{\varphi}+\xi\cdot\mathbf{z}}\dfrac{J_{n+1}^{2-r}(u\rho/a)}{K_{n+1}^{1-r}(w\rho/a)}\right] \bar{\varepsilon}_r(\rho)\sin\phi_m}{\dfrac{2ju}{aA\omega_0}\left(\dfrac{uJ_n(u)}{wK_n(w)}\right)^{1-r}\left[\xi\cdot\boldsymbol{\rho}+\xi\cdot\boldsymbol{\varphi}\tan\phi_m - jna^2\beta\dfrac{\lambda_r u^{2r-4}}{\rho s_r w^{2r-2}}\xi\cdot\mathbf{z}\right]}$$

(3.4.1)

In order to maintain the compactness of these expressions, the general step-function is being used as before, either directly (3.3.33), or indirectly through (3.3.34). Without the step-function (3.3.33), each vector of (3.4.1) may be analyzed as 2 distinct vectors, one for each of the 2 regions of the fiber. The Kronecker deltas, which could not be retained for this expression for consistency with the vector-index of the inner summation of (3.3.53), have been replaced by their equivalent vector relations,

$$\delta_{\xi\xi'} = 1 - |\xi\times\xi'| = \xi\cdot\xi'; \quad \xi'\in\{\rho,\varphi,z\},\ \boldsymbol{\xi}'\in\{\boldsymbol{\rho},\boldsymbol{\varphi},\mathbf{z}\} \tag{3.4.2}$$

These expressions are not as explicit as (3.2.35) due to the use of the composite angle (3.2.8), without which the expressions would not fit within the margins of the page, for the same font-size. Alternative, compact versions of (3.4.1) are found in **Appendix B**.

The transverse and longitudinal vector components are respectively recovered from (3.4.1) using the vector operations

$$\vec{\mathbf{V}}_T(\zeta) = \mathbf{z}\times\vec{\mathbf{V}}(\zeta)\times\mathbf{z}, \tag{3.4.3}$$

$$\vec{\mathbf{V}}_z(\zeta) = \mathbf{z}\cdot\vec{\mathbf{V}}(\zeta)\mathbf{z}. \tag{3.4.4}$$

Alternatively, the desired component(s) may be sifted using Kronecker deltas, as

$$\vec{\mathbf{V}}_T(\zeta) = \left(\delta_{\xi\boldsymbol{\rho}} + \delta_{\xi\boldsymbol{\varphi}}\right)\vec{\mathbf{V}}(\zeta), \tag{3.4.5}$$

$$\vec{\mathbf{V}}_z(\zeta) = \delta_{\xi\mathbf{z}}\vec{\mathbf{V}}(\zeta). \tag{3.4.6}$$

with $\xi\in\{\boldsymbol{\rho},\boldsymbol{\varphi},\mathbf{z}\}$ as before.

Adapting the generating function (3.2.22) to (3.4.1), the generalized EM-field vector expression (3.4.1) can be reduced to the more compact forms

$$\vec{E}_m(\rho,\varphi) = -j\beta \sum_{r=1}^{2} \sum_{\xi=\rho,\varphi,z} \xi \frac{\Lambda_{nr}\left[\rho; s|\xi\times\mathbf{z}|, \lambda_r(\xi\cdot\boldsymbol{\rho} - \xi\cdot\boldsymbol{\varphi} - \xi\cdot\mathbf{z})\right] f_r(\rho)}{\xi\cdot\boldsymbol{\rho} - \xi\cdot\boldsymbol{\varphi}\cot\phi_m + jna^2\beta \dfrac{\lambda_r u^{2r-4}}{\rho w^{2r-2}}\xi\cdot\mathbf{z}} \cos\phi_m$$

$$\vec{H}_m(\rho,\varphi) = -j\omega_0 \sum_{r=1}^{2} \sum_{\xi=\rho,\varphi,z} \xi \frac{\Lambda_{nr}\left[\rho; s_r|\xi\times\mathbf{z}|, -\lambda_r(\xi\cdot\boldsymbol{\rho} - \xi\cdot\boldsymbol{\varphi} + \xi\cdot\mathbf{z})\right] \bar{\varepsilon}_r(\rho)}{\xi\cdot\boldsymbol{\rho} + \xi\cdot\boldsymbol{\varphi}\tan\phi_m - jna^2\beta \dfrac{\lambda_r u^{2r-4}}{\rho s_r w^{2r-2}}\xi\cdot\mathbf{z}} \sin\phi_m$$

(3.4.7)

The coefficients within the brackets of the generating functions (3.2.22) are all unit-less. The vectors have been additionally subscripted with an '$m$' here, to emphasize modal dependence with respect to polarization.

Eq. (3.4.7) can be specialized to the fundamental, HE$_{11}$-mode, which is the only mode supported by the SMF, simply by setting $n = 1$ in the generating function (3.2.22), in the $s$-parameter (3.1.17), and in (3.2.8), reducing the $n$-dependent quantities to

$$\Lambda_r(\rho; \eta, \pm\lambda_r') = \frac{aA}{2u}\left(\frac{uJ_1(u)}{wK_1(w)}\right)^{r-1}\left[(1-\eta)\frac{J_0^{2-r}(u\rho/a)}{K_0^{1-r}(w\rho/a)} \pm \lambda_r'(1+\eta)\frac{J_2^{2-r}(u\rho/a)}{K_2^{1-r}(w\rho/a)}\right],$$
$$\lambda_r' = \lambda_r(\xi\cdot\boldsymbol{\rho} - \xi\cdot\boldsymbol{\varphi} \mp \xi\cdot\mathbf{z}), \quad r \in \{1,2\};$$

(3.4.8)

$$s = \frac{v^2 J_1(u) K_1(w)}{u w^2 J_1'(u) K_1(w) + u^2 w K_1'(w) J_1(u)}, \quad s_r = s\left(\frac{\beta}{k_0 n_r}\right)^2;$$

(3.4.9)

$$\phi_m = \varphi + \psi_m, \quad m \in \{1,2\}.$$

(3.4.10)

The $n$-subscript has been omitted on the LHS of (3.4.8) to minimize the burdensome notation. Thus (3.4.7) simplifies to

$$\vec{E}_m(\rho,\varphi) = -j\beta \sum_{r=1}^{2} \sum_{\xi=\rho,\varphi,z} \xi \frac{\Lambda_r\left[\rho; s|\xi\times\mathbf{z}|, \lambda_r(\xi\cdot\boldsymbol{\rho} - \xi\cdot\boldsymbol{\varphi} - \xi\cdot\mathbf{z})\right] f_r(\rho)}{\xi\cdot\boldsymbol{\rho} - \xi\cdot\boldsymbol{\varphi}\cot(\varphi+\psi_m) + ja^2\beta\dfrac{u^{2r-4}\lambda_r}{\rho w^{2r-2}}\xi\cdot\mathbf{z}} \cos(\varphi+\psi_m)$$

$$\vec{H}_m(\rho,\varphi) = -j\omega_0 \sum_{r=1}^{2} \sum_{\xi=\rho,\varphi,z} \xi \frac{\Lambda_r\left[\rho; s_r|\xi\times\mathbf{z}|, -\lambda_r(\xi\cdot\boldsymbol{\rho} - \xi\cdot\boldsymbol{\varphi} + \xi\cdot\mathbf{z})\right] \bar{\varepsilon}_r(\rho)}{\xi\cdot\boldsymbol{\rho} + \xi\cdot\boldsymbol{\varphi}\tan(\varphi+\psi_m) - ja^2\beta\dfrac{\lambda_r u^{2r-4}}{\rho s_r w^{2r-2}}\xi\cdot\mathbf{z}} \sin(\varphi+\psi_m)$$

(3.4.11)

The polarization states of a given hybrid mode, can only be distinguished through the phase factor (3.1.19) which is incorporated in (3.4.10), also used in (3.4.11), and reproduced here

$$\psi_m = (m-1)\pi/2, \quad m \in \{1,2\}. \qquad (3.4.12)$$

Using (3.4.11, 12) with $m = 1$ yields the EM-field vector profile of the $HE_{11}^x$-mode, which is the *x*-polarization,

$$\vec{E}_1(\rho,\varphi) = -j\beta \sum_{r=1}^{2} \sum_{\xi=\rho,\varphi,z} \xi \frac{\Lambda_r\left[\rho; s|\xi \times z|, \lambda_r(\xi \cdot \rho - \xi \cdot \varphi - \xi \cdot z)\right]\cos\varphi}{\xi \cdot \rho - \xi \cdot \varphi \cot\varphi + ja^2\beta \dfrac{\lambda_r u^{2r-4}}{\rho w^{2r-2}} \xi \cdot z} f_r(\rho)$$

$$\vec{H}_1(\rho,\varphi) = -j\omega_0 \sum_{r=1}^{2} \sum_{\xi=\rho,\varphi,z} \xi \frac{\Lambda_r\left[\rho; s_r|\xi \times z|, -\lambda_r(\xi \cdot \rho - \xi \cdot \varphi + \xi \cdot z)\right]\sin\varphi}{\xi \cdot \rho + \xi \cdot \varphi \tan\varphi - ja^2\beta \dfrac{\lambda_r u^{2r-4}}{\rho s_r w^{2r-2}} \xi \cdot z} \bar{\varepsilon}_r(\rho)$$

(3.4.13)

Using (3.4.11, 12) again, but with $m = 2$ for the EM-field vector of the $HE_{11}^y$-mode, which is the *y*-polarization,

$$\vec{E}_2(\rho,\varphi) = j\beta \sum_{r=1}^{2} \sum_{\xi=\rho,\varphi,z} \xi \frac{\Lambda_r\left[\rho; s|\xi \times z|, \lambda_r(\xi \cdot \rho - \xi \cdot \varphi - \xi \cdot z)\right]\sin\varphi}{\xi \cdot \rho + \xi \cdot \varphi \tan\varphi + ja^2\beta \dfrac{\lambda_r u^{2r-4}}{\rho w^{2r-1}} \xi \cdot z} f_r(\rho)$$

$$\vec{H}_2(\rho,\varphi) = -j\omega_0 \sum_{r=1}^{2} \sum_{\xi=\rho,\varphi,z} \xi \frac{\Lambda_r\left[\rho; s_r|\xi \times z|, -\lambda_r(\xi \cdot \rho - \xi \cdot \varphi + \xi \cdot z)\right]\cos\varphi}{\xi \cdot \rho - \xi \cdot \varphi \cotan\varphi - ja^2\beta \dfrac{\lambda_r u^{2r-4}}{\rho s_r w^{2r-1}} \xi \cdot z} \bar{\varepsilon}_r(\rho)$$

(3.4.14)

It can also be seen from (3.4.13, 14), that the 2 polarization modes are in "quadrature" with respect to the azimuth $\varphi$. This observation, which is actually true for any hybrid mode, may be operationally expressed as a convolution ($*$) in the azimuth, of the spatial EM-field vector $\vec{V}_m$ (3.4.7) of either *m*-th polarization with a Dirac delta-function (3.3.6), as follows:

$$\vec{V}_{3-m}(\rho,\varphi) = \delta(\varphi + (3-2m)\pi/2) * \vec{V}_m(\rho,\varphi), \quad \vec{V} \in \{\vec{E}, \vec{H}\}, \quad m \in \{1,2\}. \qquad (3.4.15)$$

Thus, given the expression of the EM-field vector of one polarization state, the other polarization state is obtained by a convolution relation.

**In the bi-complex formulation**, the generalized expression for the vecsor $\tilde{\mathbf{U}}(\zeta)$ of the EM-field of any hybrid mode, and valid over the entire cross-section of the fiber, is found by applying (3.3.53) to the generalized phasor component expressions (3.2.37), yielding

$$\tilde{\mathbf{E}}(\zeta) = \sum_{r=1}^{2} \sum_{\xi=\boldsymbol{\rho},\boldsymbol{\varphi},\mathbf{z}} \xi \frac{\left[ (1-s|\xi\times\mathbf{z}|) \frac{J_{n-1}^{2-r}(u\rho/a)}{K_{n-1}^{1-r}(w\rho/a)} + \frac{(1+s|\xi\times\mathbf{z}|)e^{j\pi r}}{\xi\bullet\boldsymbol{\rho} - \xi\bullet\boldsymbol{\varphi} - \xi\bullet\mathbf{z}} \frac{J_{n+1}^{2-r}(u\rho/a)}{K_{n+1}^{1-r}(w\rho/a)} \right] f_r(\rho)}{\frac{2ju}{aA\beta} \left( \frac{uJ_n(u)}{wK_n(w)} \right)^{1-r} \left[ \xi\bullet\boldsymbol{\rho} + i\xi\bullet\boldsymbol{\varphi} + jna^2\beta \frac{u^{2r-4}e^{j\pi r}}{\rho w^{2r-2}} \xi\bullet\mathbf{z} \right]} e^{-i(n\varphi+\psi_m)}$$

$$\tilde{\mathbf{H}}(\zeta) = \sum_{r=1}^{2} \sum_{\xi=\boldsymbol{\rho},\boldsymbol{\varphi},\mathbf{z}} \xi \frac{\left[ (1-s_r|\xi\times\mathbf{z}|) \frac{J_{n-1}^{2-r}(u\rho/a)}{K_{n-1}^{1-r}(w\rho/a)} - \frac{(1+s_r|\xi\times\mathbf{z}|)e^{j\pi r}}{\xi\bullet\boldsymbol{\rho} - \xi\bullet\boldsymbol{\varphi} + \xi\bullet\mathbf{z}} \frac{J_{n+1}^{2-r}(u\rho/a)}{K_{n+1}^{1-r}(w\rho/a)} \right] \bar{\varepsilon}_r(\rho)}{\frac{2ju}{aA\omega_0} \left( \frac{uJ_n(u)}{wK_n(w)} \right)^{1-r} \left[ \xi\bullet\boldsymbol{\rho} + i\xi\bullet\boldsymbol{\varphi} - jna^2\beta \frac{u^{2r-4}e^{j\pi r}}{\rho s_r w^{2r-2}} \xi\bullet\mathbf{z} \right]} ie^{-i(n\varphi+\psi_m)}$$

(3.4.16)

The expressions are clearly separable in the radial and azimuth-directions, by contrast to (3.4.1), because they are devoid of trigonometric functions. They can also be more explicit if need be, and shown in (3.4.16), without sacrificing font-size. Lastly, using the generating function (3.2.22) used for the complex formulation, the EM-field vecsor is in the more compact form given by

$$\tilde{\mathbf{E}}(\rho,\varphi) = -j\beta \sum_{r=1}^{2} \sum_{\xi=\boldsymbol{\rho},\boldsymbol{\varphi},\mathbf{z}} \xi \frac{\Lambda_{nr}\left[\rho; s|\xi\times\mathbf{z}|, \lambda_r(\xi\bullet\boldsymbol{\rho} - \xi\bullet\boldsymbol{\varphi} - \xi\bullet\mathbf{z})\right]}{\xi\bullet\boldsymbol{\rho} + i\xi\bullet\boldsymbol{\varphi} + jna^2\beta \frac{\lambda_r u^{2r-4}}{\rho w^{2r-2}} \xi\bullet\mathbf{z}} f_r(\rho) e^{-i(n\varphi+\psi_m)}$$

$$\tilde{\mathbf{H}}(\rho,\varphi) = -ij\omega_0 \sum_{r=1}^{2} \sum_{\xi=\boldsymbol{\rho},\boldsymbol{\varphi},\mathbf{z}} \xi \frac{\Lambda_{nr}\left[\rho; s_r|\xi\times\mathbf{z}|, -\lambda_r(\xi\bullet\boldsymbol{\rho} - \xi\bullet\boldsymbol{\varphi} + \xi\bullet\mathbf{z})\right]}{\xi\bullet\boldsymbol{\rho} + i\xi\bullet\boldsymbol{\varphi} - jna^2\beta \frac{\lambda_r u^{2r-4}}{\rho s_r w^{2r-2}} \xi\bullet\mathbf{z}} \bar{\varepsilon}_r(\rho) e^{-i(n\varphi+\psi_m)}$$

(3.4.17)

Using the specialized parameters (3.4.8-10), the generalized EM-field vecsor for the $HE_{11}$-mode is found to be

$$\tilde{\mathbf{E}}_m(\rho,\varphi) = -j\beta \sum_{r=1}^{2} \sum_{\xi=\boldsymbol{\rho},\boldsymbol{\varphi},\mathbf{z}} \xi \frac{\Lambda_r\left[\rho; s|\xi\times\mathbf{z}|, \lambda_r(\xi\bullet\boldsymbol{\rho} - \xi\bullet\boldsymbol{\varphi} - \xi\bullet\mathbf{z})\right]}{\xi\bullet\boldsymbol{\rho} + i\xi\bullet\boldsymbol{\varphi} + ja^2\beta \frac{\lambda_r u^{2r-4}}{\rho w^{2r-2}} \xi\bullet\mathbf{z}} f_r(\rho) e^{-i(\varphi+\psi_m)}$$

$$\tilde{\mathbf{H}}_m(\rho,\varphi) = -ij\omega_0 \sum_{r=1}^{2} \sum_{\xi=\boldsymbol{\rho},\boldsymbol{\varphi},\mathbf{z}} \xi \frac{\Lambda_r\left[\rho; s_r|\xi\times\mathbf{z}|, -\lambda_r(\xi\bullet\boldsymbol{\rho} - \xi\bullet\boldsymbol{\varphi} + \xi\bullet\mathbf{z})\right]}{\xi\bullet\boldsymbol{\rho} + i\xi\bullet\boldsymbol{\varphi} - ja^2\beta \frac{\lambda_r u^{2r-4}}{\rho s_r w^{2r-2}} \xi\bullet\mathbf{z}} \bar{\varepsilon}_r(\rho) e^{-i(\varphi+\psi_m)}$$

(3.4.18)

Setting $m = 1$ in (3.4.18) yields the generalized EM-field vecsor of the $HE_{11}^x$-mode:

$$\tilde{\mathbf{E}}_1(\rho,\varphi) = -j\beta \sum_{r=1}^{2} \sum_{\xi=\rho,\varphi,z} \xi \frac{\Lambda_r\left[\rho;1,s|\xi\times\mathbf{z}|,\lambda_r(\xi\bullet\rho-\xi\bullet\varphi-\xi\bullet\mathbf{z})\right]}{\xi\bullet\rho+i\xi\bullet\varphi+ja^2\beta\dfrac{\lambda_r u^{2r-4}}{\rho w^{2r-2}}\xi\bullet\mathbf{z}} f_r(\rho)e^{-i\varphi}$$

$$\tilde{\mathbf{H}}_1(\rho,\varphi) = -ij\omega_0 \sum_{r=1}^{2} \sum_{\xi=\rho,\varphi,z} \xi \frac{\Lambda_r\left[\rho;1,s_r|\xi\times\mathbf{z}|,-\lambda_r(\xi\bullet\rho-\xi\bullet\varphi+\xi\bullet\mathbf{z})\right]}{\xi\bullet\rho+i\xi\bullet\varphi-ja^2\beta\dfrac{\lambda_r u^{2r-4}}{\rho s_r w^{2r-2}}\xi\bullet\mathbf{z}} \bar{\varepsilon}_r(\rho)e^{-i\varphi}$$

(3.4.19)

The generalized EM-field vecsor of the $HE_{11}^y$-mode is obtained from (3.4.18) for $m=2$:

$$\tilde{\mathbf{E}}_2(\rho,\varphi) = ij\beta \sum_{r=1}^{2} \sum_{\xi=\rho,\varphi,z} \xi \frac{\Lambda_r\left[\rho;1,s|\xi\times\mathbf{z}|,\lambda_r(\xi\bullet\rho-\xi\bullet\varphi-\xi\bullet\mathbf{z})\right]}{\xi\bullet\rho+i\xi\bullet\varphi+ja^2\beta\dfrac{\lambda_r u^{2r-4}}{\rho w^{2r-2}}\xi\bullet\mathbf{z}} f_r(\rho)e^{-i\varphi}$$

$$\tilde{\mathbf{H}}_2(\rho,\varphi) = -j\omega_0 \sum_{r=1}^{2} \sum_{\xi=\rho,\varphi,z} \xi \frac{\Lambda_r\left[\rho;1,s_r|\xi\times\mathbf{z}|,-\lambda_r(\xi\bullet\rho-\xi\bullet\varphi+\xi\bullet\mathbf{z})\right]}{\xi\bullet\rho+i\xi\bullet\varphi-ja^2\beta\dfrac{\lambda_r u^{2r-4}}{\rho s_r w^{2r-2}}\xi\bullet\mathbf{z}} \bar{\varepsilon}_r(\rho)e^{-i\varphi}$$

(3.4.20)

Lastly, a polarization state of the EM-field vecsor $\tilde{\mathbf{U}}$ of a hybrid mode can also be related to its other polarization state, using the simple multiplicative relation

$$\tilde{\mathbf{U}}_{3-m}(\rho,\varphi) = \exp[-i(3-2m)\pi/2]\tilde{\mathbf{U}}_m(\rho,\varphi), \qquad m \in \{1,2\} \quad (3.4.21)$$

instead of the convolution (3.4.15) used for the complex method. Thus, a relation based on calculus (3.4.15) and special functions, is reduced to one based on simple complex algebra. It also holds for any hybrid mode, when expressed in the bi-complex formulation.

The **vector CTB transform** is different from the scalar CTB transform (3.1.21, 23), which is applied to each EM-field component individually. In this case, one possible form of this transform may be given by the **dot-product**:

$$\tilde{\mathbf{U}}(\zeta) = \left[\delta_{\mathbf{VE}} \frac{\rho\rho+i(\cot\phi_m)\varphi\varphi+\mathbf{zz}}{\cos\phi_m} + i\delta_{\mathbf{VH}} \frac{\rho\rho-i(\tan\phi_m)\varphi\varphi+\mathbf{zz}}{\sin\phi_m}\right]\bullet\vec{\mathbf{V}}(\zeta)e^{-i\phi_m}. \quad (3.4.22)$$

The 1st bracketed term represents the E-field vector, and activates a division by $\cos\phi_m$ for each of its components except that for the azimuth, which is instead divided by $-i\sin\phi_m$. The 2nd bracketed term represents the H-field vector, and activates a division by a $-i\sin\phi_m$ for each of its components except that of the azimuth, in which case it is divided by $\cos\phi_m$ instead. Lastly, the expression requires a dot-product with the EM-field vector (3.4.1). **The reverse, vector BTC-transform is just a real-operation with respect to i.**

## 3.5 Generalization of the electromagnetic field vector of the hybrid mode

At this juncture, the question may arise as to whether it is possible to express the electromagnetic (EM-) field **vector** of any hybrid mode using a single equation. Whether in the complex (3.4.1), or the bi-complex (3.4.16) formulation, it is observable that the EM-field vector is remarkably similar in its expressions of the electric and magnetic fields. It is indeed possible to construct such an equation for the EM-field vector, equally valid for both electric and magnetic fields, but which is considerably more cumbersome. For the **complex formulation**, (3.4.1) is generalized to:

$$\vec{V}(\zeta) = \sum_{r=1}^{2} \sum_{\xi=\rho,\varphi,z} \frac{\left[\left(1 - \frac{s_r s |\xi \times \mathbf{z}|}{s_r \delta_{VE} + s \delta_{VH}}\right) \frac{J_{n-1}^{2-r}(u\rho/a)}{K_{n-1}^{1-r}(w\rho/a)} + \lambda_r \frac{1 + \frac{s_r s |\xi \times \mathbf{z}|}{s_r \delta_{VE} + s \delta_{VH}}}{\frac{\xi \cdot (\boldsymbol{\rho} - \boldsymbol{\varphi})}{\delta_{VE} - \delta_{VH}} - \xi \cdot \mathbf{z}} \frac{J_{n+1}^{2-r}(u\rho/a)}{K_{n+1}^{1-r}(w\rho/a)}\right] f_r(\rho) \xi}{\frac{2ju}{aA}\left(\frac{uJ_n(u)}{wK_n(w)}\right)^{1-r} \left[\xi \cdot \boldsymbol{\rho} + \frac{\xi \cdot \boldsymbol{\varphi}}{\cotan(\phi_m - \frac{\pi}{2}\delta_{VE})} + j\frac{\lambda_r u^{2r-4}}{\rho w^{2r-2}} \frac{na^2 \beta \xi \cdot \mathbf{z}}{\delta_{VE} - s_r \delta_{VH}}\right] \frac{\sec(\phi_m - \frac{\pi}{2}\delta_{VH})}{\beta \delta_{VE} + \omega_0 \varepsilon_r \delta_{VH}}}$$

(3.5.1)

This vector yields the electric field vector, when $\mathbf{V} = \mathbf{E}$, and the magnetic field vector, when $\mathbf{V} = \mathbf{H}$. It is valid for either polarization of a hybrid mode, and also valid over both (core and cladding) regions of the fiber, due to the use of the general step-function $f_r(\rho)$ (3.3.33). This generalization is based on 3 main observations, that the trigonometric dependence of the electric and magnetic fields are in quadrature with respect to the azimuthal variable, that their respective *z*-components are out of phase relative to each other, and that they differ in their dielectric dependence. The azimuthal rotation is enforced when $\mathbf{V} = \mathbf{H}$, which is incorporated into the arguments of the trigonometric functions in the denominator of the summand. The sign of the *z*-component is changed when $\mathbf{V} = \mathbf{H}$, and this is reflected in the 3rd bracketed term in the denominator. The electric and magnetic field vectors also have different scaling parameters, with $\beta$ for the electric field, and with $\omega_0 \varepsilon_r$ for the magnetic field vector, which are represented in the multiplicative denominator end term. The dissimilar dielectric dependence of the electric and magnetic field vectors is exhibited in their differing *s*-parameter requirements, in both the numerator and denominator. The equation could not be attained in a compact form, without the use of the new variables (3.2.7, 8), which are not used by Okamoto [1].

Using the generating function (3.2.22), the EM-field vector (3.5.1) can be simplified to the following expression

$$\vec{V}(\zeta) = \sum_{r=1}^{2} \sum_{\xi=\rho,\varphi,z} \frac{(\beta \delta_{VE} + \omega_0 \varepsilon_r \delta_{VH}) \Lambda_{nr}\left[\rho; \frac{s_r s |\xi \times \mathbf{z}|}{s_r \delta_{VE} + s \delta_{VH}}, \lambda_r \frac{\xi \cdot (\boldsymbol{\rho} - \boldsymbol{\varphi})}{\delta_{VE} - \delta_{VH}} - \lambda_r \xi \cdot \mathbf{z}\right] f_r(\rho) \xi}{\left[\xi \cdot \boldsymbol{\rho} + \frac{\xi \cdot \boldsymbol{\varphi}}{\cotan(n\varphi + \psi_m - \frac{\pi}{2}\delta_{VH})} + j\frac{\lambda_r u^{2r-4}}{\rho w^{2r-2}} \frac{na^2 \beta \xi \cdot \mathbf{z}}{\delta_{VE} - s_r \delta_{VH}}\right] \sec(n\varphi + \psi_m - \frac{\pi}{2}\delta_{VH})}$$

(3.5.2)

Unlike (3.5.1), this equation is attained without using the new variable (3.2.8), although it no longer retains Okamoto's original Bessel function algebra.

In particular, the generalized vector for the HE$_{11}$-mode, which is common to both single-mode and multi-mode fibers, is found by setting $n = 1$ in (3.5.2),

$$\vec{V}_{11}(\zeta) = \sum_{r=1}^{2}\sum_{\xi=\rho,\varphi,z} \frac{(\beta\delta_{\mathbf{VE}} + \omega_0\varepsilon_r\delta_{\mathbf{VH}})\Lambda_r\left[\rho; \dfrac{s_r s|\xi\times\mathbf{z}|}{s_r\delta_{\mathbf{VE}} + s\delta_{\mathbf{VH}}}, \lambda_r\dfrac{\xi\cdot(\rho-\varphi)}{\delta_{\mathbf{VE}}-\delta_{\mathbf{VH}}} - \lambda_r\xi\cdot\mathbf{z}\right]f_r(\rho)\xi}{\left[\xi\cdot\rho + \dfrac{\xi\cdot\varphi}{\cotan\left(\varphi+\psi_m-\frac{\pi}{2}\delta_{\mathbf{VH}}\right)} + j\dfrac{\lambda_r u^{2r-4}}{\rho w^{2r-2}}\dfrac{a^2\beta\xi\cdot\mathbf{z}}{\delta_{\mathbf{VE}}-s_r\delta_{\mathbf{VH}}}\right]\sec\left(\varphi+\psi_m-\frac{\pi}{2}\delta_{\mathbf{VH}}\right)}$$

(3.5.3)

with the attendant specialization (3.4.8-10) for $n = 1$.

**In the bi-complex formulation**, the generalized expression for the vecsor $\tilde{\mathbf{U}}(\zeta)$ of the EM-field of any hybrid mode, for either polarization, and valid over the entire cross-section of the fiber, is found from (3.4.16) to be

$$\tilde{\mathbf{U}}(\zeta) = \sum_{r=1}^{2}\sum_{\xi=\rho,\varphi,z} \frac{\left[\left(1 - \dfrac{s_r s|\xi\times\mathbf{z}|}{s_r\delta_{\mathbf{UE}}+s\delta_{\mathbf{UH}}}\right)\dfrac{J_{n-1}^{2-r}(u\rho/a)}{K_{n-1}^{1-r}(w\rho/a)} + \lambda_r\dfrac{1+\dfrac{s_r s|\xi\times\mathbf{z}|}{s_r\delta_{\mathbf{UE}}+s\delta_{\mathbf{UH}}}}{\dfrac{\xi\cdot(\rho-\varphi)}{\delta_{\mathbf{UE}}-\delta_{\mathbf{UH}}}-\xi\cdot\mathbf{z}}\dfrac{J_{n+1}^{2-r}(u\rho/a)}{K_{n+1}^{1-r}(w\rho/a)}\right]f_r(\rho)\xi}{\dfrac{2ju}{aA}\left(\dfrac{uJ_n(u)}{wK_n(w)}\right)^{1-r}\left[\xi\cdot\rho + i\xi\cdot\varphi + j\dfrac{\lambda_r u^{2r-4}}{\rho w^{2r-2}}\dfrac{na^2\beta\xi\cdot\mathbf{z}}{\delta_{\mathbf{UE}}-s_r\delta_{\mathbf{UH}}}\right]\left(\dfrac{\delta_{\mathbf{UE}}}{\beta}-i\dfrac{\delta_{\mathbf{UH}}}{\omega_0\varepsilon_r}\right)e^{i\phi_m}}$$

(3.5.4)

The denominator of the equivalent expression (3.5.1) is much larger, due to the use of trigonometric functions. The vecsor yields the E-field expression when $\mathbf{U} = \mathbf{E}$, and that of the H-field, when $\mathbf{U} = \mathbf{H}$. Using the generating function (3.2.22), a much more compact version is found as

$$\tilde{\mathbf{U}}(\zeta) = -j\sum_{r=1}^{2}\sum_{\xi=\rho,\varphi,z} \xi\frac{\Lambda_{nr}\left[\rho; \dfrac{s_r s|\xi\times\mathbf{z}|}{s_r\delta_{\mathbf{UE}}+s\delta_{\mathbf{UH}}}, \lambda_r\dfrac{\xi\cdot(\rho-\varphi)}{\delta_{\mathbf{UE}}-\delta_{\mathbf{UH}}}-\lambda_r\xi\cdot\mathbf{z}\right]}{\left(\dfrac{\delta_{\mathbf{UE}}}{\beta}-i\dfrac{\delta_{\mathbf{UH}}}{\omega_0\varepsilon_r}\right)\left(\xi\cdot\rho - i\xi\cdot\varphi + j\dfrac{\lambda_r u^{2r-4}}{\rho w^{2r-2}}\dfrac{na^2\beta\xi\cdot\mathbf{z}}{\delta_{\mathbf{UE}}-s_r\delta_{\mathbf{UH}}}\right)}f_r(\rho)e^{-i(n\varphi+\psi_m)}$$

(3.5.5)

which can be specialized to the HE$_{11}$-mode using (3.4.8-10), yielding

$$\tilde{\mathbf{U}}_{11}(\zeta) = -j\sum_{r=1}^{2}\sum_{\xi=\rho,\varphi,z} \xi\frac{\Lambda_r\left[\rho; \dfrac{s_r s|\xi\times\mathbf{z}|}{s_r\delta_{\mathbf{UE}}+s\delta_{\mathbf{UH}}}, \lambda_r\dfrac{\xi\cdot(\rho-\varphi)}{\delta_{\mathbf{UE}}-\delta_{\mathbf{UH}}}-\lambda_r\xi\cdot\mathbf{z}\right]}{\left(\dfrac{\delta_{\mathbf{UE}}}{\beta}-i\dfrac{\delta_{\mathbf{UH}}}{\omega_0\varepsilon_r}\right)\left(\xi\cdot\rho - i\xi\cdot\varphi + j\dfrac{\lambda_r u^{2r-4}}{\rho w^{2r-2}}\dfrac{a^2\beta\xi\cdot\mathbf{z}}{\delta_{\mathbf{UE}}-s_r\delta_{\mathbf{UH}}}\right)}f_r(\rho)e^{-i(\varphi+\psi_m)}.$$

(3.5.6)

# 4. Power Integral Derivation

## 4.1 Derivation of the power integral using the complex field formulation

Okamoto's EM-field expressions for the core (3.1.1-6) and the cladding (3.1.7-12), are meant to be used with the complex field formulation. Following this approach in the derivation of power integrals can complicate analyses that involve integrations over the fiber's cross-section $(\rho, \varphi)$, which would entail trigonometric terms in relatively large, cumbersome expressions.

According to §**1**, the physical EM-field vector is obtained from its corresponding spatiotemporal vecsor by a real-operation with respect to the imaginary number j:

$$\vec{v}(r,t) = \text{Re}\left[\tilde{\mathbf{V}}(r,t)\right], \quad \vec{v} \in \{\vec{e}, \vec{h}\}, \tag{4.1.1}$$

with the corresponding vecsor given by

$$\tilde{\mathbf{V}}(r,t) = \vec{\mathbf{V}}(\zeta)e^{j(\omega_0 t - \beta z)}, \quad \mathbf{V} \in \{\mathbf{E}, \mathbf{H}\}. \tag{4.1.2}$$

It is the product of a spatial vector, and a spatiotemporal phasor represented by a complex exponential. The spatial vector may be complex in general, but is not a phasor. Each of the 3 spatial vecsor components may be found from the relevant expression from (3.3.27-32), generally stated over the entire cross-section of the fiber using (3.3.33),

$$V_\xi(\zeta) = \xi \cdot \vec{\mathbf{V}}(\zeta) = \sum_{r=1}^{2} V_{r\xi}(\zeta) f_r(\rho), \quad \xi \in \{\rho, \varphi, z\}. \tag{4.1.3}$$

The component in the summand may also be in the form of (3.2.35 or 36).

After specializing (4.1.1) to the electric and magnetic fields of the EM-field, the instantaneous, *real* Poynting vector $\vec{S}(r,t)$, defined as the power density [W·m$^{-2}$], or the intensity per unit time [J·m$^{-2}$·s$^{-1}$], is found as [1, 21]

$$\vec{S}(r,t) = \vec{e}(r,t) \times \vec{h}(r,t). \tag{4.1.4}$$

After substituting for each vector from (4.1.1), expanding the corresponding brackets and carrying out the cross-product using (2.20), there results

$$\vec{S}(r,t) = \text{Re}\left[\tilde{\mathbf{E}}(r,t)\right] \times \text{Re}\left[\tilde{\mathbf{H}}(r,t)\right] = \frac{1}{4}\begin{pmatrix} \tilde{\mathbf{E}}^*(r,t) \times \tilde{\mathbf{H}}(r,t) + \tilde{\mathbf{E}}(r,t) \times \tilde{\mathbf{H}}^*(r,t) \\ + \tilde{\mathbf{E}}(r,t) \times \tilde{\mathbf{H}}(r,t) + \tilde{\mathbf{E}}^*(r,t) \times \tilde{\mathbf{H}}^*(r,t) \end{pmatrix}. \tag{4.1.5}$$

Although their argument $(r,t)$ have been retained, the first 2 terms are clearly independent of the propagating factor $e^{j(\omega_0 t - \beta z)}$ based on the definition (4.1.2), which is not the case for the 3rd and 4th bracketed terms. After applying (4.1.2) to the 3rd and 4th terms, the expression simplifies to

$$\vec{S}(\boldsymbol{r},t) = \frac{1}{2}\text{Re}\left[\tilde{\mathbf{E}}^*(\boldsymbol{r},t)\times\tilde{\mathbf{H}}(\boldsymbol{r},t)\right] + \frac{1}{2}\text{Re}\left[\vec{\mathbf{E}}(\zeta)\times\vec{\mathbf{H}}(\zeta)\text{e}^{2\text{j}(\omega_0 t-\beta z)}\right]. \quad (4.1.6)$$

For a practical fiber-optic telecommunication system with a conventional receiver, the power contribution due to the 2nd term is actually inadmissible upon detection, as it oscillates at twice the signal's optical carrier frequency ($\omega_0/2\pi$) which is of the order of 200 THz. Practical baseband detection bandwidth is typically $\leq$ 100 GHz. Although the 2nd term could be discarded at this juncture based on practical limitations and/or the physics of reception, it is traditionally retained, and later eliminated in the derivation of the time-averaged Poynting vector [1, 21].

The *time-averaged* Poynting vector is given by the time-average bracket [1]

$$\vec{S}(\boldsymbol{r}) = \langle\vec{S}(\boldsymbol{r},t)\rangle = \lim_{T\to\infty}\frac{1}{T}\int_{-T/2}^{T/2}\vec{S}(\boldsymbol{r},t)\text{d}t, \quad \omega_0 = 2\pi/T \quad (4.1.7)$$

**which is italicized**, and not to be confused with the instantaneous Poynting vector, whereas $T$ is a temporal period derived from the angular frequency $\omega_0$. Thus,

$$\vec{S}(\boldsymbol{r}) = \frac{1}{2}\text{Re}\left[\tilde{\mathbf{E}}^*(\boldsymbol{r},t)\times\tilde{\mathbf{H}}(\boldsymbol{r},t)\right] + \frac{1}{2}\text{Re}\left[\vec{\mathbf{E}}(\zeta)\times\vec{\mathbf{H}}(\zeta)\langle\text{e}^{2\text{j}(\omega_0 t-\beta z)}\rangle\right]. \quad (4.1.8)$$

The time-average bracket resolves as follows for the 2nd term:

$$\langle\text{e}^{2\text{j}(\omega_0 t-\beta z)}\rangle = \lim_{T\to\infty}\frac{\text{e}^{-2\text{j}\beta z}}{T}\int_{-T/2}^{T/2}\text{e}^{2\text{j}\omega_0 t}\text{d}t = \lim_{T\to\infty}\frac{\text{e}^{-2\text{j}\beta z}}{T}\left[\frac{\text{e}^{2\text{j}\omega_0 t}}{2\text{j}\omega_0}\right]_{-T/2}^{T/2} = \text{e}^{-2\text{j}\beta z}\text{sinc}(2\pi) = 0. \quad (4.1.9)$$

Consequently, the time-averaged Poynting vector simplifies to

$$\vec{S}(\boldsymbol{r}) = \frac{1}{2}\text{Re}\left[\tilde{\mathbf{E}}^*(\boldsymbol{r},t)\times\tilde{\mathbf{H}}(\boldsymbol{r},t)\right] = \frac{1}{4}\left(\tilde{\mathbf{E}}^*(\boldsymbol{r},t)\times\tilde{\mathbf{H}}(\boldsymbol{r},t) + \tilde{\mathbf{E}}(\boldsymbol{r},t)\times\tilde{\mathbf{H}}^*(\boldsymbol{r},t)\right). \quad (4.1.10)$$

It is thus concluded that the time-averaged *complex* Poynting vector is given by:

$$\tilde{S}(\boldsymbol{r}) = \tilde{\mathbf{E}}^*(\boldsymbol{r},t)\times\tilde{\mathbf{H}}(\boldsymbol{r},t). \quad (4.1.11)$$

Due to the orientation of **the cylindrical coordinate system** relative to the geometry of the fiber, as well as to guided-mode propagation, the net power flow occurs in the positive *z*-direction, so that the *time-averaged* detected power in Watts (W) is given by

$$P = \int_A \vec{S}(\boldsymbol{r})\bullet\mathbf{z}\rho\text{d}\rho\text{d}\varphi. \quad (4.1.12)$$

Applying (4.1.12) to (4.1.10) yields **the most general form of the power expression**:

$$P = \frac{1}{2}\text{Re}\int \tilde{\mathbf{E}}^*(r,t)\times\tilde{\mathbf{H}}(r,t)\bullet\mathbf{dA} = \frac{1}{2}\text{Re}\int \tilde{\mathbf{E}}(r,t)\times\tilde{\mathbf{H}}^*(r,t)\bullet\mathbf{dA}, \qquad (4.1.13)$$

with the equivalence being due to the fact that the real-parts of a complex quantity and its complex-conjugate are identical. The detected power is actually the convolution of (4.1.13) with the impulse response of the detector, scaled by its responsivity, among other parameters. The assumption being made here is that the effective bandwidth of the receiver electronics is large and uniform enough to render convolution effects negligible, which is not always the case in practice. Based on (4.1.2), a simpler spatial dependence is obtained in the integrand of (4.1.13) with the cancellation of the common phasor $e^{\text{j}(\omega_0 t-\beta z)}$,

$$P = \frac{1}{2}\text{Re}\int \vec{\mathbf{E}}^*(\zeta)\times\vec{\mathbf{H}}(\zeta)\bullet\mathbf{dA} \qquad (4.1.14)$$

It is observable that the propagating power (4.1.14) is independent of both time and propagation distance *z*, which is to be expected for a longitudinally and temporally invariant, loss-free waveguide such as the ideal fiber. At any distance, the power is still effectively that at the input plane of the waveguide. Furthermore, since the vector cross-product must yield a longitudinal component to avoid a non-trivial dot-product, the integrand of (4.1.14) is constrained to the transverse components of the spatial vectors,

$$P = \frac{1}{2}\text{Re}\int \vec{\mathbf{E}}_T^*(\zeta)\times\vec{\mathbf{H}}_T(\zeta)\bullet\mathbf{dA} \qquad (4.1.15)$$

For the *m*-th polarization state of the *n*-th hybrid mode, the complex *spatial* EM-field vector is given by the pair of spatial vector equations (3.4.7),

$$\vec{\mathbf{E}}_m(\zeta) = -\text{j}\beta\sum_{r=1}^{2}\sum_{\xi=\rho,\varphi,z}\xi\frac{\Lambda_{nr}\left[\rho; s|\xi\times\mathbf{z}|,\lambda_r(\xi\bullet\rho-\xi\bullet\varphi-\xi\bullet\mathbf{z})\right]f_r(\rho)\cos(n\varphi+\psi_m)}{\xi\bullet\rho-\xi\bullet\varphi\cot(n\varphi+\psi_m)+\text{j}\left(a^2 n\beta\lambda_r u^{2r-4}/\rho w^{2r-2}\right)\xi\bullet\mathbf{z}}$$

$$\vec{\mathbf{H}}_m(\zeta) = -\text{j}\omega_0\sum_{r=1}^{2}\sum_{\xi=\rho,\varphi,z}\xi\frac{\Lambda_{nr}\left[\rho; s_r|\xi\times\mathbf{z}|,-\lambda_r(\xi\bullet\rho-\xi\bullet\varphi+\xi\bullet\mathbf{z})\right]\bar{\varepsilon}_r(\rho)\sin(n\varphi+\psi_m)}{\xi\bullet\rho+\xi\bullet\varphi\tan(n\varphi+\psi_m)-\text{j}\left(a^2 n\beta\lambda_r u^{2r-4}/\rho s_r w^{2r-2}\right)\xi\bullet\mathbf{z}}$$

$$(4.1.16)$$

which are valid over the entire cross-section of the fiber, as well as for both polarization states. The transverse vectors can be found from (4.1.16) using either (3.4.3), or (3.4.5),

$$\vec{\mathbf{E}}_{mT}(\zeta) = -\text{j}\beta\sum_{r=1}^{2}f_r(\rho)\left[\boldsymbol{\rho}\Lambda_{nr}(\rho; s,\lambda_r)\cos(n\varphi+\psi_m) - \boldsymbol{\varphi}\Lambda_{nr}(\rho; s,-\lambda_r)\sin(n\varphi+\psi_m)\right]$$

$$\vec{\mathbf{H}}_{mT}(\zeta) = -\text{j}\omega_0\sum_{r=1}^{2}\bar{\varepsilon}_r(\rho)\left[\boldsymbol{\rho}\Lambda_{nr}(\rho; s_r,-\lambda_r)\sin(n\varphi+\psi_m) + \boldsymbol{\varphi}\Lambda_{nr}(\rho; s_r,\lambda_r)\cos(n\varphi+\psi_m)\right]$$

$$(4.1.17)$$

which, with respect to (4.1.3), are evidently of the general form

$$\vec{V}_T(\zeta) = \boldsymbol{\rho} V_\rho(\zeta) + \boldsymbol{\varphi} V_\varphi(\zeta). \tag{4.1.18}$$

Based on (4.1.16 or 17), it can be surmised that the Re-operation in (4.1.13-15) is actually redundant due to the cancellation of the multiplicative imaginary number j in the integrands. Applying (4.1.18) to the electric and magnetic field vectors, and substituting the resultant vectors in (4.1.14) or equivalently (4.1.13), for the power of the *m*-th polarization state of a given *n*-th hybrid mode,

$$P_m = \frac{1}{2}\operatorname{Re}\int \tilde{\mathbf{E}}_m^*(\mathbf{r},t)\times\tilde{\mathbf{H}}_m(\mathbf{r},t)\cdot\mathbf{dA} = \frac{1}{2}\int_A\left[E_\rho^*(\zeta)H_\varphi(\zeta) - E_\varphi^*(\zeta)H_\rho(\zeta)\right]dA, \quad m\in\{1,2\} \tag{4.1.19}$$

with all EM-field components given by Okamoto's complex expressions (3.1.1-12), respectively dependent on whether the cross-sectional region is the core or the cladding of the ideal fiber, or using from the bi-regional forms (3.3.27-32), or (4.1.17) in vector form. Additionally, using the general bi-regional form (4.1.3) specialized to each RHS integrand field component in (4.1.19), the power expression expands to

$$P_m = \frac{1}{2}\sum_{r_1=1}^{2}\sum_{r_2=1}^{2}\int_A\left[E_{r_1\rho}^*(\zeta)H_{r_2\varphi}(\zeta)f_{r_1}(\rho)f_{r_2}(\rho) - E_{r_1\varphi}^*(\zeta)H_{r_2\rho}(\zeta)f_{r_1}(\rho)f_{r_2}(\rho)\right]dA. \tag{4.1.20}$$

Invoking the orthogonality relation of the step-functions (3.3.24) yields the simplification

$$P_m = \frac{1}{2}\sum_{r_1=1}^{2}\sum_{r_2=1}^{2}\int_A\left[E_{r_1\rho}^*(\rho,\varphi)H_{r_2\varphi}(\rho,\varphi) - E_{r_1\varphi}^*(\rho,\varphi)H_{r_2\rho}(\rho,\varphi)\right]f_{r_2}(\rho)\delta_{r_1 r_2}\rho d\rho d\varphi, \tag{4.1.21}$$

and after enforcing the Kronecker delta in the bracketed quantity, and disposing of a redundant *r*-subscript, the expression is reduced to

$$P_m = \frac{1}{2}\sum_{r=1}^{2}\int_0^{2\pi}\int_{\mathbb{R}^+}\left[E_{r\rho}^*(\rho,\varphi)H_{r\varphi}(\rho,\varphi) - E_{r\varphi}^*(\rho,\varphi)H_{r\rho}(\rho,\varphi)\right]f_r(\rho)\rho d\rho d\varphi, \tag{4.1.22}$$

implying that each component product is evaluated over a region of the fiber, with the core being the 1st region, and the cladding, the 2nd region. The radial integral is carried out over the entire, positive real-number line. The power integral (4.1.19) is finally resolved using (4.1.17) in §**5**, in the derivation of the orthogonality relation for the hybrid modes of an ideal fiber, where it emerges as a special, degenerate case when the 2 modes involved happen to be identical instead of being dissimilar.

## 4.2 Power derivation using the bi-complex field formulation

The previous section reviewed the derivation of power based on the complex field approach, and presented nothing new. In this section, the new, bi-complex field formulation adopted for the EM-field will be used to derive an expression for the modal power. As explained in §**1**, the spatial profile of the vector (4.1.2) may be optionally defined as the real part with respect to i, of a *spatial phasor*

$$\vec{V}(\zeta) = \text{Re}_i \, \tilde{\vec{U}}(\zeta); \quad \{\vec{V}, \tilde{\vec{U}}\} \in \{\{\vec{E}, \tilde{\vec{E}}\}, \{\vec{H}, \tilde{\vec{H}}\}\} \qquad (4.2.1)$$

with the spatial vecsor given by the separable function

$$\tilde{\vec{U}}(\zeta) = \vec{U}(\rho) e^{-i\phi_m} \qquad (4.2.2)$$

for which $\vec{U}(\rho)$ may be a vector bi-complex in both i and j, but is not a phasor. Exercising this option leads to the definition of a new, universal vecsor,

$$\hat{\vec{U}}(\boldsymbol{r},t) = \tilde{\vec{U}}(\zeta) e^{j(\omega_0 t - \beta z)} \qquad (4.2.3)$$

which is seen to be complex in both i and j, and exhibits phasors in both i and j, since the arguments of the 2 complex exponentials of (4.2.2, 3) are both variable. The phasor $e^{j(\omega_0 t - \beta z)}$ with respect to j has been introduced §**1**, and is responsible for the spatiotemporal propagation of the vector (4.2.3). The 2nd phasor $e^{-i\phi_m}$ with respect to i describes the dependence of the vector on the angle $\phi$,

$$\phi_m = n\varphi + \psi_m, \quad \psi_m = (m-1)\pi/2, \quad m \in \{1,2\}, \quad n \geq 1, \qquad (4.2.4)$$

which carries the **cylindrical azimuthal variable** $\varphi$. As in the previous section, $\boldsymbol{r}$ is short-form for $(\zeta, z)$, and $\zeta$ is short-form for the cylindrical coordinate polar couple $(\rho, \varphi)$. Each component of the vecsor (4.2.2) is of the form of (3.3.41-46) or (3.3.47-52),

$$\tilde{U}_\xi(\zeta) = \boldsymbol{\xi} \cdot \tilde{\vec{U}}(\zeta) = \sum_{r=1}^{2} \tilde{U}_{r\xi}(\zeta) f_r(\rho), \quad \xi \in \{\rho, \varphi\} \qquad (4.2.5)$$

and is a phasor, expressed using the orthogonal step-functions (3.3.33).

The general form of the *instantaneous*, *real* Poynting vector $\vec{S}(\boldsymbol{r},t)$ may be derived using (4.2.2, 3), but is more easily adapted from (4.1.5) of the previous section,

$$\vec{S}(\boldsymbol{r},t) = \text{Re}_j \left[ \text{Re}_i \hat{\vec{E}}(\boldsymbol{r},t) \right] \times \text{Re}_j \left[ \text{Re}_i \hat{\vec{H}}(\boldsymbol{r},t) \right]. \qquad (4.2.6)$$

As explained in **§2**, consecutive Re-operations conducted in this expression are commutative, as j (i) is to be treated as a constant under the operation $\text{Re}_i$ ( $\text{Re}_j$ ). The general rule (2.21) derived in §2.1 will be used multiple times here. Carrying out this rule with respect to ($\mu =$) j *first*, yields

$$\vec{\mathbf{S}}(\boldsymbol{r},t) = \frac{1}{2}\text{Re}_j\left[\text{Re}_i\,\hat{\mathbf{E}}^*(\boldsymbol{r},t)\times\text{Re}_i\,\hat{\mathbf{H}}(\boldsymbol{r},t)\right] + \frac{1}{2}\text{Re}_j\left[\text{Re}_i\,\hat{\mathbf{E}}(\boldsymbol{r},t)\times\text{Re}_i\,\hat{\mathbf{H}}(\boldsymbol{r},t)\right].$$
(4.2.7)

It should be clear that the 1st term, unlike the 2nd term, is temporally independent based on the definition (4.2.3), because it carries a single conjugation in j. The 2nd term can be re-expressed as

$$\text{Re}_i\,\hat{\mathbf{E}}(\boldsymbol{r},t)\times\text{Re}_i\,\hat{\mathbf{H}}(\boldsymbol{r},t) = \text{Re}_i\,\tilde{\mathbf{E}}(\boldsymbol{\zeta})\times\text{Re}_i\,\tilde{\mathbf{H}}(\boldsymbol{\zeta})\mathrm{e}^{2\mathrm{j}(\omega_0 t - \beta z)}$$
(4.2.8)

which effectively reduces (4.2.7) to

$$\vec{\mathbf{S}}(\boldsymbol{r},t) = \frac{1}{2}\text{Re}_j\left[\text{Re}_i\,\hat{\mathbf{E}}^*(\boldsymbol{r},t)\times\text{Re}_i\,\hat{\mathbf{H}}(\boldsymbol{r},t)\right] + \frac{1}{2}\text{Re}_j\left[\text{Re}_i\,\tilde{\mathbf{E}}(\boldsymbol{\zeta})\times\text{Re}_i\,\tilde{\mathbf{H}}(\boldsymbol{\zeta})\mathrm{e}^{2\mathrm{j}(\omega_0 t - \beta z)}\right].$$
(4.2.9)

As before, the *time-averaged* or *mean* Poynting vector is given by the time-average bracket of (4.1.7),

$$\vec{S}(\boldsymbol{r}) = \left\langle\vec{\mathbf{S}}(\boldsymbol{r},t)\right\rangle = \lim_{T\to\infty}\frac{1}{T}\int_{-T/2}^{T/2}\vec{\mathbf{S}}(\boldsymbol{r},t)\,\mathrm{d}t, \quad \omega_0 = 2\pi/T$$
(4.2.10)

yielding

$$\vec{S}(\boldsymbol{r}) = \frac{1}{2}\text{Re}_j\left[\text{Re}_i\,\hat{\mathbf{E}}^*(\boldsymbol{r},t)\times\text{Re}_i\,\hat{\mathbf{H}}(\boldsymbol{r},t)\right] + \frac{1}{2}\text{Re}_j\left\{\left[\text{Re}_i\,\tilde{\mathbf{E}}(\boldsymbol{\zeta})\times\text{Re}_i\,\tilde{\mathbf{H}}(\boldsymbol{\zeta})\right]\left\langle\mathrm{e}^{2\mathrm{j}(\omega_0 t - \beta z)}\right\rangle\right\}.$$
(4.2.11)

It was shown in the previous section, specifically with (4.1.9), that the time average of the propagator $\mathrm{e}^{\mathrm{j}(\omega_0 t - \beta z)}$ vanishes. Consequently the time-averaged Poynting vector simplifies to

$$\vec{S}(\boldsymbol{r}) = \frac{1}{2}\text{Re}_j\left[\text{Re}_i\,\hat{\mathbf{E}}^*(\boldsymbol{r},t)\times\text{Re}_i\,\hat{\mathbf{H}}(\boldsymbol{r},t)\right]$$
(4.2.12)

and after expanding the real operator with respect to i using (2.21) again, but this time with ($\mu =$) i while holding j constant,

$$\vec{S}(\boldsymbol{r}) = \frac{1}{4}\text{Re}_i\,\text{Re}_j\left[\hat{\mathbf{E}}^{\circ*}(\boldsymbol{r},t)\times\hat{\mathbf{H}}(\boldsymbol{r},t) + \hat{\mathbf{E}}^*(\boldsymbol{r},t)\times\hat{\mathbf{H}}(\boldsymbol{r},t)\right],$$
(4.2.13)

where a superscript of '°' indicates conjugation with respect to i. In its long-form, given here for the sake of completeness,

$$\vec{S}(r) = \frac{1}{16}\begin{bmatrix} \hat{\mathbf{E}}(r,t) \times \hat{\mathbf{H}}^{\circ*}(r,t) + \hat{\mathbf{E}}^{\circ}(r,t) \times \hat{\mathbf{H}}^{*}(r,t) + \hat{\mathbf{E}}^{*}(r,t) \times \hat{\mathbf{H}}^{\circ}(r,t) + \hat{\mathbf{E}}^{\circ*}(r,t) \times \hat{\mathbf{H}}(r,t) \\ + \hat{\mathbf{E}}^{\circ}(r,t) \times \hat{\mathbf{H}}^{\circ*}(r,t) + \hat{\mathbf{E}}(r,t) \times \hat{\mathbf{H}}^{*}(r,t) + \hat{\mathbf{E}}^{\circ*}(r,t) \times \hat{\mathbf{H}}^{*}(r,t) + \hat{\mathbf{E}}^{*}(r,t) \times \hat{\mathbf{H}}(r,t) \end{bmatrix}.$$

(4.2.14)

It may also be surmised from (4.2.13), that the time-averaged, *bi-complex* Poynting vecsor *in one form*, is generally given by

$$\hat{S}(r) = \hat{\mathbf{E}}^{\circ*}(r,t) \times \hat{\mathbf{H}}(r,t) + \hat{\mathbf{E}}^{*}(r,t) \times \hat{\mathbf{H}}(r,t).$$ (4.2.15)

It can be seen that if the field vectors are all independent of i, the conjugations with respect to i are rendered redundant, and the expression (4.2.13) or (4.2.14) reduces to (4.1.10) for the complex formulation:

$$\vec{S}(r) = \frac{1}{4}\text{Re}_i\,\text{Re}_j\left[\tilde{\mathbf{E}}^{\circ*}(r,t) \times \tilde{\mathbf{H}}(r,t) + \tilde{\mathbf{E}}^{*}(r,t) \times \tilde{\mathbf{H}}(r,t)\right] = \frac{1}{2}\text{Re}\left[\tilde{\mathbf{E}}^{*}(r,t) \times \tilde{\mathbf{H}}(r,t)\right].$$

(4.2.16)

As before, the *time-averaged* detected power in Watts (W) is given by

$$P = \int_A \vec{S}(r) \cdot \mathbf{dA}.$$ (4.2.17)

Applying (4.2.17) to (4.2.13) yields the most general form of the power expression:

$$P = \frac{1}{4}\text{Re}_{i,j}\int_A \left[\hat{\mathbf{E}}^{\circ*}(r,t) \times \hat{\mathbf{H}}(r,t) + \hat{\mathbf{E}}^{*}(r,t) \times \hat{\mathbf{H}}(r,t)\right] \cdot \mathbf{dA}$$ (4.2.18)

**where Re$_{i,j}$ is short-form for the sequential operation Re$_i$ Re$_j$.**

Based on (4.2.2, 3) and their evidently separable forms, it can be concluded that the 2nd bracketed term in (4.2.18) resolves to

$$\frac{1}{4}\text{Re}_{i,j}\int_A \hat{\mathbf{E}}^{*}(r,t) \times \hat{\mathbf{H}}(r,t) \cdot \mathbf{dA} = \frac{1}{4}\text{Re}_{i,j}\int_{R^+} \vec{\mathbf{E}}^{*}(\rho) \times \mathbf{H}(\rho) \cdot \mathbf{z}\rho d\rho \int_0^{2\pi} e^{-2i(n\varphi + \psi_m)}d\varphi$$

(4.2.19)

where, since $n \geq 1$ according to (4.2.4),

$$\int_0^{2\pi} e^{-2i(n\varphi + \psi_m)}d\varphi = e^{-2i\psi_m}\int_0^{2\pi} e^{-i2n\varphi}d\varphi = \frac{e^{-2i\psi_m}}{-2in}\left(e^{-i4n\pi} - 1\right) = 0.$$ (4.2.20)

Consequently, the **most general expression for power flow** simplifies from (4.2.18) to

$$P = \frac{1}{4}\operatorname{Re}_{i,j}\int_A \widehat{\mathbf{E}}^{\circ *}(\boldsymbol{r},t) \times \widehat{\mathbf{H}}(\boldsymbol{r},t)\bullet \mathbf{dA} \qquad (4.2.21)$$

which has the long form of

$$P = \frac{1}{16}\int_A \left[ \begin{array}{c} \widehat{\mathbf{E}}^{\circ *}(\boldsymbol{r},t) \times \widehat{\mathbf{H}}(\boldsymbol{r},t) + \widehat{\mathbf{E}}(\boldsymbol{r},t) \times \widehat{\mathbf{H}}^{\circ *}(\boldsymbol{r},t) \\ + \widehat{\mathbf{E}}^{*}(\boldsymbol{r},t) \times \widehat{\mathbf{H}}^{\circ}(\boldsymbol{r},t) + \widehat{\mathbf{E}}^{\circ}(\boldsymbol{r},t) \times \widehat{\mathbf{H}}^{*}(\boldsymbol{r},t) \end{array} \right]\bullet \mathbf{dA}. \qquad (4.2.22)$$

Based on (4.2.3), a simpler spatial dependence is obtained in the integrand of (4.2.21) with the cancellation of the common phasor $e^{j(\omega_0 t - \beta z)}$,

$$P = \frac{1}{4}\operatorname{Re}_{i,j}\int_A \tilde{\mathbf{E}}^{\circ *}(\boldsymbol{\zeta}) \times \tilde{\mathbf{H}}(\boldsymbol{\zeta})\bullet \mathbf{dA}. \qquad (4.2.23)$$

The conjugation with respect to j is still retained however, since an EM-field vector may still be complex in j without being a phasor. This power expression can be applied to the EM-field vecsor of the *m*-th polarization mode of the *n*-th hybrid mode in the bi-complex formulation, which is expressed by (3.4.17),

$$\tilde{\mathbf{E}}_m(\boldsymbol{\zeta}) = -\mathrm{j}\beta \sum_{r=1}^{2}\sum_{\xi=\rho,\varphi,z} \boldsymbol{\xi} \frac{\Lambda_{nr}\left[\rho;s\big|\boldsymbol{\xi}\times\mathbf{z}\big|,\lambda_r(\boldsymbol{\xi}\bullet\boldsymbol{\rho}-\boldsymbol{\xi}\bullet\boldsymbol{\varphi}-\boldsymbol{\xi}\bullet\mathbf{z})\right]}{\boldsymbol{\xi}\bullet\boldsymbol{\rho}+\mathrm{i}\boldsymbol{\xi}\bullet\boldsymbol{\varphi}+\mathrm{j}\left(a^2 n\beta\lambda_r u^{2r-4}\big/\rho w^{2r-2}\right)\boldsymbol{\xi}\bullet\mathbf{z}} f_r(\rho)\mathrm{e}^{-\mathrm{i}(n\varphi+\psi_m)}$$

$$\tilde{\mathbf{H}}_m(\boldsymbol{\zeta}) = -\mathrm{i}\mathrm{j}\omega_0 \sum_{r=1}^{2}\sum_{\xi=\rho,\varphi,z} \boldsymbol{\xi} \frac{\Lambda_{nr}\left[\rho;s_r\big|\boldsymbol{\xi}\times\mathbf{z}\big|,-\lambda_r(\boldsymbol{\xi}\bullet\boldsymbol{\rho}-\boldsymbol{\xi}\bullet\boldsymbol{\varphi}+\boldsymbol{\xi}\bullet\mathbf{z})\right]}{\boldsymbol{\xi}\bullet\boldsymbol{\rho}+\mathrm{i}\boldsymbol{\xi}\bullet\boldsymbol{\varphi}-\mathrm{j}\left(a^2 n\beta\lambda_r u^{2r-4}\big/\rho s_r w^{2r-2}\right)\boldsymbol{\xi}\bullet\mathbf{z}} \bar{\varepsilon}_r(\rho)\mathrm{e}^{-\mathrm{i}(n\varphi+\psi_m)}$$

(4.2.24)

A much simpler version of (4.2.23) is obtainable by recognizing the cancellation of the phasor $e^{-\mathrm{i}(n\varphi+\psi_m)}$ common to the EM-field vectors seen in (4.2.24). This is found by applying (4.2.2) to the electric and magnetic fields, and substituting the results into (4.2.23), yielding

$$P_m = \frac{1}{4}\operatorname{Re}_{i,j}\int_A \vec{\mathbf{E}}_{mT}^{\circ *}(\rho) \times \vec{\mathbf{H}}_{mT}(\rho)\bullet \mathbf{dA}. \qquad (4.2.25)$$

Since the integral with respect to the azimuth is now trivial, the power is immediately recognized to be independent of the polarization modal index *m*, a conclusion that is not attainable from the equivalent power expression (4.1.19) of the complex formulation, *before* carrying out the integration over the azimuth. This demonstrates one advantage of using the bi-complex formulation of the EM-field. In accordance with (3.4.3) or (3.4.5) and (4.2.24), or by mere inspection of (4.2.24), the transverse field vectors required in the integrand of (4.2.25) are given by

$$\vec{E}_{mT}(\rho) = -j\beta \sum_{r=1}^{2}\left[\boldsymbol{\rho}\Lambda_{nr}(\rho;s,\lambda_r) - i\boldsymbol{\varphi}\Lambda_{nr}(\rho;s,-\lambda_r)\right]f_r(\rho)$$

$$\vec{H}_{mT}(\rho) = -j\omega_0 \sum_{r=1}^{2}\left[i\boldsymbol{\rho}\Lambda_{nr}(\rho;s_r,-\lambda_r) + \boldsymbol{\varphi}\Lambda_{nr}(\rho;s_r,\lambda_r)\right]\bar{\varepsilon}_r(\rho)$$

(4.2.26)

which, using (4.2.2), are of the general form of

$$\vec{U}_T(\rho) = \boldsymbol{\rho}\tilde{U}_\rho(\zeta)e^{+i\phi_m} + \boldsymbol{\varphi}\tilde{U}_\rho(\zeta)e^{+i\phi_m} \qquad (4.2.27)$$

since the spatial phasor (4.2.2) is separable. The transverse EM-field vectors (4.2.26) are evidently much simpler than their counterparts (4.1.17) in the complex formulation, due to the absence of trigonometric functions. Furthermore, the imaginary number j, which only appears as a multiplicative factor in both vectors, is eliminated in the cross-product of the power integrand. The second imaginary number i is likewise eliminated, since the cross-product vanishes unless it involves the radial and azimuthal unit vectors. It is therefore concluded that *for the ideal fiber*, the two real-operations are redundant in the bi-complex power integral. Thus, (4.2.21) simplifies to the following general expression, with the help of (4.2.27),

$$P = \frac{1}{4}\mathrm{Re}_{i,j}\int_A \hat{\vec{E}}_m^{\circ*}(\vec{r},t) \times \hat{\vec{H}}_m(\vec{r},t) \cdot \mathbf{dA} = \frac{\pi}{2}\int_{\mathbb{R}^+}\left[\tilde{E}_\rho^{\circ*}(\zeta)\tilde{H}_\varphi(\zeta) - \tilde{E}_\varphi^{\circ*}(\zeta)\tilde{H}_\rho(\zeta)\right]\rho\mathrm{d}\rho$$

(4.2.28)

which is comparable to (4.19), although much simpler with respect to the integration.

Specializing (4.2.5) to each field component, and substituting it into (4.2.28),

$$P = \frac{\pi}{2}\sum_{r_1=1}^{2}\sum_{r_2=1}^{2}\int_{\mathbb{R}^+}\left[\tilde{E}_{r_1\rho}^{\circ*}(\zeta)\tilde{H}_{r_2\varphi}(\zeta)f_{r_1}(\rho)f_{r_2}(\rho) - \tilde{E}_{r_1\varphi}^{\circ*}(\zeta)\tilde{H}_{r_2\rho}(\zeta)f_{r_1}(\rho)f_{r_2}(\rho)\right]\rho\mathrm{d}\rho$$

(4.2.29)

with all conjugations omitted due to their redundancy. Invoking the orthogonality relation for step-functions (3.3.24), enforcing the Kronecker delta in the bracketed quantity, and disposing of a redundant subscript,

$$P = \frac{\pi}{2}\sum_{r=1}^{2}\int_{\mathbb{R}^+}\left[\tilde{E}_{r\rho}^{\circ*}(\rho,\varphi)\tilde{H}_{r\varphi}(\rho,\varphi) - \tilde{E}_{r\varphi}^{\circ*}(\rho,\varphi)\tilde{H}_{r\rho}(\rho,\varphi)\right]f_r(\rho)\rho\mathrm{d}\rho. \qquad (4.2.30)$$

## 5. Orthogonality of modes

As in the previous sections, the *xy*-plane is assumed to be coplanar with the cross-section of the waveguide, and with the origin in coincidence with the geometric center of the waveguide. Thus, propagation is in the longitudinal, +*z*-direction, based on a right-handed coordinate system. The waveguide is also assumed to be homogeneous, isotropic, non-magnetic, time-invariant, *z*-invariant, and devoid of physical and temporal perturbations. Lastly, it is also assumed that the power carried by reflection and radiation modes is negligibly small, which is not always the case in practice.

### 5.1 The complex formulation

As a consequence of these assumptions, and in the complex formulation (4.1.2), the *normalized* field vecsor $\bar{\tilde{\mathbf{V}}}_m(\mathbf{r},t)$ is defined as the sum of its transverse $\bar{\tilde{\mathbf{V}}}_{mT}(\mathbf{r},t)$ and longitudinal $\bar{\tilde{\mathbf{V}}}_{mz}(\mathbf{r},t)$ vecsor components,

$$\bar{\tilde{\mathbf{V}}}_m(\mathbf{r},t) = \bar{\tilde{\mathbf{V}}}_{mT}(\mathbf{r},t) + \bar{\tilde{\mathbf{V}}}_{mz}(\mathbf{r},t); \quad \mathbf{V} \in \{\mathbf{E},\mathbf{H}\}, \ m \in \{1,2,\cdots,N\} \tag{5.1.1}$$

whose transverse coordinate dependence $(\zeta)$ is decoupled from that of its temporal and longitudinal (*z*) propagation through a phase factor,

$$\bar{\tilde{\mathbf{V}}}_m(\mathbf{r},t) = \bar{\vec{\mathbf{V}}}_m(\zeta) e^{j(\omega_0 t - \beta_m z)} = \frac{\vec{\mathbf{V}}_m(\zeta)}{P_m^{1/2}} e^{j(\omega_0 t - \beta_m z)}. \tag{5.1.2}$$

The vector $\vec{\mathbf{V}}_m(\zeta)$ (or its normalized counterpart $\bar{\vec{\mathbf{V}}}_m(\zeta)$) may still be complex, without being a phasor like the LHS.

Any EM-field coupled to a multi-mode waveguide, may be expressed as an expansion of the eigenmodes of that waveguide, or more explicitly in terms of their respective EM-fields;

$$\tilde{\mathbf{E}}(\mathbf{r},t) = \sum_{m=1}^{N} C_m \bar{\tilde{\mathbf{E}}}_m(\mathbf{r},t); \quad \tilde{\mathbf{H}}(\mathbf{r},t) = \sum_{m=1}^{N} C_m \bar{\tilde{\mathbf{H}}}_m(\mathbf{r},t). \tag{5.1.3}$$

A coefficient $|C_m|^2$ may be construed as a measure of the power coupled to the *m*-th mode at the input junction of the waveguide, where the EM-field given by $\tilde{\mathbf{E}}(\mathbf{r},t)$ and $\tilde{\mathbf{H}}(\mathbf{r},t)$, is incident. The *instantaneous*, *complex* Poynting vector of the entire EM-field is given by (4.1.11),

$$\tilde{\mathbf{S}}(\mathbf{r},t) = \tilde{\mathbf{E}}^*(\mathbf{r},t) \times \tilde{\mathbf{H}}(\mathbf{r},t) \tag{5.1.4}$$

which, if using (5.1.3), would yield a double-summation. For (*N* =) 2 modes for instance, as is the case of the SMF, the above expression simplifies to the single summation

$$\tilde{\mathbf{S}}(\mathbf{r},t) = \sum_{m=1}^{2} |C_m|^2 \bar{\tilde{\mathbf{E}}}_m^*(\mathbf{r},t) \times \bar{\tilde{\mathbf{H}}}_m(\mathbf{r},t) + \sum_{m=1}^{2} C_m^* C_{3-m} \bar{\tilde{\mathbf{E}}}_m^*(\mathbf{r},t) \times \bar{\tilde{\mathbf{H}}}_{3-m}(\mathbf{r},t). \tag{5.1.5}$$

Using (5.1.4), the *time-averaged* Poynting vector was found to be (4.1.10),

$$\vec{S}(r) = \frac{1}{2}\text{Re}\left[\tilde{\mathbf{S}}(r,t)\right] = \frac{1}{4}\left[\tilde{\mathbf{E}}^*(r,t)\times\tilde{\mathbf{H}}(r,t) + \tilde{\mathbf{E}}(r,t)\times\tilde{\mathbf{H}}^*(r,t)\right]. \tag{5.1.6}$$

Since propagation was assumed to be in the +z-direction at the outset, the total *time-averaged* (*mean*) power in Watts is given by

$$P = \int_A \vec{S}(r)\cdot\mathbf{dA} \tag{5.1.7}$$

for which the surface integral is over the entire dielectric cross-section of the waveguide. It was also assumed that the waveguide is devoid of perturbations and loss, implying that the power at the exit plane of the waveguide must be identical to that at the input plane of the waveguide. In other words, the power is z-invariant, like the waveguide itself:

$$\frac{\partial P}{\partial z} = \int_A \frac{\partial \vec{S}(r,t)}{\partial z}\cdot\mathbf{dA} = 0. \tag{5.1.8}$$

Upon a substitution from (5.1.6), and the elimination of a multiplicative constant,

$$\frac{\partial}{\partial z}\int_A \left[\tilde{\mathbf{E}}^*(r,t)\times\tilde{\mathbf{H}}(r,t) + \tilde{\mathbf{E}}(r,t)\times\tilde{\mathbf{H}}^*(r,t)\right]\cdot\mathbf{dA} = 0. \tag{5.1.9}$$

After substituting the expansions (5.1.3)[11],

$$\sum_{m=1}^{N}\sum_{n=1}^{N}\frac{\partial}{\partial z}\int_A \left[C_m^*C_n\bar{\tilde{\mathbf{E}}}_m^*(r,t)\times\bar{\tilde{\mathbf{H}}}_n(r,t) + C_m C_n^*\bar{\tilde{\mathbf{E}}}_m(r,t)\times\bar{\tilde{\mathbf{H}}}_n^*(r,t)\right]\cdot\mathbf{dA} = 0.$$

$$\tag{5.1.10}$$

Since the summation indices are identical in bounds, and that addition is commutative, interchanging the modal indices of the field vectors in the 2nd term of the integrand will have no effect on (5.1.10). Doing so also renders the coefficient-products identical, which permits its factorization outside the integral,

$$\sum_{m=1}^{N}\sum_{n=1}^{N}C_m^*C_n\frac{\partial}{\partial z}\int_A \left[\bar{\tilde{\mathbf{E}}}_m^*(r,t)\times\bar{\tilde{\mathbf{H}}}_n(r,t) + \bar{\tilde{\mathbf{E}}}_n(r,t)\times\bar{\tilde{\mathbf{H}}}_m^*(r,t)\right]\cdot\mathbf{dA} = 0 \tag{5.1.11}$$

in which the summand is basically the product of 2 scalars. The RHS is satisfied for several possibilities. It may be met if the product of coefficients vanishes, or if the integral vanishes, or if both vanish simultaneously. The product of coefficients may vanish, but this is only possible in the trivial case of the absence of an EM-field

---

[11] Note that the indices *m* and *n* used in this section, should not be confused with the *m* and *n* used in **§3**, till (5.1.20)

propagation in the waveguide. Therefore, the only remaining possibility is that the derivative of the integral vanishes, which, using (5.1.2), can be stated as [22]

$$\left(\beta_m - \beta_n\right) \int_A \left[ \bar{\tilde{\mathbf{E}}}_m^*(\mathbf{r},t) \times \bar{\tilde{\mathbf{H}}}_n(\mathbf{r},t) + \bar{\tilde{\mathbf{E}}}_n(\mathbf{r},t) \times \bar{\tilde{\mathbf{H}}}_m^*(\mathbf{r},t) \right] \cdot \mathbf{dA} = 0. \tag{5.1.12}$$

**This concludes the derivation of the orthogonality relation for the complex formulation.** A few observations can be made about this relation. It can be seen that the 2nd term of the integrand is a mirror-image of its 1st term with respect to the modal indices *m* and *n*, and also with respect to complex-conjugation. Moreover, only the EM-field of the *m*-th mode is conjugated in the integrand.

The orthogonality relation is a product of 2 constants, at least one of which must vanish, leading to the following possibilities:

- When the EM-fields belong to different modes, or to adjacent waveguides, as in the case of a waveguide coupler for instance, the LHS multiplicative factor $\left(\beta_m - \beta_n\right)$ is extant, requiring that

$$\int_A \left[ \bar{\tilde{\mathbf{E}}}_m^*(\mathbf{r},t) \times \bar{\tilde{\mathbf{H}}}_n(\mathbf{r},t) + \bar{\tilde{\mathbf{E}}}_n(\mathbf{r},t) \times \bar{\tilde{\mathbf{H}}}_m^*(\mathbf{r},t) \right] \cdot \mathbf{dA} = 0, \ m \neq n. \tag{5.1.13}$$

- When the EM-fields belong to modes degenerate in the propagation constant $\beta$, the multiplicative factor $\left(\beta_m - \beta_n\right)$ vanishes, but nothing can be concluded about the integral of (5.1.12), unless explicit expressions for the field vectors are substituted directly into the integrand. This is indeed the case for the ideal fiber, and will be investigated later in this section.
- When the EM-fields belong to the polarization states of the same mode, or to a single-mode, single-polarization waveguide, the multiplicative factor $\left(\beta_m - \beta_n\right)$ vanishes. However, equating *m* = *n* in (5.1.12) yields the simplification

$$\int_A \left[ \bar{\tilde{\mathbf{E}}}_m^*(\mathbf{r},t) \times \bar{\tilde{\mathbf{H}}}_m(\mathbf{r},t) + \bar{\tilde{\mathbf{E}}}_m(\mathbf{r},t) \times \bar{\tilde{\mathbf{H}}}_m^*(\mathbf{r},t) \right] \cdot \mathbf{dA} = \frac{2}{P_m} \mathrm{Re} \int_A \tilde{\mathbf{E}}_m^*(\mathbf{r},t) \times \tilde{\mathbf{H}}_m(\mathbf{r},t) \cdot \mathbf{dA}. \tag{5.1.14}$$

For the last case, recalling the expression for power (4.1.19) from §**4.1**,

$$P_m = \frac{1}{2} \mathrm{Re} \int_A \tilde{\mathbf{E}}_m^*(\mathbf{r},t) \times \tilde{\mathbf{H}}_m(\mathbf{r},t) \cdot \mathbf{dA} \tag{5.1.15}$$

and after substituting it into (5.1.14), it is concluded that

$$\int_A \left[ \bar{\tilde{\mathbf{E}}}_m^*(\mathbf{r},t) \times \bar{\tilde{\mathbf{H}}}_m(\mathbf{r},t) + \bar{\tilde{\mathbf{E}}}_m(\mathbf{r},t) \times \bar{\tilde{\mathbf{H}}}_m^*(\mathbf{r},t) \right] \cdot \mathbf{dA} = 4. \tag{5.1.16}$$

Finally, it is possible to combine the results of (5.1.13) and (5.1.16), so that **the orthogonality relation is in its most general form given by**

$$\int_A \left[ \bar{\tilde{\mathbf{E}}}_m^*(\mathbf{r},t) \times \bar{\tilde{\mathbf{H}}}_n(\mathbf{r},t) + \bar{\tilde{\mathbf{E}}}_n(\mathbf{r},t) \times \bar{\tilde{\mathbf{H}}}_m^*(\mathbf{r},t) \right] \bullet \mathbf{dA} = 4\delta_{mn} . \qquad (5.1.17)$$

In one application of this orthogonality relation, the expansion coefficients of (5.1.3) may be found for a *propagating* EM-field described in terms of the waveguide's eigenmode expansions. For the *n*-th expansion coefficient for example, the cross-product of the *n*-th mode's electric field vecsor $\bar{\tilde{\mathbf{E}}}_n^*(\mathbf{r},t)$ is taken with that of the propagating magnetic field $\tilde{\mathbf{H}}(\mathbf{r},t)$, and is added to the cross-product of the propagating electric field vecsor $\tilde{\mathbf{E}}(\mathbf{r},t)$ with that of the *n*-th mode's magnetic field $\bar{\tilde{\mathbf{H}}}_n^*(\mathbf{r},t)$, as follows

$$\int_A \left[ \bar{\tilde{\mathbf{E}}}_n^*(\mathbf{r},t) \times \tilde{\mathbf{H}}(\mathbf{r},t) + \tilde{\mathbf{E}}(\mathbf{r},t) \times \bar{\tilde{\mathbf{H}}}_n^*(\mathbf{r},t) \right] \bullet \mathbf{dA} = \sum_{m=1}^N C_m \int_A \left[ \bar{\tilde{\mathbf{E}}}_n^*(\mathbf{r},t) \times \bar{\tilde{\mathbf{H}}}_m(\mathbf{r},t) + \bar{\tilde{\mathbf{E}}}_m(\mathbf{r},t) \times \bar{\tilde{\mathbf{H}}}_n^*(\mathbf{r},t) \right] \bullet \mathbf{dA}$$

(5.1.18)

with the RHS obtained after substituting the expansions (5.1.3) for the propagating EM-field vectors of the LHS. The RHS vanishes unless *m* is identical with *n*, in accordance with (5.1.17). Solving the RHS for the coefficient, after enforcing *m* = *n* yields

$$C_n = \frac{1}{4} \int_A \left[ \bar{\tilde{\mathbf{E}}}_n^*(\mathbf{r},t) \times \tilde{\mathbf{H}}(\mathbf{r},t) + \tilde{\mathbf{E}}(\mathbf{r},t) \times \bar{\tilde{\mathbf{H}}}_n^*(\mathbf{r},t) \right] \bullet \mathbf{dA} , \quad n \in \{1,2,...,N\}$$

(5.1.19)

sometimes more explicitly stated as, using (5.1.16) with *m* = *n*,

$$C_n = \frac{\int_A \left[ \tilde{\mathbf{E}}_n^*(\mathbf{r},t) \times \tilde{\mathbf{H}}(\mathbf{r},t) + \tilde{\mathbf{E}}(\mathbf{r},t) \times \tilde{\mathbf{H}}_n^*(\mathbf{r},t) \right] \bullet \mathbf{dA}}{\left\{ \int_A \left[ \tilde{\mathbf{E}}_n^*(\mathbf{r},t) \times \tilde{\mathbf{H}}_n(\mathbf{r},t) + \tilde{\mathbf{E}}_n(\mathbf{r},t) \times \tilde{\mathbf{H}}_n^*(\mathbf{r},t) \right] \bullet \mathbf{dA} \right\}^{1/2}} . \qquad (5.1.20)$$

**If the EM-fields belong to modes that are degenerate in the propagation constant** $\beta$, as is the case for the ideal fiber, it results in the cancellation of the complex exponential $e^{j(\omega_0 t - \beta z)}$ with its complex-conjugate in each cross-product in (5.1.17). The field vectors lose their phasor attributes in the process, and the equation can be recast solely in terms of the transverse spatial coordinates $\zeta$, which is short-form for ($\rho$, $\varphi$),

$$\int_A \left[ \bar{\tilde{\mathbf{E}}}_m^*(\zeta) \times \bar{\tilde{\mathbf{H}}}_n(\zeta) + \bar{\tilde{\mathbf{E}}}_n(\zeta) \times \bar{\tilde{\mathbf{H}}}_m^*(\zeta) \right] \bullet \mathbf{dA} = 4\delta_{mn} . \qquad (5.1.21)$$

The modal indices are also retained, since the field quantities are normalized by their respective powers, and it has not yet been established that the modal powers are identical. After specializing (5.1.1) to the electric and magnetic fields, and substituting the resultant expressions in (5.1.21),

$$\int_A \left[ \bar{\bar{\mathbf{E}}}^*_{mT}(\zeta) \times \bar{\bar{\mathbf{H}}}_{nT}(\zeta) + \bar{\bar{\mathbf{E}}}_{nT}(\zeta) \times \bar{\bar{\mathbf{H}}}^*_{mT}(\zeta) \right] \cdot d\mathbf{A} = 4\delta_{mn}. \qquad (5.1.22)$$

It is evident that this relation effectively reduces to one based on the integral

$$I_{mn} = \int_A \bar{\bar{\mathbf{E}}}^*_{mT}(\zeta) \times \bar{\bar{\mathbf{H}}}_{nT}(\zeta) \cdot d\mathbf{A} \ , \quad m, n \in \{1, 2\} \qquad (5.1.23)$$

resulting in the equivalent orthogonality expression

$$I_{mn} + I^*_{nm} = 4\delta_{mn} \qquad (5.1.24)$$

since the 2nd term is just the complex conjugate of the 1st term, with its indices interchanged. The expression may be simplified further, but not without some knowledge about the specifics of the EM-field of the fiber under consideration. However, the above expression is the simplest version of the relation (5.1.17), based on the general definition (5.1.2).

As previously stated, for the ideal fiber, the HE$_{11}$-mode is supported in 2 polarization states that are degenerate in the propagation constant $\beta$. For degenerate modes, explicit expressions for the normalized, vecsors can be substituted into (5.1.17) to verify orthogonality, or in its simplest form (5.1.22),

$$\int_A \left[ \bar{\bar{\mathbf{E}}}^*_{mT}(\zeta) \times \bar{\bar{\mathbf{H}}}_{nT}(\zeta) + \bar{\bar{\mathbf{E}}}_{nT}(\zeta) \times \bar{\bar{\mathbf{H}}}^*_{mT}(\zeta) \right] \cdot d\mathbf{A} = 4\delta_{mn}, \quad m, n \in \{1, 2\}. \qquad (5.1.25)$$

For dissimilar modes, and for $m = 1$,

$$\int_A \left( \bar{\bar{\mathbf{E}}}^*_{1T}(\zeta) \times \bar{\bar{\mathbf{H}}}_{2T}(\zeta) + \bar{\bar{\mathbf{E}}}_{2T}(\zeta) \times \bar{\bar{\mathbf{H}}}^*_{1T}(\zeta) \right) \cdot d\mathbf{A} = 0, \qquad (5.1.26)$$

which is just the re-arranged, complex-conjugate of that, for $m = 2$,

$$\int_A \left( \bar{\bar{\mathbf{E}}}^*_{2T}(\zeta) \times \bar{\bar{\mathbf{H}}}_{1T}(\zeta) + \bar{\bar{\mathbf{E}}}_{1T}(\zeta) \times \bar{\bar{\mathbf{H}}}^*_{2T}(\zeta) \right) \cdot d\mathbf{A} = 0. \qquad (5.1.27)$$

Thus, there is one orthogonality relation to verify for the SMF or the HE$_{11}$-mode,

$$\int_A \left( \bar{\bar{\mathbf{E}}}^*_m(\zeta) \times \bar{\bar{\mathbf{H}}}_{3-m}(\zeta) + \bar{\bar{\mathbf{E}}}_{3-m}(\zeta) \times \bar{\bar{\mathbf{H}}}^*_m(\zeta) \right) \cdot d\mathbf{A} = 0, \quad m \in \{1, 2\}. \qquad (5.1.28)$$

The normalized EM-field **vectors** of the *k*-th polarization of the *n*-th hybrid mode are given by (3.4.7)

$$\bar{\bar{\mathbf{E}}}_k(\zeta) = \frac{-j\beta}{P_k^{1/2}} \sum_{r=1}^{2} \sum_{\xi=\rho,\varphi,z} \xi \frac{\Lambda_{nr}\left[\rho; s|\xi \times \mathbf{z}|, \lambda_r(\xi \cdot \boldsymbol{\rho} - \xi \cdot \boldsymbol{\varphi} - \xi \cdot \mathbf{z})\right]\cos\phi_k}{\xi \cdot \boldsymbol{\rho} - \xi \cdot \boldsymbol{\varphi}\cot\phi_k + \xi \cdot \mathbf{z}\left(a^2 n\beta\lambda_r u^{2r-4}/\rho w^{2r-2}\right)} f_r(\rho)$$

$$\bar{\bar{\mathbf{H}}}_k(\zeta) = \frac{-j\omega_0}{P_k^{1/2}} \sum_{r=1}^{2} \sum_{\xi=\rho,\varphi,z} \xi \frac{\Lambda_{nr}\left[\rho; s_r|\xi \times \mathbf{z}|, -\lambda_r(\xi \cdot \boldsymbol{\rho} - \xi \cdot \boldsymbol{\varphi} + \xi \cdot \mathbf{z})\right]\sin\phi_k}{\xi \cdot \boldsymbol{\rho} + \xi \cdot \boldsymbol{\varphi}\tan\phi_k - j\xi \cdot \mathbf{z}\left(a^2 n\beta\lambda_r u^{2r-4}/\rho s_r w^{2r-2}\right)} \bar{\varepsilon}_r(\rho)$$

(5.1.29)

which depend on the azimuthal eigenvalue *n*, **not to be confused with the polarization state index *n* used previously in the discussion of modal orthogonality**. Moreover, (5.1.29) is actually complex with respect to j, and in the multiplicative factor *A* used in the generating function (3.2.22), which is defined by Okamoto [1] as a complex amplitude factor. The corresponding transverse components for the *k*-th polarization vectors are given by applying either (3.4.3) or (3.4.5) to (5.1.29), yielding

$$\bar{\bar{\mathbf{E}}}_{kT}(\zeta) = \frac{-j\beta}{P_k^{1/2}} \sum_{r=1}^{2} \left[\boldsymbol{\rho}\Lambda_r(\rho; s, \lambda_r)\cos\phi_k - \boldsymbol{\varphi}\Lambda_r(\rho; s, -\lambda_r)\sin\phi_k\right] f_r(\rho)$$

$$\bar{\bar{\mathbf{H}}}_{kT}(\zeta) = \frac{-j\omega_0}{P_k^{1/2}} \sum_{r=1}^{2} \left[\boldsymbol{\rho}\Lambda_r(\rho; s_r, -\lambda_r)\sin\phi_k + \boldsymbol{\varphi}\Lambda_r(\rho; s_r, \lambda_r)\cos\phi_k\right] \bar{\varepsilon}_r(\rho)$$

(5.1.30)

For the HE$_{11}$-mode, the only mode supported by both the SMF and the multi-mode fiber, the following functions and parameters are used, reproduced here from previous sections, with some specialized to this mode by setting $n = 1$,

$$\phi_k = \varphi + (k-1)\pi/2, \quad k \in \{1, 2\};$$  (5.1.31)

$$f_r(\rho) = \left(1 - (r-1)\delta_{\rho a}\right)\mathrm{H}\left(\frac{(2-r)a - (1-r)b - \rho}{(\rho - (r-1)a)^{-1}}\right), \quad r \in \{1, 2\};$$  (5.1.32)

$$s = \frac{v^2 J_1(u) K_1(w)}{u w^2 J_1'(u) K_1(w) + u^2 w K_1'(w) J_1(u)}, \quad s_r = s\left(\frac{\beta}{k_0 n_r}\right)^2;$$  (5.1.33)

$$\Lambda_r(\rho; \eta, \pm\lambda_r) = \frac{aA}{2u}\left(\frac{uJ_1(u)}{wK_1(w)}\right)^{r-1}\left[(1-\eta)\frac{J_0^{2-r}(u\rho/a)}{K_0^{1-r}(w\rho/a)} \pm \lambda_r(1+\eta)\frac{J_2^{2-r}(u\rho/a)}{K_2^{1-r}(w\rho/a)}\right];$$

$$\eta \in \{s, s_r\}, \quad \lambda_r = \mathrm{e}^{jr\pi}$$

(5.1.34)

$$\bar{\varepsilon}_r(\rho) = \varepsilon_r f_r(\rho)$$  (5.1.35)

The general step-function (3.3.33) has been re-expressed as (5.1.32), which will prove to be in a more useful form here. The expressions (5.1.30) are not separable in the radial and the azimuthal directions. Substituting them into (5.1.23) yields the double-summation

$$I_{mn} = \frac{\omega_0 \beta}{(P_m P_n)^{1/2}} \sum_{r_1=1}^{2} \sum_{r_2=1}^{2} \iint \left( \begin{array}{l} \Lambda^*_{r_1}(\rho; s, \lambda_{r_1}) \Lambda_{r_2}(\rho; s_{r_2}, \lambda_{r_2}) \cos\phi_m \cos\phi_n \\ + \Lambda^*_{r_1}(\rho; s, -\lambda_{r_1}) \Lambda_{r_2}(\rho; s_{r_2}, -\lambda_{r_2}) \sin\phi_m \sin\phi_n \end{array} \right) f_{r_1}(\rho) \bar{\varepsilon}_{r_2}(\rho) \rho \, d\rho \, d\varphi$$

(5.1.36)

which is reduced to the following single-summation

$$I_{mn} = \frac{\omega_0 \beta}{(P_m P_n)^{1/2}} \sum_{r=1}^{2} \iint \left( \begin{array}{l} \Lambda^*_r(\rho; s, \lambda_r) \Lambda_r(\rho; s_r, \lambda_r) \cos\phi_m \cos\phi_n \\ + \Lambda^*_r(\rho; s, -\lambda_r) \Lambda_r(\rho; s_r, -\lambda_r) \sin\phi_m \sin\phi_n \end{array} \right) \bar{\varepsilon}_r(\rho) \rho \, d\rho \, d\varphi$$

(5.1.37)

with the help of (5.1.35), and the orthogonality rule for step-functions (3.3.24),

$$f_{r_1}(\rho) \bar{\varepsilon}_{r_2}(\rho) = f_{r_1}(\rho) \varepsilon_{r_2} f_{r_2}(\rho) = \varepsilon_{r_2} f_{r_2}^2(\rho) \delta_{r_1 r_2} = \varepsilon_{r_2} f_{r_2}(\rho) \delta_{r_1 r_2} = \bar{\varepsilon}_{r_2}(\rho) \delta_{r_1 r_2}. \quad (5.1.38)$$

Consequently, with the integrands being *individually* separable in the radial and azimuthal directions,

$$I_{mn} = \frac{\omega_0 \beta}{(P_m P_n)^{1/2}} \sum_{r=1}^{2} \left( \begin{array}{l} \int_{\mathbb{R}^+} \Lambda^*_r(\rho; s, \lambda_r) \Lambda(\rho; s_r, \lambda_r) \bar{\varepsilon}_r(\rho) \rho \, d\rho \int_0^{2\pi} \cos\phi_m \cos\phi_n \, d\varphi \\ + \int_{\mathbb{R}^+} \Lambda^*_r(\rho; s, -\lambda_r) \Lambda(\rho; s_r, -\lambda_r) \bar{\varepsilon}_r(\rho) \rho \, d\rho \int_0^{2\pi} \sin\phi_m \sin\phi_n \, d\varphi \end{array} \right).$$

(5.1.39)

The trigonometric integrals are recognized as orthogonality relations that are amenable to an immediate resolution: Using product-sum trigonometric identities for the product of 2 co/sinusoidal functions, and after substituting for $\phi_k$ from (5.1.31), it is concluded that

$$\int_0^{2\pi} \cos\phi_m \cos\phi_n \, d\varphi = \int_0^{2\pi} \sin\phi_m \sin\phi_n \, d\varphi = \pi \cos\left(\tfrac{\pi}{2}(m-n)\right) = \pi \delta_{mn} \quad (5.1.40)$$

since $m$ and $n$ are restricted to the set $\{1, 2\}$. Substituting (5.1.40) into (5.1.39) yields

$$I_{mn} = \frac{\pi \omega_0 \beta \delta_{mn}}{P_m} \sum_{r=1}^{2} \int_{\mathbb{R}^+} \left( \Lambda^*_r(\rho; s, -\lambda_r) \Lambda_r(\rho; s_r, -\lambda_r) + \Lambda^*_r(\rho; s, \lambda_r) \Lambda_r(\rho; s_r, \lambda_r) \right) \bar{\varepsilon}_r(\rho) \rho \, d\rho$$

(5.1.41)

The expression is clearly symmetric with respect to the indices *m* and *n*. Using (5.1.34), each integrand term is found to be of the form

$$\frac{\Lambda_r^*(\rho;s,\pm\lambda_r)\Lambda_r(\rho;s_r,\pm\lambda_r)}{\frac{4u^2}{a^2|A|^2}\left(\frac{uJ_1(u)}{wK_1(w)}\right)^{2-2r}} = \frac{1-s}{(1-s_r)^{-1}}\frac{J_0^{4-2r}\left(\frac{u\rho}{a}\right)}{K_0^{2-2r}\left(\frac{w\rho}{a}\right)} \pm \frac{2\lambda_r}{(1-s_r s)^{-1}}\frac{J_0^{2-r}\left(\frac{w\rho}{a}\right)J_2^{2-r}\left(\frac{w\rho}{a}\right)}{K_0^{1-r}\left(\frac{w\rho}{a}\right)K_2^{1-r}\left(\frac{w\rho}{a}\right)} + \frac{1+s}{(1+s_r)^{-1}}\frac{J_2^{4-2r}\left(\frac{w\rho}{a}\right)}{K_2^{2-2r}\left(\frac{w\rho}{a}\right)}$$

(5.1.42)

and is entirely real. Consequently, substituting (5.1.41) into the orthogonality relation (5.1.24), yields

$$\frac{2\pi\omega_0\beta\delta_{mn}}{P_m}\sum_{r=1}^{2}\int_{\mathbb{R}^+}\left(\Lambda_r^*(\rho;s,-\lambda_r)\Lambda_r(\rho;s_r,-\lambda_r)+\Lambda_r^*(\rho;s,\lambda_r)\Lambda_r(\rho;s_r,\lambda_r)\right)\bar{\varepsilon}_r(\rho)\rho\,d\rho = 4\delta_{mn}$$

(5.1.43)

After omitting the redundant Kronecker-delta common to both sides of (5.1.43), and cross-multiplying by the modal power *P*, it is found that

$$P = \frac{\pi\omega_0\beta}{2}\sum_{r=1}^{2}\int_{\mathbb{R}^+}\left(\Lambda_r^*(\rho;s,-\lambda_r)\Lambda_r(\rho;s_r,-\lambda_r)+\Lambda_r^*(\rho;s,\lambda_r)\Lambda_r(\rho;s_r,\lambda_r)\right)\bar{\varepsilon}_r(\rho)\rho\,d\rho$$

(5.1.44)

which is independent of the modal (polarization) indices *m* and *n*. Substituting (5.1.32) into (5.1.35), and the resultant expression into (5.1.44) yields, after simplifying the parenthesized expression into a single term with the help of a second, *q*-summation,

$$P = \frac{\pi\omega_0\beta}{2}\sum_{r=1}^{2}\sum_{q=-1}^{1}\int_{\mathbb{R}^+}\mathrm{H}\left(\frac{a(2-r)-b(1-r)-\rho}{(\rho-a(r-1))^{-1}}\right)\left(1-(r-1)\delta_{\rho a}\right)q^2\Lambda_r^*(\rho;s,q\lambda_r)\Lambda_r(\rho;s_r,q\lambda_r)\varepsilon_r\rho\,d\rho$$

(5.1.45)

The step-function limits the radial integral to a lower bound of *a*(*r*-1), which is derived from the denominator of its argument, and to an upper bound of *a*(2-*r*)-*b*(1-*r*), which is derived from the numerator of that argument. The product of the step-function with the Kronecker delta is only non-zero for *r* = 2, and reduces the step-function to unity at $\rho = a$, since it is unity over [*a*, *b*]. Consequently, the expression reduces to the 2 integrals

$$\frac{\pi\omega_0\beta}{2}\sum_{q=-1}^{1}\sum_{r=1}^{2}q^2\left[\int_{a(r-1)}^{a(2-r)-b(1-r)}\Lambda_r^*(\rho;s,q\lambda_r)\Lambda_r(\rho;s_r,q\lambda_r)\varepsilon_r\rho\,d\rho - \frac{a\varepsilon_2}{2}\Lambda_2^*(a;s,q)\Lambda_2(a;s_2,q)\int_a^b\delta_{\rho a}\,d\rho\right]$$

(5.1.46)

The 2nd bracketed integral vanishes, as explained in §**3.3**. Then (5.1.45) simplifies to the following power expression:

$$P = \frac{\pi \omega_0 \beta}{2} \sum_{r=1}^{2} \int_{a(r-1)}^{a(2-r)-b(1-r)} \left[ \Lambda_r^*(\rho;s,-\lambda_r) \Lambda_r(\rho;s_r,-\lambda_r) + \Lambda_r^*(\rho;s,-\lambda_r) \Lambda_r(\rho;s_r,-\lambda_r) \right] \varepsilon_r \rho d\rho$$

(5.1.47)

The integrand of (5.1.47) is algebraically of the form of $(X^2\sigma_- - 2\lambda_r XY\sigma_-\sigma_+ + Y^2\sigma_+) + (X^2\sigma_- + 2\lambda_r XY\sigma_-\sigma_+ + Y^2\sigma_+)$ based on (5.1.42), and therefore simplifies to $2(X^2\sigma_- + Y^2\sigma_+)$. After multiplying through by the denominator of the LHS of (5.1.42) which is common to both terms of the integrand of (5.1.47), there results

$$P = \frac{\omega_0 \beta |A|^2 S_1}{4u^2} \sum_{r=1}^{2} \varepsilon_r \left(\frac{uJ_1(u)}{wK_1(w)}\right)^{2r-2} \int_{a(r-1)}^{a(2-r)-b(1-r)} \left( \frac{1-s}{(1-s_r)^{-1}} \frac{J_0^{4-2r}(u\rho/a)}{K_0^{2-2r}(w\rho/a)} + \frac{1+s}{(1+s_r)^{-1}} \frac{J_2^{4-2r}(u\rho/a)}{K_2^{2-2r}(w\rho/a)} \right) \rho d\rho$$

(5.1.48)

where **$S_1$ is the area of the circular core.** The integral is to be carried out over 2 *r*egions: the core for which $r = 1$, and the cladding, for which $r = 2$.

The integrand can be simplified to a single term using a *p*-**summation**, as follows:

$$P = \frac{\omega_0 \beta |A|^2 S_1}{4u^2} \sum_{r=1}^{2} \sum_{p=0}^{1} \frac{(1-s\lambda_p)\varepsilon_r}{(1-s_r\lambda_p)^{-1}} \left(\frac{uJ_1(u)}{wK_1(w)}\right)^{2r-2} \int_{a(r-1)}^{a(2-r)-b(1-r)} \frac{J_{2p}^{4-2r}(u\rho/a)}{K_{2p}^{2-2r}(w\rho/a)} \rho d\rho$$

(5.1.49)

The integral is either in terms of a Bessel function of the 1st kind, or of a modified Bessel function of the 2nd kind, but not simultaneously in both, since $r$ is constrained to being *either* 1 *or* 2 for the 2 regions of the fiber. That is, the integral is equivalent to

$$\int \frac{J_{2p}^{4-2r}(u\rho/a)}{K_{2p}^{2-2r}(w\rho/a)} \rho d\rho = \frac{\left[\int J_{2p}^2(u\rho/a) \rho d\rho\right]^{2-r}}{\left[\int K_{2p}^2(w\rho/a) \rho d\rho\right]^{1-r}} .$$

(5.1.50)

The integral (5.1.49) can be thus be carried out in closed form. It yields 2 terms in general, which may be combined into 1 generalized term using a 3rd, *q*-**summation** unrelated to that used previously in (5.1.45, 46), with the result that the power

$$P = \frac{\omega_0 \beta |A|^2 S_1}{4\pi u^2} \sum_{r=1}^{2} \sum_{p=0}^{1} \sum_{q=0}^{1} \frac{(1-s\lambda_p)\lambda_q \varepsilon_r}{(1-s_r\lambda_p)^{-1}} \left(\frac{uJ_1(u)}{wK_1(w)}\right)^{2r-2} \left[ \frac{J_{2p-q}^{2-r}(u\rho/a) J_{2p+q}^{2-r}(u\rho/a)}{K_{2p-q}^{1-r}(w\rho/a) K_{2p+q}^{1-r}(w\rho/a)} \frac{\rho^2}{2} \right]_{a(r-1)}^{a(2-r)-b(1-r)}$$

(5.1.51)

After introducing a new, unit-less variable $h$, dependent on $r$, and on another new index, $l$,

$$h_{lr} = (1-l)\big((2-r) - (b/a)(1-r)\big) + l(r-1); \quad l \in \{0,1\}, r \in \{1,2\},$$

(5.1.52)

and evaluating the bracket at the 2 integration bounds, an additional pair of terms is created, which are combined by a subtraction using a 4th, *l*-summation, finally resulting in an expression for **the total power flow per polarization state**,

$$P = \frac{\omega_0 \beta \varepsilon_1 |A|^2 S_1^2}{8\pi u^2} \sum_{r=1}^{2} \sum_{l=0}^{1} \sum_{p=0}^{1} \sum_{q=0}^{1} \frac{(1-s\lambda_p)\lambda_{l+q}}{(1-s_r\lambda_p)^{-1}} \left(\frac{u^2 \varepsilon_2}{w^2 \varepsilon_1}\right)^{r-1} \frac{h_{lr}^2 J_1^{2r-2}(u) J_{2p-q}^{2-r}(h_{lr}u) J_{2p+q}^{2-r}(h_{lr}u)}{K_1^{2r-2}(w) K_{2p-q}^{1-r}(h_{lr}w) K_{2p+q}^{1-r}(h_{lr}w)}$$

(5.1.53)

which yields 16 terms in total, some of which might be redundant. In order to explore further simplifications, the power expression is now evaluated separately for the core and the cladding.

For the core, $r = 1$, and

$$P_1 = \delta_{r1} P = \frac{\omega_0 \beta \varepsilon_1 |A|^2 S_1^2}{8\pi u^2} \sum_{l=0}^{1} \sum_{p=0}^{1} \sum_{q=0}^{1} (1-s\lambda_p)(1-s_1\lambda_p)\lambda_{l+q} h_{l1}^2 J_{2p-q}(h_{l1}u) J_{2p+q}(h_{l1}u)$$

(5.1.54)

for which the summand of (5.1.53) vanishes at $l = 1$, due to (5.1.52), but is extant for $l = 0$. **The core power flow** thus simplifies to 2 summations,

$$P_1 = \frac{\omega_0 \beta \varepsilon_1 |A|^2 S_1^2}{8\pi u^2} \sum_{p=0}^{1} \sum_{q=0}^{1} (1-s\lambda_p)(1-s_1\lambda_p)\lambda_q J_{2p-q}(u) J_{2p+q}(u).$$
(5.1.55)

For the cladding, $r = 2$, and the expression is more complicated, since neither of the integration bounds is zero, unlike the case for the core. In this case, it simplifies to

$$P_2 = \delta_{r2} P = \frac{\omega_0 \beta \varepsilon_1 |A|^2 S_1^2}{8\pi u^2} \sum_{l=0}^{1} \sum_{p=0}^{1} \sum_{q=0}^{1} \left(\frac{u^2 \varepsilon_2}{w^2 \varepsilon_1} \frac{J_1^2(u)}{K_1^2(w)}\right) \frac{(1-s\lambda_p)\lambda_{l+q}}{(1-s_2\lambda_p)^{-1}} h_{l2}^2 K_{2p-q}(h_{l2}w) K_{2p+q}(h_{l2}w).$$

(5.1.56)

Since $l$ is restricted to being either 0 or 1, while $r = 2$, then (5.1.52) can be simplified with the help of Kronecker deltas, for instance, to a new variable

$$h_{l2} = (b/a)\delta_{l0} + \delta_{l1} = (a/b)^{l-1} = \kappa_l.$$

(5.1.57)

**The cladding power flow** is therefore given by the expression

$$P_2 = \frac{\omega_0 \beta \varepsilon_2 |A|^2 S_1^2}{8\pi w^2} \frac{J_1^2(u)}{K_1^2(w)} \sum_{l=0}^{1} \sum_{p=0}^{1} \sum_{q=0}^{1} (1-s\lambda_p)(1-s_2\lambda_p)\lambda_{l+q} \kappa_l^2 K_{2p-q}(\kappa_l w) K_{2p+q}(\kappa_l w).$$

(5.1.58)

Okamoto [1] assumes the cladding radius $b$ to be comparatively infinite for the power integrals, since $b \gg a$ for a practical fiber, which with (5.1.57) leads to

$$\lim_{\kappa_0 \to \infty} \kappa_0^2 K_{2p-q}(\kappa_0 w) K_{2p+q}(\kappa_0 w) = 0 \qquad (5.1.59)$$

and therefore simplifying (5.1.58) to

$$P_2 = \frac{\omega_0 \beta \varepsilon_2 |A|^2 S_1^2}{8\pi w^2} \frac{J_1^2(u)}{K_1^2(w)} \sum_{p=0}^{1}\sum_{q=0}^{1} (1-s\lambda_p)(1-s_2\lambda_p)\lambda_{q+1} K_{2p-q}(w) K_{2p+q}(w).$$

(5.1.60)

However, the approximation need not be made since an analytical result for the cladding power is still attainable without it. The most transparent, but least compact expression of **the total power flow per polarization state for the HE$_{11}$-mode**, can be found by simply adding the power expressions (5.1.55) and (5.1.58) for the 2 regions,

$$P = P_1 + P_2 = \frac{\omega_0 \beta |A|^2 S_1^2}{8\pi} \left\{ \begin{array}{l} \displaystyle\sum_{p=0}^{1}\sum_{q=0}^{1} \frac{\varepsilon_1}{u^2}(1-s\lambda_p)(1-s_1\lambda_p)\lambda_q J_{2p-q}(u) J_{2p+q}(u) \\ + \displaystyle\sum_{p=0}^{1}\sum_{q=0}^{1}\sum_{l=0}^{1} \frac{\varepsilon_2}{w^2}(1-s\lambda_p)(1-s_2\lambda_p)\lambda_{l+q} \frac{K_{2p-q}(\kappa_l w) K_{2p+q}(\kappa_l w)}{\left(K_1(w)/\kappa_l J_1(u)\right)^2} \end{array} \right\}$$

(5.1.61)

The contributions of the core and the cladding to the total power within the brace-brackets, are identifiable by their respective *r*egional dielectric constants $\varepsilon_r$, as well as the attendant normalized frequencies. It has the alternate, function-normalized form of

$$P = \frac{\omega_0 \beta |A|^2 J_1^2(u) S_1^2}{8\pi} \sum_{p=0}^{1}\sum_{q=0}^{1}\sum_{l=0}^{1}(1-s\lambda_p)\lambda_q \left\{ \begin{array}{l} \displaystyle\frac{1}{2}\frac{\varepsilon_1}{u^2}(1-s_1\lambda_p)\frac{J_{2p-q}(u)}{J_1(u)}\frac{J_{2p+q}(u)}{J_1(u)} \\ +\lambda_l \kappa_l^2 \frac{\varepsilon_2}{w^2}(1-s_2\lambda_p)\frac{K_{2p-q}(\kappa_l w)}{K_1(\kappa_l w)}\frac{K_{2p+q}(\kappa_l w)}{K_1(\kappa_l w)} \end{array} \right\}$$

(5.1.62)

An alternative expression for **the total power flow of the HE$_{11}$-mode** is obtained by combining the core and the cladding power flows using a fourth summation in the regional index *r*, which combines the 2 terms enclosed within the brace-brackets above,

$$P = \frac{\omega_0 \varepsilon_1 \beta |A|^2 S_1^2}{8\pi u^2} \sum_{r=1}^{2}\sum_{l=0}^{1}\sum_{p=0}^{1}\sum_{q=0}^{1} \left(\frac{u^2 \varepsilon_2}{w^2 \varepsilon_1}\right)^{r-1} \frac{(1-s\lambda_p)(1-s_r\lambda_p)\lambda_q J_1^{2r-2}(u) J_{2p-q}^{2-r}(u) J_{2p+q}^{2-r}(u)}{2^{2-r} \lambda_l^{1-r} \kappa_l^{2-2r} K_1^{2r-2}(w) K_{2p-q}^{1-r}(\kappa_l w) K_{2p+q}^{1-r}(\kappa_l w)}$$

(5.1.63)

This is the most compact form for the power of the HE$_{11}$-mode, using Okamoto's original parameters [1], along with 3 new ones consisting of $\lambda$ (5.1.34), $\kappa_l$ (5.1.57), and $S_1$. The expression is also independent of the polarization state index $m$, with the implication that the 2 states carry identical powers **for an ideal fiber**. Furthermore, the total power flow for the $HE_{11}$-mode has actually been known for a long time, but the above expressions which are presented here for the first time, *and prior to any approximations*, are the most concise yet.

The power contributions due to either one of the 2 regions can also be generally sifted from (5.1.63) with the help of a Kronecker delta, as

$$P_{r_0} = \delta_{rr_0} P = \frac{\omega_0 \varepsilon_1 \beta |A|^2 S_1^2}{8\pi u^2} \sum_{l=0}^{1} \sum_{p=0}^{1} \sum_{q=0}^{1} \left(\frac{u^2 \varepsilon_2}{w^2 \varepsilon_1}\right)^{r_0-1} \frac{(1-s\lambda_p)(1-s_{r_0}\lambda_p)\lambda_q J_1^{2r_0-2}(u) J_{2p-q}^{2-r_0}(u) J_{2p+q}^{2-r_0}(u)}{2^{2-r_0} \lambda_l^{1-r_0} \kappa_l^{2-2r_0} K_1^{2r_0-2}(w) K_{2p-q}^{1-r_0}(\kappa_l w) K_{2p+q}^{1-r_0}(\kappa_l w)}$$

(5.1.64)

in which $r_0$ may be either 1 for the core, or 2 for the cladding.

If the *total* power is known by measurement for a given laser angular frequency ($\omega_0$), for instance, it is then possible to estimate the propagation constant $\beta$, and then to extract the magnitude of the complex amplitude factor, as follows from (5.1.63),

$$\frac{1}{|A|} = \left[\frac{\omega_0 \varepsilon_1 \beta S_1^2}{8\pi u^2 P} \sum_{r=1}^{2} \sum_{l=0}^{1} \sum_{p=0}^{1} \sum_{q=0}^{1} \left(\frac{u^2 \varepsilon_2}{w^2 \varepsilon_1}\right)^{r-1} \frac{(1-s\lambda_p)(1-s_r\lambda_p)\lambda_q J_1^{2r-2}(u) J_{2p-q}^{2-r}(u) J_{2p+q}^{2-r}(u)}{2^{2-r} \lambda_l^{1-r} \kappa_l^{2-2r} K_1^{2r-2}(w) K_{2p-q}^{1-r}(\kappa_l w) K_{2p+q}^{1-r}(\kappa_l w)}\right]^{1/2}$$

(5.1.65)

which also assumes knowledge of the constitutive material parameters $\varepsilon_1$ and $\varepsilon_2$, and the radial dimensions $a$ and $b$. However, the phase of the complex amplitude factor $A$ is clearly not knowable from the power relations (5.1.61 - 63).

Combining (5.1.17) with (5.1.63) results in the orthogonality relation for the polarization states of the $HE_{11}$-mode, for which it is implicitly understood that the states $m$ and $n \in \{1,2\}$,

$$\int_A \left(\tilde{\mathbf{E}}_m^*(\mathbf{r},t) \times \tilde{\mathbf{H}}_n(\mathbf{r},t) + \tilde{\mathbf{E}}_n(\mathbf{r},t) \times \tilde{\mathbf{H}}_m^*(\mathbf{r},t)\right) \cdot \mathbf{dA}$$

$$= \delta_{mn} \frac{\omega_0 \varepsilon_1 \beta |A|^2 S_1^2}{2\pi u^2} \sum_{r=1}^{2} \sum_{l=0}^{1} \sum_{p=0}^{1} \sum_{q=0}^{1} \left(\frac{u^2 \varepsilon_2}{w^2 \varepsilon_1}\right)^{r-1} \frac{(1-s\lambda_p)(1-s_r\lambda_p)\lambda_q J_1^{2r-2}(u) J_{2p-q}^{2-r}(u) J_{2p+q}^{2-r}(u)}{2^{2-r} \lambda_l^{1-r} \kappa_l^{2-2r} K_1^{2r-2}(w) K_{2p-q}^{1-r}(\kappa_l w) K_{2p+q}^{1-r}(\kappa_l w)}.$$

(5.1.66)

The integral vanishes unless the polarization states are identical, which leads to the power per polarization state.

Lastly, it should be emphasized that significantly simpler expressions for the power can be found with the help of the dispersion relation, along with the weakly-guided fiber (WGF) approximation, as shown in §**6**. In the WGF approximation, the difference between the refractive indices of the core and the cladding is assumed to be of the order of 1% [1], which holds for most practical transmission fibers.

## 5.2 The bi-complex formulation

In addition to the definition of (5.1.1), a normalized field vecsor may *optionally* be defined as the real part of a bi-complex vecsor

$$\bar{\bar{\mathbf{V}}}_m(\mathbf{r},t) = \text{Re}_i\, \bar{\bar{\mathbf{U}}}_m(\mathbf{r},t), \quad \{\mathbf{U}_m, \mathbf{V}_m\} \in \{\mathbf{E}_m, \mathbf{H}_m\}, \quad m \in \{1,2\}. \tag{5.2.1}$$

For the HE$_{11}$-mode of the ideal fiber, the *m*ode-dependent angle $\phi$ (5.1.31) is given by

$$\phi_m = \varphi + \psi_m\,;\ \psi_m = (m-1)\pi/2,\quad m \in \{1,2\}\ . \tag{5.2.2}$$

The bi-complex vecsor $\bar{\bar{\mathbf{U}}}_m(\mathbf{r},t)$ used in (5.2.1), is expressed in polar form as

$$\bar{\bar{\mathbf{U}}}_m(\mathbf{r},t) = \bar{\bar{\mathbf{U}}}_m(\boldsymbol{\zeta})e^{j(\omega_0 t - \beta_m z)} = \bar{\bar{\mathbf{U}}}_m(\rho)e^{-i\phi_m}e^{j(\omega_0 t - \beta_m z)} \tag{5.2.3}$$

and is normalized with respect to its time-averaged *m*odal power. Furthermore,

$$\bar{\bar{\mathbf{U}}}_m(\boldsymbol{\zeta}) = \bar{\bar{\mathbf{U}}}_{mT}(\boldsymbol{\zeta}) + \bar{\bar{\mathbf{U}}}_{mz}(\boldsymbol{\zeta})\ . \tag{5.2.4}$$

**This approach is currently being proffered solely for cylindrical waveguides**, such as the ideal fiber, whereas the previous approach has been widely known to be valid for any waveguide, under the assumptions previously stated at the beginning of §**5**. This is an important point that is stressed here.

The orthogonality relation for the complex formulation was found to be

$$\int_A \left[ \bar{\bar{\mathbf{E}}}_m^*(\mathbf{r},t) \times \bar{\bar{\mathbf{H}}}_n(\mathbf{r},t) + \bar{\bar{\mathbf{E}}}_n(\mathbf{r},t) \times \bar{\bar{\mathbf{H}}}_m^*(\mathbf{r},t) \right] \bullet \mathbf{dA} = 4\delta_{mn}, \tag{5.2.5}$$

which is (5.1.17). After substituting (5.2.1),

$$\int_A \left[ \text{Re}_i\, \bar{\bar{\mathbf{E}}}_m^*(\mathbf{r},t) \times \text{Re}_i\, \bar{\bar{\mathbf{H}}}_n(\mathbf{r},t) + \text{Re}_i\, \bar{\bar{\mathbf{E}}}_n(\mathbf{r},t) \times \text{Re}_i\, \bar{\bar{\mathbf{H}}}_m^*(\mathbf{r},t) \right] \bullet \mathbf{dA} = 4\delta_{mn}. \tag{5.2.6}$$

Expanding each real-operation and relocating the resultant factor of 1/4 to the RHS yields

$$\int_A \begin{bmatrix} \bar{\bar{\mathbf{E}}}_m^*(\mathbf{r},t) \times \bar{\bar{\mathbf{H}}}_n^\circ(\mathbf{r},t) + \bar{\bar{\mathbf{E}}}_n^\circ(\mathbf{r},t) \times \bar{\bar{\mathbf{H}}}_m^*(\mathbf{r},t) + \bar{\bar{\mathbf{E}}}_m^{*\circ}(\mathbf{r},t) \times \bar{\bar{\mathbf{H}}}_n(\mathbf{r},t) + \bar{\bar{\mathbf{E}}}_n(\mathbf{r},t) \times \bar{\bar{\mathbf{H}}}_m^{*\circ}(\mathbf{r},t) \\ + \bar{\bar{\mathbf{E}}}_m^*(\mathbf{r},t) \times \bar{\bar{\mathbf{H}}}_n(\mathbf{r},t) + \bar{\bar{\mathbf{E}}}_n(\mathbf{r},t) \times \bar{\bar{\mathbf{H}}}_m^*(\mathbf{r},t) + \bar{\bar{\mathbf{E}}}_m^{*\circ}(\mathbf{r},t) \times \bar{\bar{\mathbf{H}}}_n^\circ(\mathbf{r},t) + \bar{\bar{\mathbf{E}}}_n^\circ(\mathbf{r},t) \times \bar{\bar{\mathbf{H}}}_m^{*\circ}(\mathbf{r},t) \end{bmatrix} \bullet \mathbf{dA} = 16\delta_{mn}$$
$$\tag{5.2.7}$$

This is the orthogonality relation sought for the bi-complex formulation. If the field vectors are considered independent of i, the equation reduces to (5.1.17) for the complex formulation:

$$\int_A \left[ \bar{\bar{\mathbf{E}}}_m^*(\mathbf{r},t) \times \bar{\bar{\mathbf{H}}}_n(\mathbf{r},t) + \bar{\bar{\mathbf{E}}}_n(\mathbf{r},t) \times \bar{\bar{\mathbf{H}}}_m^*(\mathbf{r},t) \right] \bullet d\mathbf{A} = 4\delta_{mn} . \tag{5.2.8}$$

It is now recognized that if the phase of the EM-field is a linear function of the cylindrical azimuth as in (5.2.2), the 2nd row of terms in the integrand of (5.2.7) simplifies to

$$\int_A \left[ \begin{array}{l} \bar{\bar{\mathbf{E}}}_m^*(\mathbf{r},t) \times \bar{\bar{\mathbf{H}}}_n(\mathbf{r},t) + \bar{\bar{\mathbf{E}}}_n(\mathbf{r},t) \times \bar{\bar{\mathbf{H}}}_m^*(\mathbf{r},t) \\ +\bar{\bar{\mathbf{E}}}_m^{\circ*}(\mathbf{r},t) \times \bar{\bar{\mathbf{H}}}_n^\circ(\mathbf{r},t) + \bar{\bar{\mathbf{E}}}_n^\circ(\mathbf{r},t) \times \bar{\bar{\mathbf{H}}}_m^{\circ*}(\mathbf{r},t) \end{array} \right] \bullet d\mathbf{A} = 2\mathrm{Re}_i \int_A \left[ \bar{\bar{\mathbf{E}}}_m^*(\mathbf{r},t) \times \bar{\bar{\mathbf{H}}}_n(\mathbf{r},t) + \bar{\bar{\mathbf{E}}}_n(\mathbf{r},t) \times \bar{\bar{\mathbf{H}}}_m^*(\mathbf{r},t) \right] \bullet d\mathbf{A}$$
(5.2.9)

and resolves to

$$\int_A \left[ \bar{\bar{\mathbf{E}}}_m^*(\mathbf{r},t) \times \bar{\bar{\mathbf{H}}}_n(\mathbf{r},t) + \bar{\bar{\mathbf{E}}}_n(\mathbf{r},t) \times \bar{\bar{\mathbf{H}}}_m^*(\mathbf{r},t) \right] \bullet d\mathbf{A} =$$

$$\int_{R^+} \left[ \bar{\bar{\mathbf{E}}}_m^*(\rho) \times \bar{\bar{\mathbf{H}}}_n(\rho) e^{j(\beta_m - \beta_n)z} + \bar{\bar{\mathbf{E}}}_n(\rho) \times \bar{\bar{\mathbf{H}}}_m^*(\rho) e^{-j(\beta_m - \beta_n)z} \right] \bullet \mathbf{z} d\rho \int_0^{2\pi} e^{-i(\phi_m + \phi_n)} d\varphi$$
(5.2.10)

where, due to (5.2.2),

$$\int_0^{2\pi} e^{-i(\phi_m + \phi_n)} d\varphi = e^{-i(\psi_m + \psi_n)} \int_0^{2\pi} e^{-i2\varphi} d\varphi = \frac{e^{-i(\psi_m + \psi_n)}}{-2i} \left( e^{-i4\pi} - 1 \right) = 0 . \tag{5.2.11}$$

Consequently, the 2nd row of terms in (5.2.7), or (5.2.9), integrates to zero, so that the orthogonality relation (5.2.7) reduces to

$$\int_A \left[ \begin{array}{l} \bar{\bar{\mathbf{E}}}_m^{\circ*}(\mathbf{r},t) \times \bar{\bar{\mathbf{H}}}_n(\mathbf{r},t) + \bar{\bar{\mathbf{E}}}_n(\mathbf{r},t) \times \bar{\bar{\mathbf{H}}}_m^{\circ*}(\mathbf{r},t) \\ +\bar{\bar{\mathbf{E}}}_m^*(\mathbf{r},t) \times \bar{\bar{\mathbf{H}}}_n^\circ(\mathbf{r},t) + \bar{\bar{\mathbf{E}}}_n^\circ(\mathbf{r},t) \times \bar{\bar{\mathbf{H}}}_m^*(\mathbf{r},t) \end{array} \right] \bullet d\mathbf{A} = 16\delta_{mn} . \tag{5.2.12}$$

When the field vectors belong to the same EM-field, as is the case for a single-mode waveguide, it leads to the following expression for the modal power:

$$P_m = \frac{1}{16} \int_A \left[ \begin{array}{l} \hat{\mathbf{E}}_m^{\circ*}(\mathbf{r},t) \times \hat{\mathbf{H}}_m(\mathbf{r},t) + \hat{\mathbf{E}}_m(\mathbf{r},t) \times \hat{\mathbf{H}}_m^{\circ*}(\mathbf{r},t) \\ +\hat{\mathbf{E}}_m^*(\mathbf{r},t) \times \hat{\mathbf{H}}_m^\circ(\mathbf{r},t) + \hat{\mathbf{E}}_m^\circ(\mathbf{r},t) \times \hat{\mathbf{H}}_m^*(\mathbf{r},t) \end{array} \right] \bullet d\mathbf{A} \tag{5.2.13}$$

which is indeed (4.2.22), when applied to the $m$-th mode of the EM-field.
**If the EM-fields belong to modes that are degenerate in the propagation**

constant $\beta$, it results in a cancellation of the complex exponential $e^{j(\omega_0 t - \beta z)}$ and its conjugate in each cross-product in (5.2.7). The field vectors lose their phasor attributes with respect to j in the process, and the equation can be recast solely in terms of the transverse spatial coordinate $\zeta$, which is short-form for $(\rho, \varphi)$,

$$\int_A \left[ \bar{\tilde{\mathbf{E}}}_m^{\circ *}(\zeta) \times \bar{\tilde{\mathbf{H}}}_n(\zeta) + \bar{\tilde{\mathbf{E}}}_n(\zeta) \times \bar{\tilde{\mathbf{H}}}_m^{\circ *}(\zeta) + \bar{\tilde{\mathbf{E}}}_m^*(\zeta) \times \bar{\tilde{\mathbf{H}}}_n^\circ(\zeta) + \bar{\tilde{\mathbf{E}}}_n^\circ(\zeta) \times \bar{\tilde{\mathbf{H}}}_m^*(\zeta) \right] \cdot \mathbf{dA} = 16 \delta_{mn}$$

(5.2.14)

**which is comprised of phasors in the azimuth, or with respect to i, only**. Moreover, the field quantities involved in each cross-products must all be transverse, otherwise the scalar triple-product vanishes. This is rigorously found by substituting (5.2.4) for the electric and magnetic fields, into (5.2.14), yielding

$$\int_A \left[ \bar{\tilde{\mathbf{E}}}_{mT}^{\circ *}(\zeta) \times \bar{\tilde{\mathbf{H}}}_{nT}(\zeta) + \bar{\tilde{\mathbf{E}}}_{nT}(\zeta) \times \bar{\tilde{\mathbf{H}}}_{mT}^{\circ *}(\zeta) + \bar{\tilde{\mathbf{E}}}_{mT}^*(\zeta) \times \bar{\tilde{\mathbf{H}}}_{nT}^\circ(\zeta) + \bar{\tilde{\mathbf{E}}}_{nT}^\circ(\zeta) \times \bar{\tilde{\mathbf{H}}}_{mT}^*(\zeta) \right] \cdot \mathbf{dA} = 16 \delta_{mn}$$

(5.2.15)

and is expressible as

$$I_{mn} + I_{mn}^\circ = 16 \delta_{mn} \qquad (5.2.16)$$

which is basically the real-part with respect to i, within a factor of 2, of the modally dependent integral given by

$$I_{mn} = \int_A \left[ \bar{\tilde{\mathbf{E}}}_{mT}^{\circ *}(\zeta) \times \bar{\tilde{\mathbf{H}}}_{nT}(\zeta) + \bar{\tilde{\mathbf{E}}}_{nT}(\zeta) \times \bar{\tilde{\mathbf{H}}}_{mT}^{\circ *}(\zeta) \right] \cdot \mathbf{dA} . \qquad (5.2.17)$$

Thus, only one integral need be carried out to verify (5.2.15,16) for the ideal fiber, the result of which may not be complex in anyway, as evidenced by the RHS of these equations.

For the *k*-th polarization state of the *normalized* EM-field of the ideal fiber, given by (3.4.17) in the bi-complex formulation,

$$\bar{\tilde{\mathbf{E}}}_k(\zeta) = \frac{-j\beta}{P_k^{1/2}} \sum_{r=1}^{2} \sum_{\xi=\rho,\varphi,z} \xi \frac{\Lambda_{nr}\left[\rho; s|\xi\times\mathbf{z}|, \lambda_r\left(\xi\cdot\rho - \xi\cdot\varphi - \xi\cdot\mathbf{z}\right)\right]}{\xi\cdot\rho + i\xi\cdot\varphi + j\left(a^2 n \beta \lambda_r u^{2r-4}/\rho w^{2r-2}\right)\xi\cdot\mathbf{z}} f_r(\rho) e^{-i(n\varphi+\psi_k)}$$

$$\bar{\tilde{\mathbf{H}}}_k(\zeta) = \frac{-ij\omega_0}{P_k^{1/2}} \sum_{r=1}^{2} \sum_{\xi=\rho,\varphi,z} \xi \frac{\Lambda_{nr}\left[\rho; s_r|\xi\times\mathbf{z}|, -\lambda_r\left(\xi\cdot\rho - \xi\cdot\varphi + \xi\cdot\mathbf{z}\right)\right]}{\xi\cdot\rho + i\xi\cdot\varphi - j\left(a^2 n \beta \lambda_r u^{2r-4}/\rho s_r w^{2r-2}\right)\xi\cdot\mathbf{z}} \bar{\varepsilon}_r(\rho) e^{-i(n\varphi+\psi_k)}$$

(5.2.18)

with parameters given by (5.1.31-35), for the HE$_{11}$-mode. **The azimuthal eigenvalue *n* that appears in (5.2.18) should not be confused with the polarization state *n* used in (5.2.17), for instance.** The corresponding transverse components for the *k*-th polarization vecsors for the HE$_{11}$-mode are given by applying either (3.4.3) or (3.4.5) to (5.2.18),

$$\bar{\tilde{\mathbf{E}}}_{kT}(\zeta) = \frac{-j\beta}{P_k^{1/2}} \sum_{r=1}^{2} \left[ \boldsymbol{\rho}\Lambda_r(\rho;s,\lambda_r) - i\boldsymbol{\varphi}\Lambda_r(\rho;s,-\lambda_r) \right] f_r(\rho) e^{-i(\varphi+\psi_k)}$$

$$\bar{\tilde{\mathbf{H}}}_{kT}(\zeta) = \frac{-j\omega_0}{P_k^{1/2}} \sum_{r=1}^{2} \left[ i\boldsymbol{\rho}\Lambda_r(\rho;s_r,-\lambda_r) + \boldsymbol{\varphi}\Lambda_r(\rho;s_r,\lambda_r) \right] \bar{\varepsilon}_r(\rho) e^{-i(\varphi+\psi_k)}$$

(5.2.19)

Consequently, it is immediately found that for (5.2.17),

$$\int_A \bar{\tilde{\mathbf{E}}}_{mT}^{\circ*}(\zeta) \times \bar{\tilde{\mathbf{H}}}_{nT}(\zeta) \cdot \mathbf{dA} = \int_A \bar{\tilde{\mathbf{E}}}_{nT}(\zeta) \times \bar{\tilde{\mathbf{H}}}_{mT}^{\circ*}(\zeta) \cdot \mathbf{dA}.$$

(5.2.20)

Based on (5.2.19), the 2 integrals that constitute (5.2.17) are evidently identical for the HE$_{11}$-mode, since the dependence of the vecsors on the modal indices is limited to a phase factor, which is identical for both integrands because the conjugation is constrained to the *m*-th mode field vector in each integrand. This simplifies the modally dependent integral (5.2.17) to a single integrand

$$I_{mn} = 2 \int_A \bar{\tilde{\mathbf{E}}}_{mT}^{\circ*}(\zeta) \times \bar{\tilde{\mathbf{H}}}_{nT}(\zeta) \cdot \mathbf{dA}$$

(5.2.21)

**so that the orthogonality relation (5.2.12) for the ideal fiber simplifies further to**

$$\int_A \left[ \bar{\tilde{\mathbf{E}}}_{mT}^{\circ*}(\zeta) \times \bar{\tilde{\mathbf{H}}}_{nT}(\zeta) + \bar{\tilde{\mathbf{E}}}_{mT}^{*}(\zeta) \times \bar{\tilde{\mathbf{H}}}_{nT}^{\circ}(\zeta) \right] \cdot \mathbf{dA} = 8\delta_{mn},$$

(5.2.22)

but since the dependence on j is limited to a multiplicative factor as seen in (5.2.19), then

$$\int_A \left[ \bar{\tilde{\mathbf{E}}}_{mT}^{\circ*}(\zeta) \times \bar{\tilde{\mathbf{H}}}_{nT}(\zeta) + \bar{\tilde{\mathbf{E}}}_{mT}(\zeta) \times \bar{\tilde{\mathbf{H}}}_{nT}^{\circ*}(\zeta) \right] \cdot \mathbf{dA} = 8\delta_{mn}$$

(5.2.23)

is also valid. **It is the simplest form of the orthogonality relation (5.2.12) for the HE$_{11}$-mode in the bi-complex formulation. It is comprised of phasors in i only.** The $(\mathbf{r},t)$ dependence may be reconstructed if desired simply by adding back the redundant *z*- components to each vector, and multiplying each integrand term by unity, in the form of $e^{j(\omega_0 t - \beta z)} \cdot e^{-j(\omega_0 t - \beta z)}$, thereby reconstituting (5.2.4), since the propagation constants are identical for the 2 modes of the ideal fiber. Doing so yields the following form of (5.2.23):

$$\int_A \left[ \bar{\bar{\mathbf{E}}}_m^{\circ *}(\mathbf{r},t) \times \bar{\bar{\mathbf{H}}}_n(\mathbf{r},t) + \bar{\bar{\mathbf{E}}}_m(\mathbf{r},t) \times \bar{\bar{\mathbf{H}}}_n^{\circ *}(\mathbf{r},t) \right] \cdot d\mathbf{A} = 8\delta_{mn}. \tag{5.2.24}$$

Apart from a factor of 8 on the RHS, the expression is similar in form to (5.1.17) of the complex formulation, with the exception that the electric field vector of one of the modes, and the magnetic field vector of the other mode are each conjugated twice, instead of a single conjugation applied to the EM-field of one of the modes. As a consequence of (5.2.21, 23), (5.2.16) may be re-cast as:

$$I_{mn} + I_{mn}^{\circ *} = 16\delta_{mn}. \tag{5.2.25}$$

In order to derive an expression for the power, (5.2.19) is substituted into (5.2.21),

$$I_{mn} = \frac{2\omega_0 \beta}{(P_m P_n)^{1/2}} \sum_{r_1=1}^{2} \sum_{r_2=1}^{2} \int_{\mathbb{R}^+} \begin{pmatrix} \Lambda_{r_1}^{\circ *}(\rho;s,\lambda_r)\Lambda_{r_2}(\rho;s_r,\lambda_r) \\ +\Lambda_{r_1}^{\circ *}(\rho;s,-\lambda_r)\Lambda_{r_2}(\rho;s_r,-\lambda_r) \end{pmatrix} \bar{\varepsilon}_{r_2}(\rho) f_{r_1}(\rho) \rho d\rho \int_0^{2\pi} e^{i(\psi_m - \psi_n)} d\varphi$$

(5.2.26)

with the radial integral being over the positive real-number line, but is constrained to the cross-section of the SMF through the regionally dependent functions $\bar{\varepsilon}_{r_2}(\rho)$ (5.1.35) and $f_r(\rho)$ (5.1.32). Substituting from (5.2.2) for the phase factor, and carrying out the integral over the azimuth yields

$$I_{mn} = \frac{4\pi\omega_0 \beta}{(P_m P_n)^{1/2}} \sum_{r=1}^{2} \int_{\mathbb{R}^+} \begin{pmatrix} \Lambda_r^{\circ *}(\rho;s,\lambda_r)\Lambda_r(\rho;s_r,\lambda_r) \\ +\Lambda_r^{\circ *}(\rho;s,-\lambda_r)\Lambda_r(\rho;s_r,-\lambda_r) \end{pmatrix} \bar{\varepsilon}_r(\rho) \rho d\rho\, e^{i(m-n)\pi/2}$$

(5.2.27)

which has been simplified to a single summation with the help of (3.3.24). Substituting (5.2.27) into (5.2.25) yields

$$\frac{8\pi\omega_0 \beta}{(P_m P_n)^{1/2}} \sum_{r=1}^{2} \int_{\mathbb{R}^+} \begin{pmatrix} \Lambda_r^{\circ *}(\rho;s,\lambda_r)\Lambda_r(\rho;s_r,\lambda_r) \\ +\Lambda_r^{\circ *}(\rho;s,-\lambda_r)\Lambda_r(\rho;s_r,-\lambda_r) \end{pmatrix} \bar{\varepsilon}_r(\rho) \rho d\rho \cos\left(\tfrac{\pi}{2}(m-n)\right) = 16\delta_{mn}$$

(5.2.28)

The LHS is only non-zero for $m = n$ due to the identity

$$\cos\left(\tfrac{\pi}{2}(m-n)\right)=\delta_{mn}\ ;\quad m,n \in \{1,2\}, \tag{5.2.29}$$

which agrees with the RHS, implying that the residual radial integral together with its multiplicative factor, must numerically resolve to 16, or equivalently, that the modal power is expressible as

$$P = \frac{\pi \omega_0 \beta}{2} \sum_{r=1}^{2} \int_{\mathbb{R}^+} \left( \Lambda_r^{\circ *}(\rho;s,\lambda_r) \Lambda_r(\rho;s_r,\lambda_r) + \Lambda_r^{\circ *}(\rho;s,-\lambda_r) \Lambda_r(\rho;s_r,-\lambda_r) \right) \bar{\varepsilon}_r(\rho) \rho \, d\rho \tag{5.2.30}$$

after the elimination of redundant Kronecker delta from both sides of the equation. The expression is seen to be independent of the modal indices. However, this is the same integral (5.1.44) encountered for the complex formulation of the previous section, since the generating function (3.2.22) is entirely real. It has also been attained with less algebra and calculus, than in the previous section. Its solution, in one form, is (5.1.63).

Finally, it is concluded that in the bi-complex formulation, the orthogonality relation (5.2.24) for the polarization modes or states of the $HE_{11}$-mode (or for the SMF) is given by,

$$\int_A \left[ \widehat{\mathbf{E}}_m^{\circ *}(\mathbf{r},t) \times \widehat{\mathbf{H}}_n(\mathbf{r},t) + \widehat{\mathbf{E}}_m(\mathbf{r},t) \times \widehat{\mathbf{H}}_n^{\circ *}(\mathbf{r},t) \right] \cdot d\mathbf{A}$$
$$= \delta_{mn} \frac{\omega_0 \varepsilon_1 \beta |A|^2 S_1^2}{\pi u^2} \sum_{r=1}^{2} \sum_{l=0}^{1} \sum_{p=0}^{1} \sum_{q=0}^{1} \left(\frac{u^2 \varepsilon_2}{w^2 \varepsilon_1}\right)^{r-1} \frac{(1-s\lambda_p)(1-s_r\lambda_p)\lambda_q J_1^{2r-2}(u) J_{2p-q}^{2-r}(u) J_{2p+q}^{2-r}(u)}{2^{2-r} \lambda_l^{1-r} \kappa_l^{2-2r} K_1^{2r-2}(\kappa_1 w) K_{2p-q}^{1-r}(\kappa_l w) K_{2p+q}^{1-r}(\kappa_l w)}. \tag{5.2.31}$$

with $m$ and $n$ both $\in \{1, 2\}$. Compared to the orthogonality relation derived using the conventional complex formulation (5.1.66),

$$\int_A \left( \tilde{\mathbf{E}}_m^*(r,t) \times \tilde{\mathbf{H}}_n(r,t) + \tilde{\mathbf{E}}_n(r,t) \times \tilde{\mathbf{H}}_m^*(r,t) \right) \cdot d\mathbf{A}$$
$$= \delta_{mn} \frac{\omega_0 \varepsilon_1 \beta |A|^2 S_1^2}{2\pi u^2} \sum_{r=1}^{2} \sum_{l=0}^{1} \sum_{p=0}^{1} \sum_{q=0}^{1} \left(\frac{u^2 \varepsilon_2}{w^2 \varepsilon_1}\right)^{r-1} \frac{(1-s\lambda_p)(1-s_r\lambda_p)\lambda_q J_1^{2r-2}(u) J_{2p-q}^{2-r}(u) J_{2p+q}^{2-r}(u)}{2^{2-r} \lambda_l^{1-r} \kappa_l^{2-2r} K_1^{2r-2}(\kappa_1 w) K_{2p-q}^{1-r}(\kappa_l w) K_{2p+q}^{1-r}(\kappa_l w)}. \tag{5.2.32}$$

it is observed that the RHS differs by a factor of 2, because the RHS of (5.1.17) is smaller than that of (5.2.24) by that factor. As shown in this section, deriving the relation in the bi-complex convention requires slightly more effort to configure. However, its application to the $HE_{11}$-mode or the ideal SMF, turns out to be simpler than that of the complex formulation, because it obviates the need for trigonometric integrations, whereas the radial integrations are identical in both approaches.

## 6. Generalizations under the weakly-guided fiber (WGF) approximation

In addition to the many assumptions made about the optical fiber in §**1**, a low refractive index contrast is assumed in this section. The **index contrast** of a step-index fiber is defined as [1]

$$\Delta = \frac{(\text{NA})^2}{2n_1^2} = \frac{n_1^2 - n_2^2}{2n_1^2}. \tag{6.1}$$

The index contrast is sometimes stated as a percentage after a multiplication by 100 [1]. NA is the numerical aperture of the fiber [23], a nomenclature inherited from geometrical optics. It is not as meaningful for the optical fiber, especially a single-mode fiber, because its acceptance angle is not determined solely by its refractive indices, but requires a consideration of diffraction or physical optics. In **the WGF approximation**, the refractive indices of the fiber must meet the criterion that

$$n_1/n_2 \approx 1 \tag{6.2}$$

**while still being larger than 1**, which simplifies the index contrast to

$$\Delta \approx \frac{2n_1 \Delta n}{2n_1^2} = \frac{\Delta n}{n_1} \ll 1. \tag{6.3}$$

This is indeed the case for practical, single-mode telecommunication fibers [23]. For instance, the index contrast for the Corning SMF-28e® fiber is 0.36% [23], whereas that for the Corning ClearCurve® Multimode fiber is approximately 1% [24]. The WGF approximation is widely attributed to Snyder [25], who was the first to develop a comprehensive, approximate theory for the EM-fields of the modes of such fibers. In his original report [25], he also cites others [2, 26] whose work helped lead him to his theoretical treatment. Okamoto's definition of the index contrast (6.1) is actually twice that of Snyder's, but this dissimilarity is not significant to the following analysis.

For a propagating mode in an optical waveguide, the effective index ($n_{\text{eff}}$), which is its eigenvalue, is constrained to be smaller than the refractive index of the core, but larger than that of the cladding. Since the refractive indices of the core and cladding materials of a practical fiber are nearly identical to within 1% in accordance with (6.2, 3), the effective index itself must therefore be *almost* identical to these material indices,

$$n_{\text{eff}} \approx n_r; \quad r \in \{1, 2\}, \tag{6.4}$$

while the effective index is still constrained to being no larger than the index of the core, but no smaller than that of the cladding. An immediate consequence of this approximation is the reduction of Okamoto's *s*-parameter to a constant that approximates to positive (negative) unity for EH (HE) modes,

$$s \approx \pm 1, \tag{6.5}$$

$$s_r = s(\beta/k_0 n_r)^2 = s(n_{\text{eff}}/n_r)^2 \approx \pm 1; \quad r \in \{1, 2\}. \tag{6.6}$$

Depending on the value of the *s*-parameter under the WGF approximation, the general EM-field yields the EH- and HE-modes. Each will be considered in turn, before another generalization is carried out.

In **the complex formulation** under the WGF approximation, **the EH-mode components are obtained from (3.2.1-6) by enforcing $s \approx s_r \approx +1$**, yielding

$$E_{r\rho}(\zeta) = -j\beta\lambda_r \frac{aA}{u}\left(\frac{uJ_n(u)}{wK_n(w)}\right)^{r-1} \frac{J_{n+1}^{2-r}(u\rho/a)}{K_{n+1}^{1-r}(w\rho/a)}\cos\phi_m \qquad (6.7)$$

$$E_{r\varphi}(\zeta) = -j\beta\lambda_r \frac{aA}{u}\left(\frac{uJ_n(u)}{wK_n(w)}\right)^{r-1} \frac{J_{n+1}^{2-r}(u\rho/a)}{K_{n+1}^{1-r}(w\rho/a)}\sin\phi_m \qquad (6.8)$$

$$E_{rz}(\zeta) = A\frac{J_n^{r-1}(u)}{K_n^{r-1}(w)}\frac{J_n^{2-r}(u\rho/a)}{K_n^{1-r}(w\rho/a)}\cos\phi_m \qquad (6.9)$$

$$H_{r\rho}(\zeta) = j\omega_0\varepsilon_r\lambda_r \frac{aA}{u}\left(\frac{uJ_n(u)}{wK_n(w)}\right)^{r-1} \frac{J_{n+1}^{2-r}(u\rho/a)}{K_{n+1}^{1-r}(w\rho/a)}\sin\phi_m \qquad (6.10)$$

$$H_{r\varphi}(\zeta) = -j\omega_0\varepsilon_r\lambda_r \frac{aA}{u}\left(\frac{uJ_n(u)}{wK_n(w)}\right)^{r-1} \frac{J_{n+1}^{2-r}(u\rho/a)}{K_{n+1}^{1-r}(w\rho/a)}\cos\phi_m \qquad (6.11)$$

$$H_{rz}(\zeta) = -\frac{A\beta}{\omega_0\mu_0}\frac{J_n^{r-1}(u)}{K_n^{r-1}(w)}\frac{J_n^{2-r}(u\rho/a)}{K_n^{1-r}(w\rho/a)}\sin\phi_m \qquad (6.12)$$

**The HE-mode components are also obtained from (3.2.1-6), but with $s \approx s_r \approx -1$**,

$$E_{r\rho}(\zeta) = -j\beta \frac{aA}{u}\left(\frac{uJ_n(u)}{wK_n(w)}\right)^{r-1} \frac{J_{n-1}^{2-r}(u\rho/a)}{K_{n-1}^{1-r}(w\rho/a)}\cos\phi_m \qquad (6.13)$$

$$E_{r\varphi}(\zeta) = j\beta \frac{aA}{u}\left(\frac{uJ_n(u)}{wK_n(w)}\right)^{r-1} \frac{J_{n-1}^{2-r}(u\rho/a)}{K_{n-1}^{1-r}(w\rho/a)}\sin\phi_m \qquad (6.14)$$

$$E_{rz}(\zeta) = A\frac{J_n^{r-1}(u)}{K_n^{r-1}(w)}\frac{J_n^{2-r}(u\rho/a)}{K_n^{1-r}(w\rho/a)}\cos\phi_m \qquad (6.15)$$

$$H_{r\rho}(\zeta) = -j\omega_0\varepsilon_r \frac{aA}{u}\left(\frac{uJ_n(u)}{wK_n(w)}\right)^{r-1} \frac{J_{n-1}^{2-r}(u\rho/a)}{K_{n-1}^{1-r}(w\rho/a)}\sin\phi_m \qquad (6.16)$$

$$H_{r\varphi}(\zeta) = -j\omega_0\varepsilon_r \frac{aA}{u}\left(\frac{uJ_n(u)}{wK_n(w)}\right)^{r-1} \frac{J_{n-1}^{2-r}(u\rho/a)}{K_{n-1}^{1-r}(w\rho/a)}\cos\phi_m \qquad (6.17)$$

$$H_{rz}(\zeta) = \frac{A\beta}{\omega_0\mu_0}\frac{J_n^{r-1}(u)}{K_n^{r-1}(w)}\frac{J_n^{2-r}(u\rho/a)}{K_n^{1-r}(w\rho/a)}\sin\phi_m \qquad (6.18)$$

The variable $\zeta$ is being used as short-form for the polar couple $(\rho, \varphi)$, as in previous sections. The components of EH- and HE-modes can all be generalized by 6 expressions:

$$E_{r\rho}(\zeta) = -j\beta\lambda_r^{(1+s)/2}\frac{aA}{u}\left(\frac{uJ_n(u)}{wK_n(w)}\right)^{r-1}\frac{J_{n+s}^{2-r}(u\rho/a)}{K_{n+s}^{1-r}(w\rho/a)}\cos\phi_m \tag{6.19}$$

$$E_{r\varphi}(\zeta) = -j\beta s\lambda_r^{(1+s)/2}\frac{aA}{u}\left(\frac{uJ_n(u)}{wK_n(w)}\right)^{r-1}\frac{J_{n+s}^{2-r}(u\rho/a)}{K_{n+s}^{1-r}(w\rho/a)}\sin\phi_m \tag{6.20}$$

$$E_{rz}(\zeta) = A\frac{J_n^{r-1}(u)}{K_n^{r-1}(w)}\frac{J_n^{2-r}(u\rho/a)}{K_n^{1-r}(w\rho/a)}\cos\phi_m \tag{6.21}$$

$$H_{r\rho}(\zeta) = j\omega_0\varepsilon_r s\lambda_r^{(1+s)/2}\frac{aA}{u}\left(\frac{uJ_n(u)}{wK_n(w)}\right)^{r-1}\frac{J_{n+s}^{2-r}(u\rho/a)}{K_{n+s}^{1-r}(w\rho/a)}\sin\phi_m \tag{6.22}$$

$$H_{r\varphi}(\zeta) = -j\omega_0\varepsilon_r\lambda_r^{(1+s)/2}\frac{aA}{u}\left(\frac{uJ_n(u)}{wK_n(w)}\right)^{r-1}\frac{J_{n+s}^{2-r}(u\rho/a)}{K_{n+s}^{1-r}(w\rho/a)}\cos\phi_m \tag{6.23}$$

$$H_{rz}(\zeta) = -\frac{A\beta s}{\omega_0\mu_0}\frac{J_n^{r-1}(u)}{K_n^{r-1}(w)}\frac{J_n^{2-r}(u\rho/a)}{K_n^{1-r}(w\rho/a)}\sin\phi_m \tag{6.24}$$

which yields a total of 24 equations, 6 for the EH-modes (6.7-12), obtained when $s = +1$, and another 6 for the HE-modes (6.13-18), obtained when $s = -1$. Each set of 6 equations represents 6 equations for the EM-field over the core when $r = 1$, and another 6 for the EM-field over the cladding when $r = 2$.

A further simplification is obtained upon consideration of the dispersion relation under the WGF approximation, which simplifies to [1]:

$$\frac{J_n'(u)}{uJ_n(u)} + \frac{K_n'(w)}{wK_n(w)} = \pm n\frac{v^2}{u^2w^2} \tag{6.25}$$

and which is valid for **$n \geq 1$**. After applying the recurrence relations [1] of the Bessel functions to eliminate the radial derivatives of the Bessel functions on the LHS of (6.25), the following dispersion relations are *respectively* found for the EH- and HE-modes:

$$\frac{J_{n+1}(u)}{uJ_n(u)} = -\frac{K_{n+1}(w)}{wK_n(w)}, \tag{6.26}$$

$$\frac{J_{n-1}(u)}{uJ_n(u)} = +\frac{K_{n-1}(w)}{wK_n(w)}, \tag{6.27}$$

which may be combined into one general expression for both EH- and HE-modes, as

$$\frac{uJ_n(u)}{wK_n(w)} = -s\frac{J_{n+s}(u)}{K_{n+s}(w)}, \quad n \geq 1, \ s \in \{-1,1\}. \tag{6.28}$$

Considering the generalized components (6.19-24) and the dispersion relation (6.28) together, it leads to yet another generalization of these components that results in a simpler set of 6 equations, after recognizing that

$$(-s)^{r-1}\lambda_r^{(1+s)/2} = -s, \quad r \in \{1,2\}, s \in \{-1,1\}, \tag{6.29}$$

yielding

$$E_{r\rho}(\zeta;s) = j\beta\frac{aA}{u}s\frac{J_{n+s}^{r-1}(u)}{K_{n+s}^{r-1}(w)}\frac{J_{n+s}^{2-r}(u\rho/a)}{K_{n+s}^{1-r}(w\rho/a)}\cos\phi_m \tag{6.30}$$

$$E_{r\varphi}(\zeta;s) = j\beta\frac{aA}{u}\frac{J_{n+s}^{r-1}(u)}{K_{n+s}^{r-1}(w)}\frac{J_{n+s}^{2-r}(u\rho/a)}{K_{n+s}^{1-r}(w\rho/a)}\sin\phi_m \tag{6.31}$$

$$E_{rz}(\zeta;s) = A\frac{J_n^{r-1}(u)}{K_n^{r-1}(w)}\frac{J_n^{2-r}(u\rho/a)}{K_n^{1-r}(w\rho/a)}\cos\phi_m \tag{6.32}$$

$$H_{r\rho}(\zeta;s) = -j\omega_0\varepsilon_r\frac{aA}{u}\frac{J_{n+s}^{r-1}(u)}{K_{n+s}^{r-1}(w)}\frac{J_{n+s}^{2-r}(u\rho/a)}{K_{n+s}^{1-r}(w\rho/a)}\sin\phi_m \tag{6.33}$$

$$H_{r\varphi}(\zeta;s) = j\omega_0\varepsilon_r\frac{aA}{u}s\frac{J_{n+s}^{r-1}(u)}{K_{n+s}^{r-1}(w)}\frac{J_{n+s}^{2-r}(u\rho/a)}{K_{n+s}^{1-r}(w\rho/a)}\cos\phi_m \tag{6.34}$$

$$H_{rz}(\zeta;s) = -\frac{A\beta s}{\omega_0\mu_0}\frac{J_n^{r-1}(u)}{K_n^{r-1}(w)}\frac{J_n^{2-r}(u\rho/a)}{K_n^{1-r}(w\rho/a)}\sin\phi_m \tag{6.35}$$

The 6 equations (6.30-35) can all be summarized as the 2 compact expressions

$$E_{r\xi}(\zeta;n,s) = j\frac{aA}{us}\frac{J_{n+s}^{r-1}(u)}{K_{n+s}^{r-1}(w)}\frac{(\delta_{\xi\rho}+\delta_{\xi\varphi})\beta\frac{J_{n+s}^{2-r}(u\rho/a)}{K_{n+s}^{1-r}(w\rho/a)}+\delta_{\xi z}\frac{u}{a}\frac{J_n^{2-r}(u\rho/a)}{K_n^{1-r}(w\rho/a)}}{\delta_{\xi\rho}\sec\phi_m + s\delta_{\xi\varphi}\cosec\phi_m + js\delta_{\xi z}\sec\phi_m}$$

$$H_{r\xi}(\zeta;n,s) = -j\frac{aA}{u\omega_0}\frac{J_{n+s}^{r-1}(u)}{K_{n+s}^{r-1}(w)}\frac{(\delta_{\xi\rho}+\delta_{\xi\varphi})\omega_0^2\varepsilon_r\frac{J_{n+s}^{2-r}(u\rho/a)}{K_{n+s}^{1-r}(w\rho/a)}+\delta_{\xi z}\frac{u\beta}{a\mu_0}\frac{J_n^{2-r}(u\rho/a)}{K_n^{1-r}(w\rho/a)}}{\delta_{\xi\rho}\cosec\phi_m - s\delta_{\xi\varphi}\sec\phi_m + js\delta_{\xi z}\cosec\phi_m}$$

$$\tag{6.36}$$

As before, the best compactness is arguably achieved by employing a quotient approach, which confines the radial terms to the numerator, and the trigonometric terms to the denominator. The expressions for the EH-modes are found by setting $s = +1$, and those for the HE-modes, by setting $s = -1$, as is the case for (6.30-35).

It is also possible to construct the three-dimensional, cylindrical coordinates vectors of the EM-field using these expressions, and (3.3.53), which yields

$$\vec{E}(\zeta;n,s) = j\frac{aA}{us}\sum_{r=1}^{2}\sum_{\xi=\rho,\varphi,z}\xi\frac{J_{n+s}^{r-1}(u)}{K_{n+s}^{r-1}(w)}\frac{|\xi\times\mathbf{z}|\beta\frac{J_{n+s}^{2-r}(u\rho/a)}{K_{n+s}^{1-r}(w\rho/a)} + \xi\cdot\mathbf{z}\frac{u}{a}\frac{J_{n}^{2-r}(u\rho/a)}{K_{n}^{1-r}(w\rho/a)}}{\xi\cdot\boldsymbol{\rho}\sec\phi_m + s\xi\cdot\boldsymbol{\varphi}\csc\phi_m + js\xi\cdot\mathbf{z}\sec\phi_m}f_r(\rho)$$

$$\vec{H}(\zeta;n,s) = -j\frac{aA}{u\omega_0}\sum_{r=1}^{2}\sum_{\xi=\rho,\varphi,z}\xi\frac{J_{n+s}^{r-1}(u)}{K_{n+s}^{r-1}(w)}\frac{|\xi\times\mathbf{z}|\omega_0^2\varepsilon_r\frac{J_{n+s}^{2-r}(u\rho/a)}{K_{n+s}^{1-r}(w\rho/a)} + \xi\cdot\mathbf{z}\frac{u\beta}{a\mu_0}\frac{J_{n}^{2-r}(u\rho/a)}{K_{n}^{1-r}(w\rho/a)}}{\xi\cdot\boldsymbol{\rho}\csc\phi_m - s\xi\cdot\boldsymbol{\varphi}\sec\phi_m + js\xi\cdot\mathbf{z}\csc\phi_m}f_r(\rho)$$

(6.37)

More horizontally compact expressions can be attained, by relocating the spatially independent cofactor to the denominator, as seen in **Appendix B**.

The HE$_{11}$-mode components under the WGF approximation for instance, are then found by recalling the composite angle (3.2.8), and setting $n = 1$, and $s = -1$ in (6.36):

$$E_{r\xi}(\zeta;1,-1) = -j\frac{aA}{u}\frac{J_0^{r-1}(u)}{K_0^{r-1}(w)}\frac{(\delta_{\xi\rho}+\delta_{\xi\varphi})\beta\frac{J_0^{2-r}(u\rho/a)}{K_0^{1-r}(w\rho/a)} + \delta_{\xi z}\frac{u}{a}\frac{J_1^{2-r}(u\rho/a)}{K_1^{1-r}(w\rho/a)}}{\delta_{\xi\rho}\sec(\varphi+\psi_m) - \delta_{\xi\varphi}\csc(\varphi+\psi_m) - j\delta_{\xi z}\sec(\varphi+\psi_m)}$$

$$H_{r\xi}(\zeta;1,-1) = -j\frac{aA}{u}\frac{J_0^{r-1}(u)}{K_0^{r-1}(w)}\frac{(\delta_{\xi\rho}+\delta_{\xi\varphi})\omega_0\varepsilon_r\frac{J_0^{2-r}(u\rho/a)}{K_0^{1-r}(w\rho/a)} + \delta_{\xi z}\frac{u\beta}{a\mu_0\omega_0}\frac{J_1^{2-r}(u\rho/a)}{K_1^{1-r}(w\rho/a)}}{\delta_{\xi\rho}\csc(\varphi+\psi_m) + \delta_{\xi\varphi}\sec(\varphi+\psi_m) - j\delta_{\xi z}\csc(\varphi+\psi_m)}$$

(6.38)

with the corresponding EM-field vector, which are found from (6.37), or constructed using (6.38), given by the pair of equations

$$\vec{E}(\zeta;1,-1) = \frac{aA}{ju}\sum_{r=1}^{2}\sum_{\xi=\rho,\varphi,z}\xi\frac{J_0^{r-1}(u)}{K_0^{r-1}(w)}\frac{\left[|\xi\times\mathbf{z}|\beta\frac{J_0^{2-r}(u\rho/a)}{K_0^{1-r}(w\rho/a)} + \xi\cdot\mathbf{z}\frac{u}{a}\frac{J_1^{2-r}(u\rho/a)}{K_1^{1-r}(w\rho/a)}\right]f_r(\rho)}{\xi\cdot\boldsymbol{\rho}\sec(\varphi+\psi_m) - \xi\cdot\boldsymbol{\varphi}\csc(\varphi+\psi_m) - j\xi\cdot\mathbf{z}\sec(\varphi+\psi_m)}$$

$$\vec{H}(\zeta;1,-1) = \frac{aA}{ju}\sum_{r=1}^{2}\sum_{\xi=\rho,\varphi,z}\xi\frac{J_0^{r-1}(u)}{K_0^{r-1}(w)}\frac{\left[|\xi\times\mathbf{z}|\omega_0\varepsilon_r\frac{J_0^{2-r}(u\rho/a)}{K_0^{1-r}(w\rho/a)} + \xi\cdot\mathbf{z}\frac{u\beta}{a\omega_0\mu_0}\frac{J_1^{2-r}(u\rho/a)}{K_1^{1-r}(w\rho/a)}\right]f_r(\rho)}{\xi\cdot\boldsymbol{\rho}\csc(\varphi+\psi_m) + \xi\cdot\boldsymbol{\varphi}\sec(\varphi+\psi_m) - j\xi\cdot\mathbf{z}\csc(\varphi+\psi_m)}$$

(6.39)

**The bi-complex generalization** is most easily obtained by applying the CTB transform (3.1.21 or 22) to (6.30-35),

$$\tilde{E}_{r\rho}(\zeta;s) = j\beta \frac{aA}{u} s \frac{J_{n+s}^{r-1}(u)}{K_{n+s}^{r-1}(w)} \frac{J_{n+s}^{2-r}(u\rho/a)}{K_{n+s}^{1-r}(w\rho/a)} e^{-i\phi_m} \tag{6.40}$$

$$\tilde{E}_{r\varphi}(\zeta;s) = ij\beta \frac{aA}{u} \frac{J_{n+s}^{r-1}(u)}{K_{n+s}^{r-1}(w)} \frac{J_{n+s}^{2-r}(u\rho/a)}{K_{n+s}^{1-r}(w\rho/a)} e^{-i\phi_m} \tag{6.41}$$

$$\tilde{E}_{rz}(\zeta;s) = A \frac{J_n^{r-1}(u)}{K_n^{r-1}(w)} \frac{J_n^{2-r}(u\rho/a)}{K_n^{1-r}(w\rho/a)} e^{-i\phi_m} \tag{6.42}$$

$$\tilde{H}_{r\rho}(\zeta;s) = -ij\omega_0\varepsilon_r \frac{aA}{u} \frac{J_{n+s}^{r-1}(u)}{K_{n+s}^{r-1}(w)} \frac{J_{n+s}^{2-r}(u\rho/a)}{K_{n+s}^{1-r}(w\rho/a)} e^{-i\phi_m} \tag{6.43}$$

$$\tilde{H}_{r\varphi}(\zeta;s) = j\omega_0\varepsilon_r \frac{aA}{u} s \frac{J_{n+s}^{r-1}(u)}{K_{n+s}^{r-1}(w)} \frac{J_{n+s}^{2-r}(u\rho/a)}{K_{n+s}^{1-r}(w\rho/a)} e^{-i\phi_m} \tag{6.44}$$

$$\tilde{H}_{rz}(\zeta;s) = -i\frac{A\beta s}{\omega_0\mu_0} \frac{J_n^{r-1}(u)}{K_n^{r-1}(w)} \frac{J_n^{2-r}(u\rho/a)}{K_n^{1-r}(w\rho/a)} e^{-i\phi_m} \tag{6.45}$$

and can be summarized as the 2 generalized scalar equations,

$$\tilde{E}_{r\xi}(\zeta;n,s) = j\frac{aA}{us} \frac{J_{n+s}^{r-1}(u)}{K_{n+s}^{r-1}(w)} \left[ (\delta_{\xi\rho} + is\delta_{\xi\varphi})\beta \frac{J_{n+s}^{2-r}(u\rho/a)}{K_{n+s}^{1-r}(w\rho/a)} - j\delta_{\xi z} \frac{us}{a} \frac{J_n^{2-r}(u\rho/a)}{K_n^{1-r}(w\rho/a)} \right] e^{-i\phi_m}$$

$$\tilde{H}_{r\xi}(\zeta;n,s) = -j\frac{aA}{u} \frac{J_{n+s}^{r-1}(u)}{K_{n+s}^{r-1}(w)} \left[ (\delta_{\xi\rho} + is\delta_{\xi\varphi})\omega_0\varepsilon_r \frac{J_{n+s}^{2-r}(u\rho/a)}{K_{n+s}^{1-r}(w\rho/a)} - j\delta_{\xi z} \frac{us\beta}{a\mu_0\omega_0} \frac{J_n^{2-r}(u\rho/a)}{K_n^{1-r}(w\rho/a)} \right] ie^{-i\phi_m}$$

(6.46)

The generalized vecsor equations are found using (6.46) and (3.3.53), and are given by

$$\tilde{\mathbf{E}}(\zeta;n,s) = \frac{jaA}{us} \sum_{r=1}^{2} \frac{J_{n+s}^{r-1}(u)}{K_{n+s}^{r-1}(w)} \left[ (\boldsymbol{\rho} + is\boldsymbol{\varphi})\beta \frac{J_{n+s}^{2-r}(u\rho/a)}{K_{n+s}^{1-r}(w\rho/a)} - j\mathbf{z}\frac{us}{a} \frac{J_n^{2-r}(u\rho/a)}{K_n^{1-r}(w\rho/a)} \right] f_r(\rho) e^{-i\phi_m}$$

$$\tilde{\mathbf{H}}(\zeta;n,s) = \frac{aA}{ju} \sum_{r=1}^{2} \frac{J_{n+s}^{r-1}(u)}{K_{n+s}^{r-1}(w)} \left[ (\boldsymbol{\rho} + is\boldsymbol{\varphi})\omega_0\varepsilon_r \frac{J_{n+s}^{2-r}(u\rho/a)}{K_{n+s}^{1-r}(w\rho/a)} - j\mathbf{z}\frac{us\beta}{a\mu_0\omega_0} \frac{J_n^{2-r}(u\rho/a)}{K_n^{1-r}(w\rho/a)} \right] f_r(\rho) ie^{-i\phi_m}$$

(6.47)

In the **bi-complex formulation**, the generalized scalar (6.46) and vecsor equations (6.47) are vertically more compact, and with fewer summations in the latter, relative to the equivalent equations (6.36) and (6.37) in the **complex formulation**.

For the HE$_{11}$-mode **in the bi-complex formulation** and under the WGF approximation, the generalized scalar equations are found with $n = 1$, and $s = -1$ in (6.46):

$$\tilde{E}_{r\xi}(\zeta;1,-1) = -j\frac{aA}{u}\frac{J_0^{r-1}(u)}{K_0^{r-1}(w)}\left[(\delta_{\xi\rho} - i\delta_{\xi\varphi})\beta\frac{J_0^{2-r}(u\rho/a)}{K_0^{1-r}(w\rho/a)} + j\delta_{\xi z}\frac{u}{a}\frac{J_1^{2-r}(u\rho/a)}{K_1^{1-r}(w\rho/a)}\right]e^{-i(\varphi+\psi_m)}$$

$$\tilde{H}_{r\xi}(\zeta;1,-1) = -j\frac{aA}{u\omega_0}\frac{J_0^{r-1}(u)}{K_0^{r-1}(w)}\left[(\delta_{\xi\rho} - i\delta_{\xi\varphi})\omega_0^2\varepsilon_r\frac{J_0^{2-r}(u\rho/a)}{K_0^{1-r}(w\rho/a)} + j\delta_{\xi z}\frac{u\beta}{a\mu_0}\frac{J_1^{2-r}(u\rho/a)}{K_1^{1-r}(w\rho/a)}\right]ie^{-i(\varphi+\psi_m)}$$

(6.48)

with corresponding generalized vecsor equations

$$\tilde{\mathbf{E}}(\zeta;1,-1) = \frac{aA}{ju}\sum_{r=1}^{2}\frac{J_0^{r-1}(u)}{K_0^{r-1}(w)}\left[(\boldsymbol{\rho} - i\boldsymbol{\varphi})\beta\frac{J_0^{2-r}(u\rho/a)}{K_0^{1-r}(w\rho/a)} + j\mathbf{z}\frac{u}{a}\frac{J_1^{2-r}(u\rho/a)}{K_1^{1-r}(w\rho/a)}\right]f_r(\rho)e^{-i(\varphi+\psi_m)}$$

$$\tilde{\mathbf{H}}(\zeta;1,-1) = \frac{aA}{ju}\sum_{r=1}^{2}\frac{J_0^{r-1}(u)}{K_0^{r-1}(w)}\left[(\boldsymbol{\rho} - i\boldsymbol{\varphi})\omega_0\varepsilon_r\frac{J_0^{2-r}(u\rho/a)}{K_0^{1-r}(w\rho/a)} + j\mathbf{z}\frac{u\beta}{a\omega_0\mu_0}\frac{J_1^{2-r}(u\rho/a)}{K_1^{1-r}(w\rho/a)}\right]f_r(\rho)ie^{-i(\varphi+\psi_m)}$$

(6.49)

Of most importance are **the transverse vectors** of the EM-field, because they are used in the power and orthogonality expressions. They are found by applying either (3.4.3) or (3.4.5) to the vector expressions developed in this section, since propagation was assumed to be in the $z$-direction at the outset. In the **complex formulation** for instance, the transverse vectors are given by

$$\vec{\mathbf{E}}_{mT}(\zeta;n,s) = j\frac{aA}{u}\beta\sum_{r=1}^{2}\frac{J_{n+s}^{r-1}(u)}{K_{n+s}^{r-1}(w)}\frac{J_{n+s}^{2-r}(u\rho/a)}{K_{n+s}^{1-r}(w\rho/a)}f_r(\rho)\left[s\boldsymbol{\rho}\cos\phi_m + \boldsymbol{\varphi}\sin\phi_m\right] \quad (6.50)$$

$$\vec{\mathbf{H}}_{mT}(\zeta;n,s) = -j\frac{aA}{u}\omega_0\sum_{r=1}^{2}\frac{J_{n+s}^{r-1}(u)}{K_{n+s}^{r-1}(w)}\frac{J_{n+s}^{2-r}(u\rho/a)}{K_{n+s}^{1-r}(w\rho/a)}\varepsilon_r f_r(\rho)\left[\boldsymbol{\rho}\sin\phi_m - s\boldsymbol{\varphi}\cos\phi_m\right] \quad (6.51)$$

whereas in the **bi-complex formulation**, the transverse vecsors are given by

$$\tilde{\mathbf{E}}_{mT}(\zeta;n,s) = j\frac{aA}{u}\beta\sum_{r=1}^{2}\frac{J_{n+s}^{r-1}(u)}{K_{n+s}^{r-1}(w)}\frac{J_{n+s}^{2-r}(u\rho/a)}{K_{n+s}^{1-r}(w\rho/a)}f_r(\rho)(s\boldsymbol{\rho} + i\boldsymbol{\varphi})e^{-i\phi_m} \quad (6.52)$$

$$\tilde{\mathbf{H}}_{mT}(\zeta;n,s) = -j\frac{aA}{u}\omega_0\sum_{r=1}^{2}\frac{J_{n+s}^{r-1}(u)}{K_{n+s}^{r-1}(w)}\frac{J_{n+s}^{2-r}(u\rho/a)}{K_{n+s}^{1-r}(w\rho/a)}\varepsilon_r f_r(\rho)(\boldsymbol{\rho} + is\boldsymbol{\varphi})ie^{-i\phi_m} \quad (6.53)$$

which are sufficiently compact to allow expansion of the $r$-summations, if desired. This is more difficult for (6.50, 51), due to the use of trigonometric functions.

**The total power flow for the HE$_{11}$-mode** is also amenable to simplification under the WGF approximation. Most generally, it was found in §**5.1** to be, **before any approximation**,

$$P = \frac{\omega_0 \varepsilon_1 \beta |A|^2 S_1^2}{8\pi u^2} \sum_{r=1}^{2} \sum_{l=0}^{1} \sum_{p=0}^{1} \sum_{q=0}^{1} \left(\frac{u^2 \varepsilon_2}{w^2 \varepsilon_1}\right)^{r-1} \frac{(1-s\lambda_p)(1-s_r\lambda_p)\lambda_q J_1^{2r-2}(u) J_{2p-q}^{2-r}(u) J_{2p+q}^{2-r}(u)}{2^{2-r} \lambda_l^{1-r} \kappa_l^{2-2r} K_1^{2r-2}(w) K_{2p-q}^{1-r}(\kappa_l w) K_{2p+q}^{1-r}(\kappa_l w)}.$$

(6.54)

**Under the WGF approximation** (6.5, 6), it is concluded that the product

$$(1-s\lambda_p)(1-s_r\lambda_p) \approx (1+\lambda_p)^2 = 4(1-p). \tag{6.55}$$

The additional application of (6.2, 27) leads to the following simplification for the group of constants under the composite summand,

$$\left(\frac{u^2 \varepsilon_2}{w^2 \varepsilon_1}\right)^{r-1} \frac{J_1^{2r-2}(u)}{K_1^{2r-2}(w)} = \left(\frac{\varepsilon_2}{\varepsilon_1} \frac{u^2 J_1^2(u)}{w^2 K_1^2(w)}\right)^{r-1} \approx \left(\frac{J_0^2(u)}{K_0^2(w)}\right)^{r-1}. \tag{6.56}$$

Substituting (6.55, 56) into (6.54), the total power flow is found to be reduced to the following simpler expression,

$$P \approx \frac{\omega_0 \varepsilon_1 \beta |A|^2 S_1^2}{2\pi u^2} \sum_{r=1}^{2} \sum_{l=0}^{1} \sum_{q=0}^{1} \left(\frac{J_0^2(u)}{K_0^2(w)}\right)^{r-1} \lambda_q \frac{\lambda_q^{2-r} J_q^{4-2r}(u)}{2^{2-r} \lambda_l^{1-r} \kappa_l^{2-2r} K_q^{2-2r}(\kappa_l w)}. \tag{6.57}$$

It can be recast as

$$P = \frac{\omega_0 \beta \varepsilon_1 |A|^2 J_0^2(u) S_1^2}{2\pi u^2} \sum_{r=1}^{2} \sum_{l=0}^{1} \sum_{q=0}^{1} \left(\frac{J_q^2(u)}{2 J_0^2(u)}\right)^{2-r} \left(\lambda_{l+q} \frac{\kappa_l^2 K_q^2(\kappa_l w)}{K_0^2(w)}\right)^{r-1}, \tag{6.58}$$

and since it is true in general that

$$\sum_{r=1}^{2} \sum_{l=0}^{1} \sum_{q=0}^{1} X_q^{2-r} Y_q^{r-1} = \sum_{r=1}^{2} \left[\sum_{l=0}^{1} \sum_{q=0}^{1} X_q\right]^{2-r} \left[\sum_{l=0}^{1} \sum_{q=0}^{1} Y_q\right]^{r-1} \tag{6.59}$$

then

$$P = \frac{\omega_0 \beta \varepsilon_1 |A|^2 J_0^2(u) S_1^2}{2\pi u^2} \sum_{r=1}^{2} \left[\sum_{l=0}^{1} \sum_{q=0}^{1} \frac{J_q^2(u)}{2 J_0^2(u)}\right]^{2-r} \left[\sum_{l=0}^{1} \sum_{q=0}^{1} \lambda_{l+q} \frac{\kappa_l^2 K_q^2(\kappa_l w)}{K_0^2(w)}\right]^{r-1}. \tag{6.60}$$

Applying the dispersion relation (6.27) to the 1st bracket, and subsequently expanding both brackets results in the expression

$$P = \frac{\omega_0 \beta \varepsilon_1 |A|^2 J_0^2(u) S_1^2}{2\pi u^2} \sum_{r=1}^{2} \left[ 1 + \frac{w^2}{u^2} \frac{K_1^2(u)}{K_0^2(u)} \right]^{2-r} \left[ \frac{\kappa_0^2 K_0^2(\kappa_0 w)}{K_0^2(w)} - \frac{\kappa_0^2 K_1^2(\kappa_0 w)}{K_0^2(w)} - 1 + \frac{K_1^2(w)}{K_0^2(w)} \right]^{1-r} \tag{6.61}$$

which simplifies to

$$P = \frac{\omega_0 \beta \varepsilon_1 |A|^2 J_0^2(u) S_1^2}{2\pi u^2} \frac{K_1^2(u)}{K_0^2(u)} \left[ 1 + \frac{w^2}{u^2} + \frac{\kappa_0^2 K_0^2(\kappa_0 w)}{K_1^2(w)} - \frac{\kappa_0^2 K_1^2(\kappa_0 w)}{K_1^2(w)} \right]. \tag{6.62}$$

**This is the total power flow for the HE$_{11}$-mode under the WGF approximation.**

The power expression (6.62) may be additionally simplified by adopting another approximation. Recalling from §**5.1** the new variable

$$\kappa_l = (b/a)^{1-l}, \quad l \in \{0,1\}, \tag{6.63}$$

then after invoking the **infinite-cladding approximation**, which assumes that the cladding outer radius $b$ is much larger than its inner radius $a$, it is found that

$$\lim_{\kappa_0 \to \infty} \kappa_0^2 K_q^2(\kappa_0 w) = 0, \quad \forall q \tag{6.64}$$

so that the total power flow for the HE$_{11}$-mode is reduced to

$$P = \frac{\omega_0 \beta \varepsilon_1 |A|^2 J_0^2(u) S_1^2}{2\pi u^2} \frac{K_1^2(w)}{K_0^2(w)} \left( 1 + \frac{w^2}{u^2} \right). \tag{6.65}$$

After applying to (6.65), the Pythagorean relation (3.1.16) that connects the normalized transverse frequencies to the fiber's $v$-number,

$$v^2 = u^2 + w^2, \tag{6.66}$$

there results

$$P = \frac{\omega_0 \beta \varepsilon_1 v^2 |A|^2 J_0^2(u) S_1^2}{2\pi u^4} \frac{K_1^2(w)}{K_0^2(w)} \tag{6.67}$$

which is Snyder's form for the total power flow [25], in Okamoto's nomenclature [1]. This expression may be used on the RHS of the orthogonality relations (5.1.66) and (5.2.31) for the polarization states of the HE$_{11}$-mode, instead of the exact version (6.54).

## 7. Summary and conclusions

The 6 complex EM-field components of a hybrid mode in **the core region** of an ideal fiber, geometrically described by $\rho \in [0,a]$, were found to be [1, 2]

$$E_\rho(\rho,\varphi) = -j\beta \frac{aA}{2u}\left[(1-s)J_{n-1}\left(\frac{u}{a}\rho\right) - (1+s)J_{n+1}\left(\frac{u}{a}\rho\right)\right]\cos(n\varphi+\psi_m)$$

$$E_\varphi(\rho,\varphi) = j\beta \frac{aA}{2u}\left[(1-s)J_{n-1}\left(\frac{u}{a}\rho\right) + (1+s)J_{n+1}\left(\frac{u}{a}\rho\right)\right]\sin(n\varphi+\psi_m)$$

$$E_z(\rho,\varphi) = AJ_n\left(\frac{u}{a}\rho\right)\cos(n\varphi+\psi_m) \qquad (7.1)$$

$$H_\rho(\rho,\varphi) = -j\omega_0\varepsilon_1 \frac{aA}{2u}\left[(1-s_1)J_{n-1}\left(\frac{u}{a}\rho\right) + (1+s_1)J_{n+1}\left(\frac{u}{a}\rho\right)\right]\sin(n\varphi+\psi_m)$$

$$H_\varphi(\rho,\varphi) = -j\omega_0\varepsilon_1 \frac{aA}{2u}\left[(1-s_1)J_{n-1}\left(\frac{u}{a}\rho\right) - (1+s_1)J_{n+1}\left(\frac{u}{a}\rho\right)\right]\cos(n\varphi+\psi_m)$$

$$H_z(\rho,\varphi) = -\frac{\beta s}{\omega_0\mu_0}AJ_n\left(\frac{u}{a}\rho\right)\sin(n\varphi+\psi_m)$$

whereas the 6 complex EM-field components of a hybrid mode in **the cladding region** of the ideal fiber, geometrically described by $\rho \in (a,b]$, are given by [1, 2],

$$E_\rho(\rho,\varphi) = -j\beta \frac{aA}{2w}\left(\frac{J_n(u)}{K_n(w)}\right)\left[(1-s)K_{n-1}\left(\frac{w}{a}\rho\right) + (1+s)K_{n+1}\left(\frac{w}{a}\rho\right)\right]\cos(n\varphi+\psi_m)$$

$$E_\varphi(\rho,\varphi) = j\beta \frac{aA}{2w}\left(\frac{J_n(u)}{K_n(w)}\right)\left[(1-s)K_{n-1}\left(\frac{w}{a}\rho\right) - (1+s)K_{n+1}\left(\frac{w}{a}\rho\right)\right]\sin(n\varphi+\psi_m)$$

$$E_z(\rho,\varphi) = A\frac{J_n(u)}{K_n(w)}K_n\left(\frac{w}{a}\rho\right)\cos(n\varphi+\psi_m) \qquad (7.2)$$

$$H_\rho(\rho,\varphi) = -j\omega_0\varepsilon_2 \frac{aA}{2w}\left(\frac{J_n(u)}{K_n(w)}\right)\left[(1-s_2)K_{n-1}\left(\frac{w}{a}\rho\right) - (1+s_2)K_{n+1}\left(\frac{w}{a}\rho\right)\right]\sin(n\varphi+\psi_m)$$

$$H_\varphi(\rho,\varphi) = -j\omega_0\varepsilon_2 \frac{aA}{2w}\left(\frac{J_n(u)}{K_n(w)}\right)\left[(1-s_2)K_{n-1}\left(\frac{w}{a}\rho\right) + (1+s_2)K_{n+1}\left(\frac{w}{a}\rho\right)\right]\cos(n\varphi+\psi_m)$$

$$H_z(\rho,\varphi) = -\frac{\beta s}{\omega_0\mu_0}A\frac{J_n(u)}{K_n(w)}K_n\left(\frac{w}{a}\rho\right)\sin(n\varphi+\psi_m)$$

A vector in any of the 3 directions in cylindrical coordinates, in either of the 2 regions, may be obtainable from the list of (7.1-2) using

$$\vec{V}_\xi(r,t) = \hat{\xi}\,\mathrm{Re}\left[V_\xi(\rho,\varphi)e^{j(\omega_0 t - \beta z)}\right]\;;\quad V \in \{E,H\},\;\xi \in \{\rho,\varphi,z\} \qquad (7.3)$$

with the complete vector, in either of the 2 regions of the fiber, constructed as

$$\vec{V}(r,t) = \vec{V}_\rho(r,t) + \vec{V}_\varphi(r,t) + \vec{V}_z(r,t)\;;\quad \vec{V} \in \{\vec{E},\vec{H}\},\;\rho \in [0,b], \qquad (7.4)$$

with **r** being short-form for ($\rho$, $\varphi$, $z$). These expressions are in the **complex formulation**, and are in terms of the parameters [1]

$$u = a\left(k_0^2 n_1^2 - \beta^2\right)^{1/2} \tag{7.5}$$

$$w = a\left(\beta^2 - k_0^2 n_2^2\right)^{1/2} \tag{7.6}$$

$$k_0 = \omega_0/c \tag{7.7}$$

$$v^2 = u^2 + w^2 \tag{7.8}$$

$$s = \frac{n\,v^2 J_n(u) K_n(w)}{u w^2 J'_n(u) K_n(w) + u^2 w K'_n(w) J_n(u)}, \quad n \geq 1,\ v > 0 \tag{7.9}$$

$$s_r = s\left(\beta/k_0 n_r\right)^2 ;\ r \in \{1, 2\} \tag{7.10}$$

$$\psi_m = (m-1)\pi/2,\ m \in \{1, 2\} \tag{7.11}$$

all of which are non-dimensional, with the exception of the wave-number $k_0$. The identification of the recurrent parameter $s$ (7.9) in the EM-field components (7.1, 2) considerably simplifies their expression. For a single-mode fiber (SMF) or the HE$_{11}$-mode, the azimuthal eigenvalue $n$ is identical with unity, whereas its parameter $v$ (7.8) is constrained to being less than 2.405, but bigger than 0. Otherwise and most generally, the fiber supports 2 types of *bound* modes, the transverse, which are the TE and TM modes, and the hybrid, which are the HE- and EH-modes. The 2 polarization states of a hybrid mode are actually indistinguishable in the radial coordinate, but are identified using the azimuthal phase factor $\psi_m$ (7.11), which yields, for $m = 1$, the $x$-polarization, and for $m = 2$, the $y$-polarization.

The derivation of (7.1, 2) is based on several assumptions, which collectively render the fiber "**ideal**". Geometrically, the ideal fiber is circular in cross-section, with an assumed eccentricity of zero, and is spatially and temporally invariant along its entire, perfectly linear, length. The cross-section of the fiber is assumed to be co-incidental with the $xy$-plane of a right-handed coordinate system, so that the electromagnetic (EM) field propagation is along the positive $z$-direction. The cross-section is comprised of 2 circularly contiguous regions: **the core**, which is the **1st region**, with a radius $a$, and a refractive index of $n_1$, and **the cladding**, which is the **2nd region**, with a radial width of $(b - a) \gg a$, and a smaller refractive index of $n_2$. **This is the regional designation to which this report adheres.** The fiber being considered is thus a single-step, step-index, perfectly cylindrical waveguide. Lastly, the ideal fiber is constrained to being a linear, a homogeneous, and a non-magnetic, isotropic, loss-free waveguide. The gestalt of these assumptions render a hybrid mode degenerate in the propagation constant $\beta$ with respect to its 2 polarization states, and decouples the spatial transverse EM-field profile, from the propagation or $z$-direction.

**One goal of this report is to investigate alternative, even more compact approaches for analytical descriptions of the component-resolved EM-field than those (7.1, 2) presented in [1, 2]. Another is to explore the possibility of expressing (7.4) explicitly in terms of all the components (7.1, 2) in compact, analytical equations, and to examine their utility in the derivation of modal powers and orthogonality.**

After adopting a regional parameter, and a composite angle[12],

$$\lambda_\alpha = e^{j\pi\alpha}, \ \alpha \in \mathbb{R} \tag{7.12}$$

$$\phi_m = n\varphi + \psi_m, \ m \in \{1,2\}, \ n \in \mathbb{Z}^+ \tag{7.13}$$

an explicit description is found in §**3.2**, of the EM-field of a hybrid mode, over the entire cross-section of the ideal fiber, **using just 6 compact equations**, instead of the 12 in Okamoto's nomenclature (7.1, 2), and is valid for both the HE- and EH-modes:

$$E_{r\rho}(\zeta) = -j\beta \frac{aA}{2u}\left(\frac{uJ_n(u)}{wK_n(w)}\right)^{r-1}\left[(1-s)\frac{J_{n-1}^{2-r}(u\rho/a)}{K_{n-1}^{1-r}(w\rho/a)} + \lambda_r(1+s)\frac{J_{n+1}^{2-r}(u\rho/a)}{K_{n+1}^{1-r}(w\rho/a)}\right]\cos\phi_m \tag{7.14}$$

$$E_{r\varphi}(\zeta) = j\beta \frac{aA}{2u}\left(\frac{uJ_n(u)}{wK_n(w)}\right)^{r-1}\left[(1-s)\frac{J_{n-1}^{2-r}(u\rho/a)}{K_{n-1}^{1-r}(w\rho/a)} - \lambda_r(1+s)\frac{J_{n+1}^{2-r}(u\rho/a)}{K_{n+1}^{1-r}(w\rho/a)}\right]\sin\phi_m \tag{7.15}$$

$$E_{rz}(\zeta) = A\frac{J_n^{r-1}(u)}{K_n^{r-1}(w)}\frac{J_n^{2-r}(u\rho/a)}{K_n^{1-r}(w\rho/a)}\cos\phi_m \tag{7.16}$$

$$H_{r\rho}(\zeta) = -j\omega_0\varepsilon_r \frac{aA}{2u}\left(\frac{uJ_n(u)}{wK_n(w)}\right)^{r-1}\left[(1-s_r)\frac{J_{n-1}^{2-r}(u\rho/a)}{K_{n-1}^{1-r}(w\rho/a)} - \lambda_r(1+s_r)\frac{J_{n+1}^{2-r}(u\rho/a)}{K_{n+1}^{1-r}(w\rho/a)}\right]\sin\phi_m \tag{7.17}$$

$$H_{r\varphi}(\zeta) = -j\omega_0\varepsilon_r \frac{aA}{2u}\left(\frac{uJ_n(u)}{wK_n(w)}\right)^{r-1}\left[(1-s_r)\frac{J_{n-1}^{2-r}(u\rho/a)}{K_{n-1}^{1-r}(w\rho/a)} + \lambda_r(1+s_r)\frac{J_{n+1}^{2-r}(u\rho/a)}{K_{n+1}^{1-r}(w\rho/a)}\right]\cos\phi_m \tag{7.18}$$

$$H_{rz}(\zeta) = -\frac{A\beta s}{\omega_0\mu_0}\frac{J_n^{r-1}(u)}{K_n^{r-1}(w)}\frac{J_n^{2-r}(u\rho/a)}{K_n^{1-r}(w\rho/a)}\sin\phi_m \tag{7.19}$$

**The nomenclature (7.1, 2) of [1, 2], which is comprised of 12 equations, is recovered simply by setting in (7.14-19), $r = 1$ for the core, which yields (7.1), or $r = 2$, for the cladding, which yields (7.2), therefore resulting in all 12 equations for the entire cross-section of the fiber.** Expressions (7.14-7.19) are generalized scalar versions of (7.1, 2). The components are still being presented in a format of a look-up table however, as in (7.1, 2), which may be the most preferable approach. A more analytical, compact alternative is however explored.

In §**3.2**, it is shown that the regional *z*-components of the electric and magnetic fields (7.16, 19) have the alternative, but less compact forms of

---

[12] Note that $\lambda$, which is extensively used throughout this report, is not to be confused with the wavelength of the EM-field, which is not used anywhere in this report. Instead of the wavelength, this report consistently uses the angular frequency $\omega_0$.

$$E_{rz}(\zeta) = -\frac{\rho A}{2anu} \frac{\lambda_r w^{2r-2}}{u^{2r-4} \sec\phi_m} \left(\frac{uJ_n(u)}{wK_n(w)}\right)^{r-1} \left[\frac{J_{n-1}^{2-r}(u\rho/a)}{K_{n-1}^{1-r}(w\rho/a)} - \lambda_r \frac{J_{n+1}^{2-r}(u\rho/a)}{K_{n+1}^{1-r}(w\rho/a)}\right],$$

$$H_{rz}(\zeta) = \frac{\omega_0}{\beta} \frac{\rho A}{2anu} \frac{\lambda_r s_r \varepsilon_r w^{2r-2}}{u^{2r-4} \text{cosec}\,\phi_m} \left(\frac{uJ_n(u)}{wK_n(w)}\right)^{r-1} \left[\frac{J_{n-1}^{2-r}(u\rho/a)}{K_{n-1}^{1-r}(w\rho/a)} - \lambda_r \frac{J_{n+1}^{2-r}(u\rho/a)}{K_{n+1}^{1-r}(w\rho/a)}\right].$$
(7.20)

The multiplicative factor of these components has been reformulated using (7.7, 10). In this form, the longitudinal components are more amenable to generalization than in their original form in (7.16, 19). Considering both the transverse (7.14, 15, 17, 18) and the longitudinal components (7.20) together, it is deduced that a general, *explicit* expression for *any* $\xi$-component of the EM-field of any hybrid mode, in either of the 2 regions of the fiber, can be efficiently given by just a pair of equations, **termed the complex generalized scalar (CGS) equations**,

$$E_{r\xi}(\rho,\varphi) = \frac{\left[(1-s+s\delta_{\xi z})\dfrac{J_{n-1}^{2-r}(u\rho/a)}{K_{n-1}^{1-r}(w\rho/a)} + \dfrac{(1+s-s\delta_{\xi z})e^{j\pi r}}{\delta_{\xi\rho} - \delta_{\xi\varphi} - \delta_{\xi z}} \dfrac{J_{n+1}^{2-r}(u\rho/a)}{K_{n+1}^{1-r}(w\rho/a)}\right]\cos(n\varphi+\psi_m)}{\dfrac{2ju}{aA\beta}\left(\dfrac{uJ_n(u)}{wK_n(w)}\right)^{1-r}\left[\delta_{\xi\rho} - \delta_{\xi\varphi}\cot(n\varphi+\psi_m) + jna^2\beta \dfrac{u^{2r-4}e^{j\pi r}}{\rho w^{2r-2}}\delta_{\xi z}\right]}$$

$$H_{r\xi}(\rho,\varphi) = \frac{\left[(1-s_r+s_r\delta_{\xi z})\dfrac{J_{n-1}^{2-r}(u\rho/a)}{K_{n-1}^{1-r}(w\rho/a)} - \dfrac{(1+s_r-s_r\delta_{\xi z})e^{j\pi r}}{\delta_{\xi\rho} - \delta_{\xi\varphi} + \delta_{\xi z}} \dfrac{J_{n+1}^{2-r}(u\rho/a)}{K_{n+1}^{1-r}(w\rho/a)}\right]\varepsilon_r\sin(n\varphi+\psi_m)}{\dfrac{2ju}{aA\omega_0}\left(\dfrac{uJ_n(u)}{wK_n(w)}\right)^{1-r}\left[\delta_{\xi\rho} + \delta_{\xi\varphi}\tan(n\varphi+\psi_m) - jna^2\beta \dfrac{u^{2r-4}e^{j\pi r}}{\rho s_r w^{2r-2}}\delta_{\xi z}\right]}$$
(7.21)

for $\xi \in \{\rho,\varphi,z\}$. Alternative forms are presented in **APPENDIX B**. This generalization is made possible by the use of the Kronecker delta. **The 2 equations may replace all 12 equations of the original nomenclature (7.1, 2), or all 6 generalized scalar equations (7.14-19)**. Equations (7.21) are meant to be taken together, as the $\xi$-component of the electromagnetic field of a hybrid mode. Any component in either of the 2 cross-sectional regions of the fiber can be easily recovered, operationally with the help of Kronecker deltas, as follows:

$$V_{r'\xi'}(\zeta) = \delta_{rr'}\delta_{\xi\xi'} V_{r\xi}(\zeta); \quad V \in \{E,H\},\ r' \in \{1,2\},\ \xi' \in \{\rho,\varphi,z\} \tag{7.22}$$

since any term in (7.21) involving a Kronecker delta whose argument is different from $\xi'$ is immediately extinguished.

Upon a cursory examination of the EM-field components (7.14-19), **it is concluded** that the following function, found in §**3.2** and which makes use of $\lambda$ (7.12), is common to all transverse (7.14, 15, 17, 18) and longitudinal (7.20) components,

$$\Lambda_{nr}(\rho;\eta,\lambda_r) = \frac{aA}{2u}\left(\frac{uJ_n(u)}{wK_n(w)}\right)^{r-1}\left[(1-\eta)\frac{J_{n-1}^{2-r}(u\rho/a)}{K_{n-1}^{1-r}(w\rho/a)} + \lambda_r(1+\eta)\frac{J_{n+1}^{2-r}(u\rho/a)}{K_{n+1}^{1-r}(w\rho/a)}\right].$$

(7.23)

It is termed **the generating function**. It results in compact versions of (7.14 - 19), depending on whether $r = 1$ (for the core) or $r = 2$ (for the cladding),

$$\begin{aligned}
E_{r\rho}(\zeta) &= -j\beta\,\Lambda_{nr}(\rho;s,\lambda_r)\cos\phi_m \\
E_{r\varphi}(\zeta) &= j\beta\,\Lambda_{nr}(\rho;s,-\lambda_r)\sin\phi_m \\
E_{rz}(\zeta) &= -\beta\frac{\rho\lambda_r w^{2r-2}}{a^2\beta nu^{2r-4}}\Lambda_{nr}(\rho;0,-\lambda_r)\cos\phi_m \\
H_{r\rho}(\zeta) &= -j\omega_0\,\Lambda_{nr}(\rho;s_r,-\lambda_r)\varepsilon_r\sin\phi_m \\
H_{r\varphi}(\zeta) &= -j\omega_0\,\Lambda_{nr}(\rho;s_r,\lambda_r)\varepsilon_r\cos\phi_m \\
H_{rz}(\zeta) &= \omega_0\frac{\rho\lambda_r s_r w^{2r-2}}{a^2\beta nu^{2r-4}}\Lambda_{nr}(\rho;0,-\lambda_r)\varepsilon_r\sin\phi_m
\end{aligned}$$

(7.24)

Thus, (7.21) may be significantly simplified using the generating function (7.23), and yielding the more compact versions

$$E_{r\xi}(\rho,\varphi) = \beta\,\frac{\Lambda_{nr}\left[\rho;s(1-\delta_{\xi z}),\lambda_r(\delta_{\xi\rho}-\delta_{\xi\varphi}-\delta_{\xi z})\right]\cos(n\varphi+\psi_m)}{j\delta_{\xi\rho}-j\delta_{\xi\varphi}\cot(n\varphi+\psi_m)-na^2\beta\dfrac{\lambda_r u^{2r-4}}{\rho w^{2r-2}}\delta_{\xi z}}$$

$$H_{r\xi}(\rho,\varphi) = \omega_0\,\frac{\Lambda_{nr}\left[\rho;s_r(1-\delta_{\xi z}),-\lambda_r(\delta_{\xi\rho}-\delta_{\xi\varphi}+\delta_{\xi z})\right]\sin(n\varphi+\psi_m)}{j\delta_{\xi\rho}+j\delta_{\xi\varphi}\tan(n\varphi+\psi_m)+na^2\beta\dfrac{\lambda_r u^{2r-4}}{\rho s_r w^{2r-2}}\delta_{\xi z}}\varepsilon_r$$

(7.25)

As shown in §**3.3**, it is also possible to reduce the 12 equations of (7.1, 2) using (7.14-19) in an analytical, *bi-regional* expression

$$V_\xi(\zeta) = \sum_{r=1}^{2}V_{r\xi}(\zeta)f_r(\rho), \quad V \in \{E,H\},\ \xi \in \{\rho,\varphi,z\},$$

(7.26)

which makes use of the orthogonal Heaviside step-functions discussed in §**3.3**, that geometrically represents the **circular core for $r = 1$**, and **the annular cladding, for $r = 2$**,

$$f_r(\rho) = (2^{2-r}-\delta_{\rho a}^{r-1})\mathrm{H}\!\left(\frac{a-\rho}{(\rho-b)^{1-r}}\right);\quad r \in \{1,2\}.$$

(7.27)

and extend the regional definition of each component in (7.1, 2), or in (7.14-19) to *both* the core and the cladding, thereby reducing the *total* number of equations from 12 to 7. There are indeed just 6 equations in the new nomenclature (7.14-19), but the total number of equations is still 12 over the entire cross-section of the fiber, like (7.1, 2). Adopting the bi-regional approach (7.26) lowers this total to 7. In (7.26), $V_{r\xi}(\zeta)$ may represent the $\xi$-component in (7.21, 25), or (7.1) when $r = 1$, and (7.2) when $r = 2$.

Using (7.21), (7.12, 13), and the equivalence between the Kronecker delta and the vector-products

$$\delta_{\xi\xi'} = 1 - |\boldsymbol{\xi} \times \boldsymbol{\xi}'| = \boldsymbol{\xi} \cdot \boldsymbol{\xi}', \quad \xi' \in \{\rho, \varphi, z\} \tag{7.28}$$

it is shown in **§3.4** that an expression for **the vector** $\vec{\mathbf{V}}(\zeta)$ (7.4) of the *entire* EM-field of any hybrid mode, and valid over the entire cross-section of the fiber, may be concisely stated as the **complex vector equations**,

$$\vec{\mathbf{E}}(\zeta) = \sum_{r=1}^{2} \sum_{\xi=\rho,\varphi,z} \boldsymbol{\xi} \frac{\left[(1-s|\boldsymbol{\xi}\times\mathbf{z}|)\dfrac{J_{n-1}^{2-r}(u\rho/a)}{K_{n-1}^{1-r}(w\rho/a)} + \dfrac{\lambda_r(1+s|\boldsymbol{\xi}\times\mathbf{z}|)}{\boldsymbol{\xi}\cdot\boldsymbol{\rho}-\boldsymbol{\xi}\cdot\boldsymbol{\varphi}-\boldsymbol{\xi}\cdot\mathbf{z}}\dfrac{J_{n+1}^{2-r}(u\rho/a)}{K_{n+1}^{1-r}(w\rho/a)}\right]f_r(\rho)\cos\phi_m}{\dfrac{2u}{aA\beta}\left(\dfrac{uJ_n(u)}{wK_n(w)}\right)^{1-r}\left[j\boldsymbol{\xi}\cdot\boldsymbol{\rho}-j\boldsymbol{\xi}\cdot\boldsymbol{\varphi}\cot\phi_m + na^2\beta\dfrac{\lambda_r u^{2r-4}}{\rho w^{2r-2}}\boldsymbol{\xi}\cdot\mathbf{z}\right]}$$

$$\vec{\mathbf{H}}(\zeta) = \sum_{r=1}^{2} \sum_{\xi=\rho,\varphi,z} \boldsymbol{\xi} \frac{\left[(1-s_r|\boldsymbol{\xi}\times\mathbf{z}|)\dfrac{J_{n-1}^{2-r}(u\rho/a)}{K_{n-1}^{1-r}(w\rho/a)} - \dfrac{\lambda_r(1+s_r|\boldsymbol{\xi}\times\mathbf{z}|)}{\boldsymbol{\xi}\cdot\boldsymbol{\rho}-\boldsymbol{\xi}\cdot\boldsymbol{\varphi}+\boldsymbol{\xi}\cdot\mathbf{z}}\dfrac{J_{n+1}^{2-r}(u\rho/a)}{K_{n+1}^{1-r}(w\rho/a)}\right]\bar{\varepsilon}_r(\rho)\sin\phi_m}{\dfrac{2u}{aA\omega_0}\left(\dfrac{uJ_n(u)}{wK_n(w)}\right)^{1-r}\left[j\boldsymbol{\xi}\cdot\boldsymbol{\rho}+j\boldsymbol{\xi}\cdot\boldsymbol{\varphi}\tan\phi_m - na^2\beta\dfrac{\lambda_r u^{2r-4}}{\rho s_r w^{2r-2}}\boldsymbol{\xi}\cdot\mathbf{z}\right]}$$

(7.29)

which are obtained by summing each of the 2 components of (7.21), over the 2 regions of the fiber's cross-section using (7.26, 27), and over the three cylindrical coordinate vectors {$\boldsymbol{\rho}$, $\boldsymbol{\varphi}$, $\mathbf{z}$}, using the $\xi$-summation. The permittivity $\bar{\varepsilon}_r(\rho)$ is the product of the regional permittivity $\varepsilon_r$ with the step-function (7.27). **It can be concluded that the field vectors are mutually orthogonal in the azimuth $\varphi$, since the trigonometric functions in one vector are rotated by $\pi/2$ radians relative to those in the other vector.** This may not be so obvious from the 12 equations (7.1, 2). A contracted version of (7.29) can be obtained by taking the alpha-numeric double-sum of (7.25), yielding

$$\vec{\mathbf{E}}(\rho,\varphi) = -j\beta \sum_{r=1}^{2} \sum_{\xi=\rho,\varphi,z} \boldsymbol{\xi} \frac{\Lambda_{nr}\left[\rho; s|\boldsymbol{\xi}\times\mathbf{z}|, \lambda_r(\boldsymbol{\xi}\cdot\boldsymbol{\rho}-\boldsymbol{\xi}\cdot\boldsymbol{\varphi}-\boldsymbol{\xi}\cdot\mathbf{z})\right]\cos(n\varphi+\psi_m)}{\boldsymbol{\xi}\cdot\boldsymbol{\rho}-\boldsymbol{\xi}\cdot\boldsymbol{\varphi}\cot(n\varphi+\psi_m) + jna^2\beta\dfrac{\lambda_r u^{2r-4}}{\rho w^{2r-2}}\boldsymbol{\xi}\cdot\mathbf{z}} f_r(\rho)$$

$$\vec{\mathbf{H}}(\rho,\varphi) = -j\omega_0 \sum_{r=1}^{2} \sum_{\xi=\rho,\varphi,z} \boldsymbol{\xi} \frac{\Lambda_{nr}\left[\rho; s_r|\boldsymbol{\xi}\times\mathbf{z}|, -\lambda_r(\boldsymbol{\xi}\cdot\boldsymbol{\rho}-\boldsymbol{\xi}\cdot\boldsymbol{\varphi}+\boldsymbol{\xi}\cdot\mathbf{z})\right]\sin(n\varphi+\psi_m)}{\boldsymbol{\xi}\cdot\boldsymbol{\rho}+\boldsymbol{\xi}\cdot\boldsymbol{\varphi}\tan(n\varphi+\psi_m) - jna^2\beta\dfrac{\lambda_r u^{2r-4}}{\rho s_r w^{2r-2}}\boldsymbol{\xi}\cdot\mathbf{z}} \bar{\varepsilon}_r(\rho)$$

(7.30)

These equations may be specialized to the HE$_{11}$-mode, which is the only mode common to both single-mode and multi-mode fibers, simply by setting $n = 1$, in (7.9, 13, 23),

$$\vec{E}_m(\rho,\varphi) = -j\beta \sum_{r=1}^{2} \sum_{\xi=\rho,\varphi,z} \xi \frac{\Lambda_r\left[\rho; s|\xi \times z|, \lambda_r(\xi \cdot \rho - \xi \cdot \varphi - \xi \cdot z)\right] \cos(\varphi + \psi_m)}{\xi \cdot \rho - \xi \cdot \varphi \cot(\varphi + \psi_m) + ja^2\beta \frac{\lambda_r u^{2r-4}}{\rho w^{2r-2}} \xi \cdot z} f_r(\rho)$$

$$\vec{H}_m(\rho,\varphi) = -j\omega_0 \sum_{r=1}^{2} \sum_{\xi=\rho,\varphi,z} \xi \frac{\Lambda_r\left[\rho; s_r|\xi \times z|, -\lambda_r(\xi \cdot \rho - \xi \cdot \varphi + \xi \cdot z)\right] \sin(\varphi + \psi_m)}{\xi \cdot \rho + \xi \cdot \varphi \tan(\varphi + \psi_m) - ja^2\beta \frac{\lambda_r u^{2r-4}}{\rho s_r w^{2r-2}} \xi \cdot z} \bar{\varepsilon}_r(\rho)$$

(7.31)

The vectors have been additionally subscripted with an $m$ to emphasize their dependence on the modal phase factor $\psi_m$ (7.11), which determines the 2 polarization states.

Using (7.31) with $m = 1$ yields the spatial profile of the HE$_{11}^x$-mode,

$$\vec{E}_1(\rho,\varphi) = -j\beta \sum_{r=1}^{2} \sum_{\xi=\rho,\varphi,z} \xi \frac{\Lambda_r\left[\rho; s|\xi \times z|, \lambda_r(\xi \cdot \rho - \xi \cdot \varphi - \xi \cdot z)\right] \cos\varphi}{\xi \cdot \rho - \xi \cdot \varphi \cot\varphi + ja^2\beta \frac{\lambda_r u^{2r-4}}{\rho w^{2r-2}} \xi \cdot z} f_r(\rho)$$

$$\vec{H}_1(\rho,\varphi) = -j\omega_0 \sum_{r=1}^{2} \sum_{\xi=\rho,\varphi,z} \xi \frac{\Lambda_r\left[\rho; s_r|\xi \times z|, -\lambda_r(\xi \cdot \rho - \xi \cdot \varphi + \xi \cdot z)\right] \sin\varphi}{\xi \cdot \rho + \xi \cdot \varphi \tan\varphi - ja^2\beta \frac{\lambda_r u^{2r-4}}{\rho s_r w^{2r-2}} \xi \cdot z} \bar{\varepsilon}_r(\rho)$$

(7.32)

and using (7.31) again with $m = 2$, yields the spatial profile of the HE$_{11}^y$-mode,

$$\vec{E}_2(\rho,\varphi) = j\beta \sum_{r=1}^{2} \sum_{\xi=\rho,\varphi,z} \xi \frac{\Lambda_r\left[\rho; s|\xi \times z|, \lambda_r(\xi \cdot \rho - \xi \cdot \varphi - \xi \cdot z)\right] \sin\varphi}{\xi \cdot \rho + \xi \cdot \varphi \tan\varphi + ja^2\beta \frac{\lambda_r u^{2r-4}}{\rho w^{2r-1}} \xi \cdot z} f_r(\rho)$$

$$\vec{H}_2(\rho,\varphi) = -j\omega_0 \sum_{r=1}^{2} \sum_{\xi=\rho,\varphi,z} \xi \frac{\Lambda_r\left[\rho; s_r|\xi \times z|, -\lambda_r(\xi \cdot \rho - \xi \cdot \varphi + \xi \cdot z)\right] \cos\varphi}{\xi \cdot \rho - \xi \cdot \varphi \cotan\varphi - ja^2\beta \frac{\lambda_r u^{2r-4}}{\rho s_r w^{2r-1}} \xi \cdot z} \bar{\varepsilon}_r(\rho)$$

(7.33)

It can also be seen from (7.32, 33), that the 2 polarization states are in "quadrature" with respect to the azimuth $\varphi$. This observation may be operationally expressed as the convolution ($*$) of the EM-field vector $\vec{V}_m$ of either $m$-th mode, with a Dirac delta-function,

$$\vec{V}_{3-m}(\boldsymbol{r},t) = \delta(\varphi + (3-2m)\pi/2) * \vec{V}_m(\boldsymbol{r},t) \;, \quad \vec{V} \in \{\vec{E}, \vec{H}\}, \; m \in \{1,2\}. \quad (7.34)$$

Thus, knowing the expressions for the EM-field vectors of one of the polarization modes, yields those for the EM-field vectors of the other, via an angular rotation operationally expressed as the above convolution. **This relation actually holds for any hybrid mode**.

**Segre's bi-complex convention** [3, 4], introduced in **§2**, makes use of 2 distinct imaginary numbers, i and j, that conform to the following rules

$$i^2 = -1; \quad j^2 = -1; \quad ij = ji \neq -1; \\ i^\circ = -i; \quad j^* = -j, \tag{7.35}$$

**and which uses different conjugation superscripts**. It should be emphasized that according to this definition, bi-complex numbers are not quaternions [7], since the product ij is defined as being commutative according to (7.35). Then for any 2 complex numbers,

$$z_1 = x_1 + i y_1, \quad \{x_1, y_1\} \in \mathbb{R} \\ z_2 = x_2 + j y_2, \quad \{x_2, y_2\} \in \mathbb{R} \tag{7.36}$$

it was shown in **§2** that sequential real-operations with respect to i and j are commutative,

$$\operatorname{Re}_i \operatorname{Re}_j [z_1 z_2] = \operatorname{Re}_j \operatorname{Re}_i [z_1 z_2] \tag{7.37}$$

**or with i (j) considered to be a real constant under a real-operation in j (i)**.

It was also shown that this treatment applies to complex conjugation,

$$(z_1 z_2)^{\circ *} = z_1^\circ z_2^* \tag{7.38}$$

**or that conjugation with respect to i (j) treats j (i) as a real constant.**

Furthermore, it was shown that **the composite real-part of a bi-complex number is identical with that of its conjugate**,

$$\operatorname{Re}_i \operatorname{Re}_j \left[ (z_1 z_2)^{\circ *} \right] = \operatorname{Re}_i \operatorname{Re}_j \left[ z_1 z_2 \right] \tag{7.39}$$

Since the imaginary number j was already in use in the nomenclature (7.1, 2), the 2nd imaginary number i is introduced to alleviate some of the difficulty associated with using trigonometric functions. This is the motivation for the adoption of the bi-complex convention in this report. Doing so reduces the EM-field components in (7.1, 2) to phasors in the cylindrical azimuth $\varphi$, and renders them bi-complex quantities.

It is possible to convert the complex EM-field components ($V_\xi(\zeta)$) given by (7.1, 2), or equations (7.14-19) derived in this report, to phasors ($\tilde{U}_\xi(\zeta)$) in the new bi-complex convention, using the following complex-to-bicomplex (CTB) transformation

$$\tilde{U}_\xi(\rho,\varphi) = \left[ \frac{(\delta_{\xi\rho} + \delta_{\xi z})\delta_{VE} + \delta_{\xi\varphi}\delta_{VH}}{\cos(n\varphi + \psi_m)} + i \frac{\delta_{\xi\varphi}\delta_{VE} + (\delta_{\xi\rho} + \delta_{\xi z})\delta_{VH}}{\sin(n\varphi + \psi_m)} \right] V_\xi(\rho,\varphi) e^{-i(n\varphi + \psi_m)} \tag{7.40}$$

which can be described as a division by a $\cos(n\varphi + \psi_m)$ if the component carries this function (which is true for either the $\rho$- and $z$-components of the $E$-field, or the $\varphi$-component of the $H$-field), OR a division by $-i\sin(n\varphi + \psi_m)$ if the component carries a $\sin(n\varphi + \psi_m)$ function (which is true for either the $\varphi$-component of the $E$-field, or the $\rho$- and $z$-components of the $H$-field). Lastly, the result is multiplied by the spatial phasor regardless of the trigonometric dependence of the component. The transformation can also be observationally summarized as the replacement of any instance of a $\cos\phi_m$ in (7.1, 2) with $e^{-i\phi_m}$, whereas any instance of a $\sin\phi_m$ in (7.1, 2), is replaced by $ie^{-i\phi_m}$. **The sign of a EM-field component in (7.1, 2) is preserved under this transformation, due to the judicious selection of a negative exponent in the phase factor** $e^{-i\phi_m}$. Although the resultant quantity $\tilde{U}_\xi(\zeta)$ is bi-complex, it is only a phasor with respect to the cylindrical azimuth $\varphi$, through the use of (7.13). Lastly, and as is shown in §**3.1**, (7.40) is not unique.

Applying the transformation (7.40) to (7.1, 2), then the 6 components of (7.14 - 19) may be reduced to the following 6 bi-complex phasor equations, which yield 12 components simply by setting $r = 1$ for the core, and $r = 2$ for the cladding,

$$\tilde{E}_{r\rho}(\rho,\varphi) = -j\beta\frac{aA}{2u}\left(\frac{uJ_n(u)}{wK_n(w)}\right)^{r-1}\left[(1-s)\frac{J_{n-1}^{2-r}(u\rho/a)}{K_{n-1}^{1-r}(w\rho/a)} + \lambda_r(1+s)\frac{J_{n+1}^{2-r}(u\rho/a)}{K_{n+1}^{1-r}(w\rho/a)}\right]e^{-i(\varphi+\psi_m)} \tag{7.41}$$

$$\tilde{E}_{r\varphi}(\rho,\varphi) = ij\beta\frac{aA}{2u}\left(\frac{uJ_n(u)}{wK_n(w)}\right)^{r-1}\left[(1-s)\frac{J_{n-1}^{2-r}(u\rho/a)}{K_{n-1}^{1-r}(w\rho/a)} - \lambda_r(1+s)\frac{J_{n+1}^{2-r}(u\rho/a)}{K_{n+1}^{1-r}(w\rho/a)}\right]e^{-i(\varphi+\psi_m)} \tag{7.42}$$

$$\tilde{E}_{rz}(\rho,\varphi) = A\frac{J_n^{r-1}(u)}{K_n^{r-1}(w)}\frac{J_n^{2-r}(u\rho/a)}{K_n^{1-r}(w\rho/a)}e^{-i(\varphi+\psi_m)} \tag{7.43}$$

$$\tilde{H}_{r\rho}(\rho,\varphi) = -ij\omega_0\frac{aA}{2u}\left(\frac{uJ_n(u)}{wK_n(w)}\right)^{r-1}\left[(1-s_r)\frac{J_{n-1}^{2-r}(u\rho/a)}{K_{n-1}^{1-r}(w\rho/a)} - \lambda_r(1+s_r)\frac{J_{n+1}^{2-r}(u\rho/a)}{K_{n+1}^{1-r}(w\rho/a)}\right]\varepsilon_r e^{-i(\varphi+\psi_m)} \tag{7.44}$$

$$\tilde{H}_{r\varphi}(\rho,\varphi) = -j\omega_0\frac{aA}{2u}\left(\frac{uJ_n(u)}{wK_n(w)}\right)^{r-1}\left[(1-s_r)\frac{J_{n-1}^{2-r}(u\rho/a)}{K_{n-1}^{1-r}(w\rho/a)} + \lambda_r(1+s_r)\frac{J_{n+1}^{2-r}(u\rho/a)}{K_{n+1}^{1-r}(w\rho/a)}\right]\varepsilon_r e^{-i(\varphi+\psi_m)} \tag{7.45}$$

$$\tilde{H}_{rz}(\rho,\varphi) = -i\frac{A\beta s}{\omega_0\mu_0}\frac{J_n^{r-1}(u)}{K_n^{r-1}(w)}\frac{J_n^{2-r}(u\rho/a)}{K_n^{1-r}(w\rho/a)}e^{-i(\varphi+\psi_m)} \tag{7.46}$$

They are more explicit than their complex counterparts (7.14-19) with respect to the azimuth, for the same level compactness. To return the bi-complex components above to those (7.14-19) found using the complex formulation, the following relation is used:

$$V_{r\xi}(\zeta) = \mathrm{Re}_i\, \tilde{U}_{r\xi}(\zeta). \tag{7.47}$$

A *real* propagation vector in an *r*-th region is obtainable using the list (7.41-46), with

$$\vec{V}_r(r,t) = \sum_{\xi=\rho,\varphi,z} \xi\, \mathrm{Re}_j\left[\mathrm{Re}_i\, \tilde{U}_{r\xi}(\zeta)\mathrm{e}^{j(\omega_0 t-\beta z)}\right]\,;\quad \tilde{U}\in\{\tilde{E},\tilde{H}\},\ \xi\in\{\rho,\varphi,z\},\ r\in\{1,2\} \tag{7.48}$$

which requires 2 real-operations, one to recover the azimuthal trigonometric functions, and another to recover the trigonometric function used for longitudinal propagation.

It is also possible to reduce the 12 equations of (7.41 - 46) to just 6, using (7.27) in the analytical expression

$$\tilde{U}_\xi(\zeta) = \sum_{r=1}^{2} \tilde{U}_{r\xi}(\zeta) f_r(\rho),\quad \tilde{U}\in\{\tilde{E},\tilde{H}\},\ \xi\in\{\rho,\varphi,z\}, \tag{7.49}$$

which extends the regional definition of each component in (7.41-46) to both the core and the cladding,

Applying (7.40) to (7.20) yields the *z*-components of the EM-field in the alternative formulation

$$\tilde{E}_{rz}(\zeta) = -\frac{\rho A}{2anu}\frac{\lambda_r w^{2r-2}}{u^{2r-4}}\left(\frac{uJ_n(u)}{wK_n(w)}\right)^{r-1}\left[\frac{J_{n-1}^{2-r}(u\rho/a)}{K_{n-1}^{1-r}(w\rho/a)} - \lambda_r\frac{J_{n+1}^{2-r}(u\rho/a)}{K_{n+1}^{1-r}(w\rho/a)}\right]\mathrm{e}^{-i\phi_m}$$

$$\tilde{H}_{rz}(\zeta) = \mathrm{i}\frac{\omega_0}{\beta}\frac{\rho A}{2anu}\frac{\lambda_r \varepsilon_r s_r w^{2r-2}}{u^{2r-4}}\left(\frac{uJ_n(u)}{wK_n(w)}\right)^{r-1}\left[\frac{J_{n-1}^{2-r}(u\rho/a)}{K_{n-1}^{1-r}(w\rho/a)} - \lambda_r\frac{J_{n+1}^{2-r}(u\rho/a)}{K_{n+1}^{1-r}(w\rho/a)}\right]\mathrm{e}^{-i\phi_m}$$

(7.50)

The transverse (7.41, 42, 44, 45) and the longitudinal components (7.50) can together be summarized in a general, *explicit* expression for *any* component of the EM-field of any hybrid mode in either of the 2 regions of the fiber, termed the **bi-complex generalized scalar (BGS) equations,**

$$\tilde{E}_{r\xi}(\rho,\varphi) = \frac{\left[(1-s+s\delta_{\xi z})\dfrac{J_{n-1}^{2-r}(u\rho/a)}{K_{n-1}^{1-r}(w\rho/a)} + \dfrac{(1+s-s\delta_{\xi z})\mathrm{e}^{j\pi r}}{\delta_{\xi\rho}-\delta_{\xi\varphi}-\delta_{\xi z}}\dfrac{J_{n+1}^{2-r}(u\rho/a)}{K_{n+1}^{1-r}(w\rho/a)}\right]\mathrm{e}^{-i(n\varphi+\psi_m)}}{\dfrac{2u}{aA\beta}\left(\dfrac{uJ_n(u)}{wK_n(w)}\right)^{1-r}\left[\mathrm{j}\delta_{\xi\rho}+\mathrm{ij}\delta_{\xi\varphi}+na^2\beta\dfrac{u^{2r-4}\mathrm{e}^{j\pi r}}{\rho w^{2r-2}}\delta_{\xi z}\right]}$$

$$\tilde{H}_{r\xi}(\rho,\varphi) = \frac{\left[(1-s_r+s_r\delta_{\xi z})\dfrac{J_{n-1}^{2-r}(u\rho/a)}{K_{n-1}^{1-r}(w\rho/a)} - \dfrac{(1+s_r-s_r\delta_{\xi z})\mathrm{e}^{j\pi r}}{\delta_{\xi\rho}-\delta_{\xi\varphi}+\delta_{\xi z}}\dfrac{J_{n+1}^{2-r}(u\rho/a)}{K_{n+1}^{1-r}(w\rho/a)}\right]\varepsilon_r\mathrm{i}\mathrm{e}^{-i(n\varphi+\psi_m)}}{\dfrac{2u}{aA\omega_0}\left(\dfrac{uJ_n(u)}{wK_n(w)}\right)^{1-r}\left[\mathrm{j}\delta_{\xi\rho}+\mathrm{ij}\delta_{\xi\varphi}-na^2\beta\dfrac{u^{2r-4}\mathrm{e}^{j\pi r}}{\rho s_r w^{2r-2}}\delta_{\xi z}\right]}$$

(7.51)

with $\xi \in \{\rho, \varphi, z\}$ as before. Any phasor (7.41-46) can be recovered by setting $r$ for the region of interest, and selecting the relevant $\xi$-component, which is operationally identical to (7.22). Compact versions of (7.41, 42, 44, 45, 50) are also found using the generating function (7.23),

$$\begin{aligned}
\tilde{E}_{r\rho}(\zeta) &= -j\beta \Lambda_{nr}(\rho; s, \lambda_r) e^{-i\phi_m} \\
\tilde{E}_{r\varphi}(\zeta) &= ij\beta \Lambda_{nr}(\rho; s, -\lambda_r) e^{-i\phi_m} \\
\tilde{E}_{rz}(\zeta) &= -\beta \frac{\lambda_r w^{2r-2}}{a^2 \beta n u^{2r-4}} \Lambda_{nr}(\rho; 0, -\lambda_r) \rho e^{-i\phi_m} \\
\tilde{H}_{r\rho}(\zeta) &= -ij\omega_0 \Lambda_{nr}(\rho; s_r, -\lambda_r) \varepsilon_r e^{-i\phi_m} \\
\tilde{H}_{r\varphi}(\zeta) &= -j\omega_0 \Lambda_{nr}(\rho; s_r, \lambda_r) \varepsilon_r e^{-i\phi_m} \\
\tilde{H}_{rz}(\zeta) &= i\omega_0 \frac{\lambda_r s_r w^{2r-2}}{a^2 \beta n u^{2r-4}} \Lambda_{nr}(\rho; 0, -\lambda_r) \rho \varepsilon_r e^{-i\phi_m}
\end{aligned} \quad (7.52)$$

They can also be turned into bi-regional expressions using the sum (7.49).

Using the new, bi-complex field formulation (7.52), the generalized, compact scalar expressions for the EM-field of a hybrid mode are deduced to be

$$\tilde{E}_{r\zeta}(\rho, \varphi) = -j\beta \frac{\Lambda_{nr}\left[\rho; s(1-\delta_{\xi z}), \lambda_r(\delta_{\xi\rho} - \delta_{\xi\varphi} - \delta_{\xi z})\right]}{\delta_{\xi\rho} + i\delta_{\xi\varphi} + jna^2\beta \frac{\lambda_r u^{2r-4}}{\rho w^{2r-2}} \delta_{\xi z}} e^{-i(n\varphi + \psi_m)}$$

$$\tilde{H}_{r\zeta}(\rho, \varphi) = -j\omega_0 \frac{\Lambda_{nr}\left[\rho; s_r(1-\delta_{\xi z}), -\lambda_r(\delta_{\xi\rho} - \delta_{\xi\varphi} + \delta_{\xi z})\right]}{\delta_{\xi\rho} + i\delta_{\xi\varphi} - jna^2\beta \frac{\lambda_r u^{2r-4}}{\rho s_r w^{2r-2}} \delta_{\xi z}} \varepsilon_r i e^{-i(n\varphi + \psi_m)}$$

(7.53)

The EM-field **vecsor** (which is a vector *and* a phasor) may be easily constructed by taking the vector sum of (7.51) over the vector-index $\xi \in \{\boldsymbol{\rho}, \boldsymbol{\varphi}, \mathbf{z}\}$, followed by a 2nd sum over the fiber's 2 regions, resulting in the **bi-complex vecsor equations**,

$$\tilde{\mathbf{E}}(\zeta) = \sum_{r=1}^{2} \sum_{\xi = \boldsymbol{\rho}, \boldsymbol{\varphi}, \mathbf{z}} \xi \frac{\left[(1-s|\xi \times \mathbf{z}|)\frac{J_{n-1}^{2-r}(u\rho/a)}{K_{n-1}^{1-r}(w\rho/a)} + \frac{(1+s|\xi \times \mathbf{z}|)e^{j\pi r}}{\xi \cdot \boldsymbol{\rho} - \xi \cdot \boldsymbol{\varphi} - \xi \cdot \mathbf{z}} \frac{J_{n+1}^{2-r}(u\rho/a)}{K_{n+1}^{1-r}(w\rho/a)}\right] f_r(\rho)}{\frac{2u}{aA\beta}\left(\frac{uJ_n(u)}{wK_n(w)}\right)^{1-r} \left[j\xi \cdot \boldsymbol{\rho} + ij\xi \cdot \boldsymbol{\varphi} - na^2\beta \frac{u^{2r-4}e^{j\pi r}}{\rho w^{2r-2}} \xi \cdot \mathbf{z}\right]} e^{-i(n\varphi + \psi_m)}$$

$$\tilde{\mathbf{H}}(\zeta) = \sum_{r=1}^{2} \sum_{\xi = \boldsymbol{\rho}, \boldsymbol{\varphi}, \mathbf{z}} \xi \frac{\left[(1-s_r|\xi \times \mathbf{z}|)\frac{J_{n-1}^{2-r}(u\rho/a)}{K_{n-1}^{1-r}(w\rho/a)} - \frac{(1+s_r|\xi \times \mathbf{z}|)e^{j\pi r}}{\xi \cdot \boldsymbol{\rho} - \xi \cdot \boldsymbol{\varphi} + \xi \cdot \mathbf{z}} \frac{J_{n+1}^{2-r}(u\rho/a)}{K_{n+1}^{1-r}(w\rho/a)}\right] \bar{\varepsilon}_r(\rho)}{\frac{2u}{aA\omega_0}\left(\frac{uJ_n(u)}{wK_n(w)}\right)^{1-r} \left[j\xi \cdot \boldsymbol{\rho} + ij\xi \cdot \boldsymbol{\varphi} + na^2\beta \frac{u^{2r-4}e^{j\pi r}}{\rho s_r w^{2r-2}} \xi \cdot \mathbf{z}\right]} i e^{-i(n\varphi + \psi_m)}$$

(7.54)

Examining (7.51) and (7.54), **it is concluded** that the bi-complex formulation has resulted in expressions that are simpler than their counterparts (7.21) and (7.29) in the complex formulation, due to the elimination of the trigonometric functions. Moreover **and unlike (7.21) and (7.29)**, the bi-complex expressions are separable in the radial and azimuthal variables, and the E-field and H-field quantities are also functionally identical. Furthermore, the magnetic field vector carries an additional multiplicative imaginary number **i** (shown in **bold-type** only for emphasis), confirming that, for the same $m$-th polarization, this vector is orthogonal to the electric field vector with respect to the azimuth $\varphi$. This well-known, gross relation is perhaps difficult to deduce immediately from the 12, component-resolved expressions, such as (7.1, 2) or (7.41-46). More compact versions of (7.54) are possible with the use of the generating function (7.23),

$$\tilde{\mathbf{E}}(\rho,\varphi) = -j\beta \sum_{r=1}^{2}\sum_{\xi=\rho,\varphi,z} \xi \frac{\Lambda_{nr}\left[\rho;s|\xi\times\mathbf{z}|,\lambda_r(\xi\cdot\boldsymbol{\rho}-\xi\cdot\boldsymbol{\varphi}-\xi\cdot\mathbf{z})\right]}{\xi\cdot\boldsymbol{\rho}+i\xi\cdot\boldsymbol{\varphi}+jna^2\beta\dfrac{\lambda_r u^{2r-4}}{\rho w^{2r-2}}\xi\cdot\mathbf{z}} f_r(\rho)e^{-i(n\varphi+\psi_m)}$$

$$\tilde{\mathbf{H}}(\rho,\varphi) = -j\omega_0 \sum_{r=1}^{2}\sum_{\xi=\rho,\varphi,z} \xi \frac{\Lambda_{nr}\left[\rho;s_r|\xi\times\mathbf{z}|,-\lambda_r(\xi\cdot\boldsymbol{\rho}-\xi\cdot\boldsymbol{\varphi}+\xi\cdot\mathbf{z})\right]}{\xi\cdot\boldsymbol{\rho}+i\xi\cdot\boldsymbol{\varphi}-jna^2\beta\dfrac{\lambda_r u^{2r-4}}{\rho s_r w^{2r-2}}\xi\cdot\mathbf{z}} \bar{\varepsilon}_r(\rho)\mathbf{i}e^{-i(n\varphi+\psi_m)}$$

(7.55)

Bi-complex vecsor expressions for the $HE_{11}$-mode are found by setting $n = 1$ in (7.9) and (7.13), and after the substitution of (7.11) into (7.55),

$$\tilde{\mathbf{E}}_m(\rho,\varphi) = -j\beta \sum_{r=1}^{2}\sum_{\xi=\rho,\varphi,z} \xi \frac{\Lambda_r\left[\rho;s|\xi\times\mathbf{z}|,\lambda_r(\xi\cdot\boldsymbol{\rho}-\xi\cdot\boldsymbol{\varphi}-\xi\cdot\mathbf{z})\right]}{\xi\cdot\boldsymbol{\rho}+i\xi\cdot\boldsymbol{\varphi}+ja^2\beta\dfrac{\lambda_r u^{2r-4}}{\rho w^{2r-2}}\xi\cdot\mathbf{z}} f_r(\rho)e^{-i(\varphi+(m-1)\pi/2)}$$

$$\tilde{\mathbf{H}}_m(\rho,\varphi) = -j\omega_0 \sum_{r=1}^{2}\sum_{\xi=\rho,\varphi,z} \xi \frac{\Lambda_r\left[\rho;s_r|\xi\times\mathbf{z}|,-\lambda_r(\xi\cdot\boldsymbol{\rho}-\xi\cdot\boldsymbol{\varphi}+\xi\cdot\mathbf{z})\right]}{\xi\cdot\boldsymbol{\rho}+i\xi\cdot\boldsymbol{\varphi}-ja^2\beta\dfrac{\lambda_r u^{2r-4}}{\rho s_r w^{2r-2}}\xi\cdot\mathbf{z}} \bar{\varepsilon}_r(\rho)\mathbf{i}e^{-i(\varphi+(m-1)\pi/2)}$$

(7.56)

The $x$-polarization vecsors of the $HE_{11}$-mode is found by setting $m = 1$,

$$\tilde{\mathbf{E}}_1(\rho,\varphi) = -j\beta \sum_{r=1}^{2}\sum_{\xi=\rho,\varphi,z} \xi \frac{\Lambda_r\left[\rho;s|\xi\times\mathbf{z}|,\lambda_r(\xi\cdot\boldsymbol{\rho}-\xi\cdot\boldsymbol{\varphi}-\xi\cdot\mathbf{z})\right]}{\xi\cdot\boldsymbol{\rho}+i\xi\cdot\boldsymbol{\varphi}+ja^2\beta\dfrac{\lambda_r u^{2r-4}}{\rho w^{2r-2}}\xi\cdot\mathbf{z}} f_r(\rho)e^{-i\varphi}$$

$$\tilde{\mathbf{H}}_1(\rho,\varphi) = -\mathbf{i}j\omega_0 \sum_{r=1}^{2}\sum_{\xi=\rho,\varphi,z} \xi \frac{\Lambda_r\left[\rho;s_r|\xi\times\mathbf{z}|,-\lambda_r(\xi\cdot\boldsymbol{\rho}-\xi\cdot\boldsymbol{\varphi}+\xi\cdot\mathbf{z})\right]}{\xi\cdot\boldsymbol{\rho}+i\xi\cdot\boldsymbol{\varphi}-ja^2\beta\dfrac{\lambda_r u^{2r-4}}{\rho s_r w^{2r-2}}\xi\cdot\mathbf{z}} \bar{\varepsilon}_r(\rho)e^{-i\varphi}$$

(7.57)

whereas the *y*-polarization vecsors of the $HE_{11}$-mode is found by setting $m = 2$,

$$\tilde{\mathbf{E}}_2(\rho,\varphi) = \mathrm{ij}\beta \sum_{r=1}^{2} \sum_{\xi=\rho,\varphi,z} \xi \frac{\Lambda_r\left[\rho;s|\xi\times\mathbf{z}|,\lambda_r(\xi\cdot\boldsymbol{\rho}-\xi\cdot\boldsymbol{\varphi}-\xi\cdot\mathbf{z})\right]}{\xi\cdot\boldsymbol{\rho}+\mathrm{i}\xi\cdot\boldsymbol{\varphi}+\mathrm{j}a^2\beta\dfrac{\lambda_r u^{2r-4}}{\rho w^{2r-2}}\xi\cdot\mathbf{z}} f_r(\rho)\mathrm{e}^{-\mathrm{i}\varphi}$$

$$\tilde{\mathbf{H}}_2(\rho,\varphi) = -\mathrm{j}\omega_0 \sum_{r=1}^{2} \sum_{\xi=\rho,\varphi,z} \xi \frac{\Lambda_r\left[\rho;s_r|\xi\times\mathbf{z}|,-\lambda_r(\xi\cdot\boldsymbol{\rho}-\xi\cdot\boldsymbol{\varphi}+\xi\cdot\mathbf{z})\right]}{\xi\cdot\boldsymbol{\rho}+\mathrm{i}\xi\cdot\boldsymbol{\varphi}-\mathrm{j}a^2\beta\dfrac{\lambda_r u^{2r-4}}{\rho s_r w^{2r-2}}\xi\cdot\mathbf{z}} \bar{\varepsilon}_r(\rho)\mathrm{e}^{-\mathrm{i}\varphi}$$

(7.58)

**It is thus concluded** that in the bi-complex formulation, the *x*- and the *y*-polarizations of the $HE_{11}$-mode are identical within an imaginary number (i), and are much simpler and more compact than their counterparts (7.32, 33) in the complex formulation. This observation is also true for any hybrid mode in the bi-complex formulation, since the expressions (7.56) have not yet been specialized to either the HE- or the EH-modes. Lastly, a polarization mode of the EM-field vecsor $\tilde{\mathbf{U}}_m$ of any hybrid mode can also be related to its other polarization mode using the simple multiplicative relation

$$\tilde{\mathbf{U}}_{3-m}(\mathbf{r},t) = \exp\left[-\mathrm{i}(3-2m)\pi/2\right]\tilde{\mathbf{U}}_m(\mathbf{r},t), \qquad m \in \{1,2\} \tag{7.59}$$

instead of the convolution (7.34) used for the basic complex formulation. **Thus, a relation (7.34) based on calculus and special functions, is reduced to one (7.59) based on simple algebra, in the bi-complex formulation.**

Even more compact expressions than those presented in this report are also possible, since the generating function (7.23) can be re-expressed as the compact sum

$$\Lambda_{nr}(\rho;\eta,\lambda_r) = \frac{aA}{2u}\left(\frac{uJ_n(u)}{wK_n(w)}\right)^{r-1} \sum_{q=-1}^{1} \frac{q^2(1+q\eta)}{\lambda_r^{(1+q)/2}} \frac{J_{n+q}^{2-r}(u\rho/a)}{K_{n+q}^{1-r}(w\rho/a)}. \tag{7.60}$$

However, the summand no longer retains the algebraic aspects of the nomenclature due to [1, 2] given by (7.1, 2), sacrificing some of its elegance and/or clarity.

In §6, the **weakly-guided fiber (WGF) approximation** is used to simplify many of the expressions developed in this report. For most practical fibers of index contrast no larger than approximately 1%, the effective index of the propagating mode becomes *almost* identical to the constitutive material indices $n_r$,

$$n_{\mathrm{eff}} \approx n_r; \quad r \in \{1,2\}, \tag{7.61}$$

while the effective index is still constrained to being no larger than the index of the core, but no smaller than that of the cladding. As consequences of this approximation,

$$s \approx -1, \tag{7.62}$$

$$s_r = s(\beta/k_0 n_r)^2 = s(n_{\mathrm{eff}}/n_r)^2 \approx -1; \quad r \in \{1,2\}. \tag{7.63}$$

Then the EM-field components or generalized scalar equations are, when under the WGF approximation, *respectively* reduced to, in the **complex** (7.64) and the **bi-complex** (7.65) formulations,

$$E_{r\xi}(\zeta;n,s) = j\frac{aA}{us}\frac{J_{n+s}^{r-1}(u)}{K_{n+s}^{r-1}(w)}\frac{\left(\delta_{\xi\rho}+\delta_{\xi\varphi}\right)\beta\frac{J_{n+s}^{2-r}(u\rho/a)}{K_{n+s}^{1-r}(w\rho/a)}+\delta_{\xi z}\frac{u}{a}\frac{J_{n}^{2-r}(u\rho/a)}{K_{n}^{1-r}(w\rho/a)}}{\delta_{\xi\rho}\sec\phi_m + s\delta_{\xi\varphi}\csc\phi_m + js\delta_{\xi z}\sec\phi_m}$$

$$H_{r\xi}(\zeta;n,s) = -j\frac{aA}{u\omega_0}\frac{J_{n+s}^{r-1}(u)}{K_{n+s}^{r-1}(w)}\frac{\left(\delta_{\xi\rho}+\delta_{\xi\varphi}\right)\omega_0^2\varepsilon_r\frac{J_{n+s}^{2-r}(u\rho/a)}{K_{n+s}^{1-r}(w\rho/a)}+\delta_{\xi z}\frac{u\beta}{a\mu_0}\frac{J_{n}^{2-r}(u\rho/a)}{K_{n}^{1-r}(w\rho/a)}}{\delta_{\xi\rho}\csc\phi_m - s\delta_{\xi\varphi}\sec\phi_m + js\delta_{\xi z}\csc\phi_m}$$

(7.64)

$$\tilde{E}_{r\xi}(\zeta;n,s) = j\frac{aA}{us}\frac{J_{n+s}^{r-1}(u)}{K_{n+s}^{r-1}(w)}\left[\left(\delta_{\xi\rho}+is\delta_{\xi\varphi}\right)\beta\frac{J_{n+s}^{2-r}(u\rho/a)}{K_{n+s}^{1-r}(w\rho/a)} - j\delta_{\xi z}\frac{us}{a}\frac{J_{n}^{2-r}(u\rho/a)}{K_{n}^{1-r}(w\rho/a)}\right]e^{-i\phi_m}$$

$$\tilde{H}_{r\xi}(\zeta;n,s) = -j\frac{aA}{u\omega_0}\frac{J_{n+s}^{r-1}(u)}{K_{n+s}^{r-1}(w)}\left[\left(\delta_{\xi\rho}+is\delta_{\xi\varphi}\right)\omega_0^2\varepsilon_r\frac{J_{n+s}^{2-r}(u\rho/a)}{K_{n+s}^{1-r}(w\rho/a)} - j\delta_{\xi z}\frac{us\beta}{a\mu_0}\frac{J_{n}^{2-r}(u\rho/a)}{K_{n}^{1-r}(w\rho/a)}\right]ie^{-i\phi_m}$$

(7.65)

Furthermore, the generalized vectorial expressions in the **complex** (7.29) and **bi-complex** (7.54) formulations, *respectively* reduce to (7.66) and (7.67),

$$\vec{E}(\zeta;n,s) = j\frac{aA}{us}\sum_{r=1}^{2}\sum_{\xi=\rho,\varphi,z}\xi\frac{J_{n+s}^{r-1}(u)}{K_{n+s}^{r-1}(w)}\frac{|\xi\times z|\beta\frac{J_{n+s}^{2-r}(u\rho/a)}{K_{n+s}^{1-r}(w\rho/a)}+\xi\cdot z\frac{u}{a}\frac{J_{n}^{2-r}(u\rho/a)}{K_{n}^{1-r}(w\rho/a)}}{\xi\cdot\rho\sec\phi_m + s\xi\cdot\varphi\csc\phi_m + js\xi\cdot z\sec\phi_m}f_r(\rho)$$

$$\vec{H}(\zeta;n,s) = -j\frac{aA}{u\omega_0}\sum_{r=1}^{2}\sum_{\xi=\rho,\varphi,z}\xi\frac{J_{n+s}^{r-1}(u)}{K_{n+s}^{r-1}(w)}\frac{|\xi\times z|\omega_0^2\varepsilon_r\frac{J_{n+s}^{2-r}(u\rho/a)}{K_{n+s}^{1-r}(w\rho/a)}+\xi\cdot z\frac{u\beta}{a\mu_0}\frac{J_{n}^{2-r}(u\rho/a)}{K_{n}^{1-r}(w\rho/a)}}{\xi\cdot\rho\csc\phi_m - s\xi\cdot\varphi\sec\phi_m + js\xi\cdot z\csc\phi_m}f_r(\rho)$$

(7.66)

$$\tilde{\vec{E}}(\zeta;n,s) = \frac{jaA}{us}\sum_{r=1}^{2}\frac{J_{n+s}^{r-1}(u)}{K_{n+s}^{r-1}(w)}\left[(\rho+is\varphi)\beta\frac{J_{n+s}^{2-r}(u\rho/a)}{K_{n+s}^{1-r}(w\rho/a)} - jz\frac{us}{a}\frac{J_{n}^{2-r}(u\rho/a)}{K_{n}^{1-r}(w\rho/a)}\right]f_r(\rho)e^{-i\phi_m}$$

$$\tilde{\vec{H}}(\zeta;n,s) = \frac{aA}{ju}\sum_{r=1}^{2}\frac{J_{n+s}^{r-1}(u)}{K_{n+s}^{r-1}(w)}\left[(\rho+is\varphi)\omega_0\varepsilon_r\frac{J_{n+s}^{2-r}(u\rho/a)}{K_{n+s}^{1-r}(w\rho/a)} - jz\frac{us\beta}{a\mu_0\omega_0}\frac{J_{n}^{2-r}(u\rho/a)}{K_{n}^{1-r}(w\rho/a)}\right]f_r(\rho)ie^{-i\phi_m}$$

(7.67)

Alternative, more horizontally compact expressions are presented in **Appendix B**. **It can be concluded that** in the bi-complex formulation (7.65, 67), the electric and magnetic fields are functionally identical in every respect, and not just with respect to the radial component, which is not the case for the complex formulation (7.64, 66). The bi-complex expressions are also much more vertically compact, since they are devoid of trigonometry. By contrast, the complex expressions must be presented as large quotients in order to fit within the margins of the page, when using the same font-size.

In §**4.1**, it is found that for **the complex formulation**, the power of the $m$-th polarization state of a given mode is given by

$$P_m = \frac{1}{2} \operatorname{Re} \int \tilde{\mathbf{E}}_m^*(\mathbf{r},t) \times \tilde{\mathbf{H}}_m(\mathbf{r},t) \cdot d\mathbf{A} \qquad (7.68)$$

whereas in §**4.2**, it is found that using **the bi-complex formulation** for the same EM-field yields the equivalent expression

$$P_m = \frac{1}{4} \operatorname{Re}_{i,j} \int_A \tilde{\mathbf{E}}_m^{\circ *}(\mathbf{r},t) \times \tilde{\mathbf{H}}_m(\mathbf{r},t) \cdot d\mathbf{A} \qquad (7.69)$$

which requires 2 real-operations as well as 2 complex-conjugations, with respect to both i and j. Over the course of the derivation of the orthogonality relation for the polarization states of the $HE_{11}$-mode in §**5.1**, it was found that

$$(P_m P_n)^{1/2} = \frac{\omega_0 \beta}{2} \sum_{r=1}^{2} \left( \begin{array}{l} \int \Lambda_r^*(\rho;s,-\lambda_r) \Lambda_r(\rho;s_r,-\lambda_r) \bar{\varepsilon}_r(\rho) \rho d\rho \int_0^{2\pi} \sin\phi_m \sin\phi_n d\varphi \\ + \int \Lambda_r^*(\rho;s,\lambda_r) \Lambda_r(\rho;s_r,\lambda_r) \bar{\varepsilon}_r(\rho) \rho d\rho \int_0^{2\pi} \cos\phi_m \cos\phi_n d\varphi \end{array} \right)$$
(7.70)

which requires the resolution of trigonometric integrals, whereas in the bi-complex formulation**,** which is found in §**5.2**, the same expression has the simpler form of

$$(P_m P_n)^{1/2} = \frac{\omega_0 \beta}{2} \sum_{r=1}^{2} \int_{\mathbb{R}^+} \left( \begin{array}{l} \Lambda_r^{\circ *}(\rho;s,\lambda_r) \Lambda_r(\rho;s_r,\lambda_r) \\ + \Lambda_r^{\circ *}(\rho;s,-\lambda_r) \Lambda_r(\rho;s_r,-\lambda_r) \end{array} \right) \bar{\varepsilon}_r(\rho) \rho d\rho \int_0^{2\pi} \cos(\psi_m - \psi_n) d\varphi .$$
(7.71)

The generating function (7.23) is *a priori* entirely real. The radial integral is thus identical for both approaches. However, it is clear that (7.71) yields a much simpler overall expression, since the integrand in the azimuth is actually independent of the azimuth based on (7.11), resulting in a trivial integral of a constant, and **the conclusion** that the power flow is modally independent with respect to polarization, but this is not immediately obvious from (7.70) derived using the complex formulation. The azimuthal integrals vanish, unless the polarization indices $m$ and $n$ are identical, yielding the following expression for the power of the $HE_{11}$-mode, per polarization state,

$$P = \frac{\pi \omega_0 \beta}{2} \sum_{r=1}^{2} \int_{\mathbb{R}^+} \left[ \Lambda_r(\rho; s, -\lambda_r) \Lambda_r(\rho; s_r, -\lambda_r) + \Lambda_r(\rho; s, \lambda_r) \Lambda_r(\rho; s_r, \lambda_r) \right] \bar{\varepsilon}_r(\rho) \rho \, d\rho.$$

(7.72)

For the HE$_{11}$-mode, substituting (7.23) in the integrand of (7.72), with $n = 1$, and carrying out the integrals, a compact, analytical power expression is attained, and **derived in this report for the first time**, in §**5.1**:

$$P = \frac{\omega_0 \varepsilon_1 \beta |A|^2 S_1^2}{16 \pi u^2} \sum_{r=1}^{2} \sum_{l=0}^{1} \sum_{p=0}^{1} \sum_{q=0}^{1} \left( \frac{u^2 \varepsilon_2}{w^2 \varepsilon_1} \right)^{r-1} \frac{(1 - s\lambda_p)(1 - s_r \lambda_p) \lambda_q J_1^{2r-2}(u) J_{2p-q}^{2-r}(u) J_{2p+q}^{2-r}(u)}{(2\lambda_l \kappa_l^2)^{1-r} K_1^{2r-2}(w) K_{2p-q}^{1-r}(\kappa_l w) K_{2p+q}^{1-r}(\kappa_l w)},$$

(7.73)

**which is obtained prior to any approximation**. It may be used in the orthogonality relations derived for the ideal fiber using the 2 formulations. The compactness of the expression is attained by locating the core power kernel in the numerator of the summand, and the cladding power kernel, in its denominator. They are respectively activated when the regional index $r$ is 1 and 2. New parameters, beyond those of Okamoto's [1], have also been adopted to help contribute to the compactness of the expression. Among the new parameters are $S_1$, which is the surface area of the core, and

$$\kappa_l = (a/b)^{l-1},$$

(7.74)

a non-dimensional parameter in terms of the ratio of the inner and outer radii of the cladding, as well as $\lambda_\alpha$ (7.12). The power contribution due to the core and the cladding are easily found by respectively setting $r_0$ to either 1 or 2, otherwise more formally stated with the help of the Kronecker delta as

$$P_{r_0} = \delta_{r r_0} P = \frac{\omega_0 \varepsilon_1 \beta |A|^2 S_1^2}{16 \pi u^2} \sum_{l=0}^{1} \sum_{p=0}^{1} \sum_{q=0}^{1} \left( \frac{u^2 \varepsilon_2}{w^2 \varepsilon_1} \right)^{r_0 - 1} \frac{(1 - s\lambda_p)(1 - s_{r_0} \lambda_p) \lambda_q J_1^{2r_0 - 2}(u) J_{2p-q}^{2-r_0}(u) J_{2p+q}^{2-r_0}(u)}{(2\lambda_l \kappa_l^2)^{1-r_0} K_1^{2r_0 - 2}(w) K_{2p-q}^{1-r_0}(\kappa_l w) K_{2p+q}^{1-r_0}(\kappa_l w)}.$$

(7.75)

The WGF approximation (7.61) may also be applied to the derivation of the power expression using either (7.66) or (7.67), although it is simpler just to find it directly from (7.73) instead, as it is available, and as shown in §**6**: It is found that the quadruple sum (7.73) reduces to the following simple expression

$$P = \frac{\omega_0 \beta \varepsilon_1 |A|^2 J_0^2(u) S_1^2}{2 \pi u^2} \frac{K_1^2(u)}{K_0^2(u)} \left[ 1 + \frac{w^2}{u^2} + \frac{\kappa_0^2 K_0^2(\kappa_0 w)}{K_1^2(w)} - \frac{\kappa_0^2 K_1^2(\kappa_0 w)}{K_1^2(w)} \right].$$

(7.76)

Making the further assumption that the cladding outer radius ($b$) is infinite relative to its inner radius ($a$), which eliminates the bracketed 3rd and 4th terms in the process, there results the well-known expression [1, 25],

$$P = \frac{\omega_0 \beta \varepsilon_1 v^2 |A|^2 J_0^2(u) S_1^2}{2\pi u^4} \frac{K_1^2(w)}{K_0^2(w)}. \tag{7.77}$$

**The orthogonality relation for the polarization states** of the HE$_{11}$-mode in an ideal fiber using **the complex formulation** is derived in §**5.1**, in terms of cross-products of the vecsors of the EM-fields of the 2 polarization modes, and with $n$, and $m$ being restricted to either 1 or 2 for those modes,

$$\int_A \left[ \tilde{\mathbf{E}}_m^*(\mathbf{r},t) \times \tilde{\mathbf{H}}_n(\mathbf{r},t) + \tilde{\mathbf{E}}_n(\mathbf{r},t) \times \tilde{\mathbf{H}}_m^*(\mathbf{r},t) \right] \cdot \mathbf{dA}$$
$$= \delta_{mn} \frac{\omega_0 \varepsilon_1 \beta |A|^2 S_1^2}{4\pi u^2} \sum_{r=1}^{2} \sum_{l,p,q=0}^{1} \left( \frac{u^2 \varepsilon_2}{w^2 \varepsilon_1} \right)^{r-1} \frac{(1-s\lambda_p)(1-s_r\lambda_p)\lambda_q J_1^{2r-2}(u) J_{2p-q}^{2-r}(u) J_{2p+q}^{2-r}(u)}{(2\lambda_l \kappa_l^2)^{1-r} K_1^{2r-2}(w) K_{2p-q}^{1-r}(\kappa_l w) K_{2p+q}^{1-r}(\kappa_l w)} \tag{7.78}$$

Its RHS is identical with the power expression (7.73) derived herein, within a factor of 4. The RHS power expression has been simplified by combining the summations in $l$, $p$, and $q$, all of which have the same range, into a single summation. This relation reduces as follows, under the WGF and the infinite-cladding approximations,

$$\int_A \left( \tilde{\mathbf{E}}_m^*(\mathbf{r},t) \times \tilde{\mathbf{H}}_n(\mathbf{r},t) + \tilde{\mathbf{E}}_n(\mathbf{r},t) \times \tilde{\mathbf{H}}_m^*(\mathbf{r},t) \right) \cdot \mathbf{dA} = \delta_{mn} \frac{2\omega_0 \beta \varepsilon_1 v^2 |A|^2 S_1^2}{\pi u^4} J_0^2(u) \frac{K_1^2(w)}{K_0^2(w)}. \tag{7.79}$$

The orthogonality relation is also derived using **the bi-complex formulation** in §**5.2**. It is found that for the polarization modes of the HE$_{11}$-mode in an ideal fiber,

$$\int_A \left[ \widehat{\mathbf{E}}_m^{\circ*}(\mathbf{r},t) \times \widehat{\mathbf{H}}_n(\mathbf{r},t) + \widehat{\mathbf{E}}_m(\mathbf{r},t) \times \widehat{\mathbf{H}}_n^{\circ*}(\mathbf{r},t) \right] \cdot \mathbf{dA}$$
$$= \delta_{mn} \frac{\omega_0 \varepsilon_1 \beta |A|^2 S_1^2}{2\pi u^2} \sum_{r=1}^{2} \sum_{l,p,q=0}^{1} \left( \frac{u^2 \varepsilon_2}{w^2 \varepsilon_1} \right)^{r-1} \frac{(1-s\lambda_p)(1-s_r\lambda_p)\lambda_q J_1^{2r-2}(u) J_{2p-q}^{2-r}(u) J_{2p+q}^{2-r}(u)}{(2\lambda_l \kappa_l^2)^{1-r} K_1^{2r-2}(w) K_{2p-q}^{1-r}(\kappa_l w) K_{2p+q}^{1-r}(\kappa_l w)} \tag{7.80}$$

which also uses (7.73), but whose LHS requires conjugation with respect to both i and j. It reduces to the following expression under the WGF and the infinite-cladding approximations,

$$\int_A \left( \widehat{\mathbf{E}}_m^{\circ*}(\mathbf{r},t) \times \widehat{\mathbf{H}}_n(\mathbf{r},t) + \widehat{\mathbf{E}}_m(\mathbf{r},t) \times \widehat{\mathbf{H}}_n^{\circ*}(\mathbf{r},t) \right) \cdot \mathbf{dA} = \delta_{mn} \frac{4\omega_0 \beta \varepsilon_1 v^2 |A|^2 S_1^2}{\pi u^4} J_0^2(u) \frac{K_1^2(w)}{K_0^2(w)}. \tag{7.81}$$

To succinctly summarize, it has been shown in this report **for the first time**, and for the hybrid modes of an ideal optical fiber,
- That the 12 general components of the EM-field can be reduced to just 6 generalized components by employing a regional parameter $r$,
- That these 6 generalized components are reducible to 2 compact expressions only,
- That it is possible to find an explicit, concise pair of vector equations for the entire EM-field using the 2, generalized component expressions,
- That bi-complex mathematics can result in more compact versions of the standard component-resolved expressions, as well as those of the new expressions, compared to ones attained using the conventional complex approach, which is heavily trigonometric,
- That expression compactness is attainable in the conventional complex approach, by employing a quotient configuration, which confines the radial terms to the numerator, and most of the trigonometric terms to the denominator,
- That in the bi-complex formulation, the electric and magnetic fields are functionally separable in every respect, which expedites power computations
- That in the bi-complex formulation, the electric and magnetic fields are functionally identical in every respect (with the exception of a few constants) and not just with respect to the radial component, which is not the case for the conventional complex formulation. Since they are devoid of trigonometry, the bi-complex expressions are also much more vertically compact, in contrast to the complex expressions,
- That a compact expression for the total power is attainable for the $HE_{11}$-mode prior to any approximation, and is more easily found with bi-complex mathematics,
- That a compact, explicit expression for the orthogonality relation for the 2 polarization states of the $HE_{11}$-mode is possible using the power relation derived,
- That the generalized approach can be adapted to the well-known weakly-guided fiber approximation, validated by reproducing previously published power expressions due to Snyder [25].

Lastly, although many of the results could probably be derived using commercially available software packages such as Mathematica® (or [WolframAlpha.com](WolframAlpha.com)), Maple®, or Maxima, casting the results into their final compact forms as shown in (7.78, 80) for instance, is not possible with these packages, nor would reducing the number of EM-field components from twelve (7.1, 2) to six, as in (7.14-19), or from six to two, as in (7.21), for instance.

**This version of the paper differs from the previously published version, as follows**:
1. Placed greater emphasis on the fact that the original expressions for the hybrid modes of an optical fiber, are actually due to Snitzer [2], while acknowledging the contributions made by Okamoto [1]
2. Clarified that generalized EM-field expressions such as (3.2.35, 37), which carry Kronecker deltas in both numerators and denominators, are not meant to be decomposed into smaller quotients, which would result in more cumbersome versions of these expressions. Instead, (3.2.35, 37) are meant to be evaluated term-wise, once the variable $\xi$ is specified.

# APPENDIX A

The Kronecker delta is frequently used throughout this report to help render compact large expressions. It may be expressed as a mapping from the space $\mathbb{R}$ onto itself, or as $\delta_{pq}: \mathbb{R} \to \mathbb{R}$. It is conventionally defined with respect to the scalars $p$ and $q$ as

$$\delta_{pq} = \delta[p-q] = \begin{cases} 1, & p = q \\ 0, & p \neq q \end{cases}. \tag{A.1}$$

The subscript "$pq$" is suggestive of a scalar product of 2 variables, which is not accurate, since the above definition is equally valid when $q$ is zero, for instance. **The subscript "$pq$" is actually meant to be taken as short-form for the subtraction "$p - q$"**. A better expression would make use of "$p, q$" or "$p - q$" in the subscript, instead of "$pq$", although at some sacrifice of compactness. **This consideration becomes more critical when $q$ is a negative integer, which however, is not the case anywhere in this report**.

A clarification is presented in this section, when the argument of the Kronecker delta is comprised of three-dimensional vectors, or in $\mathbb{R}^3$, as in §**3.3 - 3.5**, which is given by the mapping $\delta_{\mathbf{XY}}: \mathbb{R}^3 \to \mathbb{R}$. For 2 such spatial vectors $\mathbf{X}$ and $\mathbf{Y}$,

$$\delta_{\mathbf{XY}} = \begin{cases} 1, & \mathbf{X} = \mathbf{Y} \\ 0, & \mathbf{X} \neq \mathbf{Y} \end{cases}. \tag{A.2}$$

Due to notational misuse however, the delta subscript may be misconstrued as a dyadic product or a tensor, which is certainly not the case here. For 2 such vectors to be identical, their constituent components must also be so, for each of the 3 directions,

$$X_i = Y_i, \quad i \in \{1, 2, 3\}. \tag{A.3}$$

It is thus concluded that (A.2) is equivalent to the following statement,

$$\delta_{\mathbf{XY}} = \delta_{X_1 Y_1} \delta_{X_2 Y_2} \delta_{X_3 Y_3} \tag{A.4}$$

which states that the LHS vanishes if the vector components are dissimilar for any given direction. This relation is independent of the coordinate system used.

For any 2 coordinate **basis-vectors** $\mathbf{x} = [x_1, x_2, x_3]^\mathrm{T}$ and $\mathbf{y} = [y_1, y_2, y_3]^\mathrm{T}$ in $\mathbb{R}^3$, it is true that in various notations, the inner product yields

$$\langle x | y \rangle = \mathbf{x}^\mathrm{T} \mathbf{y} = \begin{cases} 1, & \mathbf{x} = \mathbf{y} \\ 0, & \mathbf{x} \neq \mathbf{y} \end{cases}. \tag{A.5}$$

It is valid for any orthogonal coordinate system, such as rectangular or cylindrical coordinates. Therefore, it should be obvious that the Kronecker delta with respect to the vectors $\mathbf{x}$ and $\mathbf{y}$ may also be expressed as

$$\delta_{xy} = \mathbf{x}^T \mathbf{y}. \tag{A.6}$$

In any orthogonal coordinate system in $\mathbb{R}^3$, any 2 of the 3 basis-vectors used to describe the vector space in that system obey the following relation,

$$\mathbf{x}_i \bullet \mathbf{y}_j + |\mathbf{x}_i \times \mathbf{y}_j| = 1, \quad i, j \in \{1, 2, 3\} \tag{A.7}$$

which can be used to define a Kronecker delta for such vectors that represent a coordinate system. In this expression, *i* and *j* may be either identical or dissimilar. Since the dot-product is equivalent to taking the vector inner product (A.5, 6), it is deduced that

$$\delta_{\mathbf{x}_i \mathbf{y}_j} = \mathbf{x}_i \bullet \mathbf{y}_j = \begin{cases} 1, & \mathbf{x}_i = \mathbf{y}_j \\ 0, & \mathbf{x}_i \neq \mathbf{y}_j \end{cases}. \tag{A.8}$$

Substituting (A.8) into (A.7), it is also found that

$$\delta_{\mathbf{x}_i \mathbf{y}_j} = 1 - |\mathbf{x}_i \times \mathbf{y}_j|. \tag{A.9}$$

Consequently, **the dual of the Kronecker delta,** which by contrast to (A.8), is only non-zero when the vectors are dissimilar, is given by

$$|\mathbf{x}_i \times \mathbf{y}_j| = 1 - \delta_{\mathbf{x}_i \mathbf{y}_j} = \begin{cases} 1, & \mathbf{x}_i \neq \mathbf{y}_j \\ 0, & \mathbf{x}_i = \mathbf{y}_j \end{cases} \tag{A.10}$$

which can be more compact than using an algebraic expression that involves the Kronecker delta, especially if it requires parentheses, and/or when frequently used in a large expression.

The general scalar expressions (3.2.35-38), as well as the general vector expressions (3.4.1, 7, 16, 17) heavily rely on some form of the Kronecker delta. Since the Kronecker delta can also be viewed as a generalized discrete function, it may be replaced by one of many analytic functions, when desired. For the regional Kronecker delta used to identify either the core or the cladding of the optical fiber, it is defined as

$$\delta_{rr'} = \frac{3 - r' - r}{3 - 2r'} = \begin{cases} 0, & \text{if } r \neq r' \\ 1, & \text{if } r = r' \end{cases}, \quad r, r' \in \{1, 2\} \tag{A.11}$$

which is analytic since it is its own Taylor series. For $r \neq r'$, the sum of *r* and *r'* is identical with 3, and the expression vanishes as desired. Otherwise, it reduces to unity. It should be stressed that the domain of *this* Kronecker delta is restricted to just 2 elements as shown in (A.11). The coordinate-dependent Kronecker delta is slightly more elaborate, because its domain is larger. It may be expressed by an analytic, quadratic function,

$$\delta_{\xi\xi'} = \frac{\xi'' - \xi}{\xi'' - \xi'} \frac{\xi''' - \xi}{\xi''' - \xi'} = \begin{cases} 0, & \text{if } \xi \neq \xi' \\ 1, & \text{if } \xi = \xi' \end{cases}, \quad \xi \in \{\xi', \xi'', \xi'''\}. \tag{A.12}$$

For instance, for the *z*-coordinate in the cylindrical coordinate system, $\delta_{\xi z}$ is implemented with $\xi'' = \rho$ and $\xi''' = \varphi$, **or** with $\xi'' = \varphi$ and $\xi''' = \rho$, since the expression is symmetric with respect to $\xi''$ and $\xi'''$. Using the first instance, it is defined as

$$\delta_{\xi z} = \frac{\xi - \rho}{z - \rho} \frac{\xi - \varphi}{z - \varphi} = \begin{cases} 0, & \text{if } \xi \neq z \\ 1, & \text{if } \xi = z \end{cases}, \qquad \xi \in \{\rho, \varphi, z\}. \tag{A.13}$$

This approach, although useful, is limited to symbolic, *discrete*, algebraic applications. Differentiation of (A.11-13) is at best, not meaningful.

On the real number line $\mathbb{R}$, the Kronecker delta may also be expressed algebraically in terms of 2 Heaviside step-functions, as follows,

$$\delta_{tt_0} = H(t - t_0) + H(t_0 - t) - 1. \tag{A.14}$$

The first step-function vanishes for all *t* *bigger* than or equal to $t_0$, whereas the second step-function vanishes for all *t* *less* than or equal to $t_0$. Consequently, the step-functions additively evaluate to 2 at $t = t_0$, but to 1 everywhere else. The RHS thus reduces to the Kronecker delta upon subtraction of unity, as shown in (A.14). Each of the 3 terms on the RHS actually represents a Schwartz distribution. Furthermore, it should be clear that (A.14) may bear no equivalence to the previously discussed versions of the Kronecker delta in this appendix, which are of little to no utility in differential algebra, unlike (A.14).

**The Leibniz derivative of the Kronecker delta is zero everywhere over its domain, except where its argument is zero, for which such a derivative is *undefined***. The derivative, can however also be found using (A.14) in the distributional sense, with a test-function $\psi$ of the required attributes, as explained in §**3.3**,

$$\langle \delta'_{tt_0}, \psi \rangle = \langle H'(t - t_0), \psi \rangle + \langle H'(t_0 - t), \psi \rangle - \langle 1', \psi \rangle. \tag{A.15}$$

Each bracket is then evaluated using integration by parts (IBP), under the assumptions that $t_0 \in \Omega$ for the first two terms, and that a **test-function and all its derivatives vanish at infinity**[13],

$$\langle \delta'_{tt_0}, \psi \rangle = -\int_{-\infty}^{\infty} H(t - t_0)\psi'(t)dt - \int_{-\infty}^{\infty} H(t_0 - t)\psi'(t)dt + \int_{-\infty}^{\infty} \psi'(t)dt \tag{A.16}$$

using (3.3.5.1), and after the elimination of IBP by-products. Upon a simplification,

$$\langle \delta'_{tt_0}, \psi \rangle = -\int_{t_0}^{\infty} \psi'(t)dt - \int_{-\infty}^{t_0} \psi'(t)dt + [\psi(t)]_{-\infty}^{\infty} = \psi(t_0) - \psi(t_0) + 0 = 0. \tag{A.17}$$

**It is thus concluded that the derivative of the Kronecker delta vanishes *everywhere*, in the distributional sense**[14].

---

[13] Perhaps the best example of such a test function, is the non-centered Gaussian distribution exp(-$(t - t_0)^2/T$)/$\sqrt{(\pi T)}$
[14] It can be shown at Wolframalpha.com that the input text "differentiate KroneckerDelta(x, a) with respect to x" yields zero, as one confirmation of (A.17)

**The integral of the Kronecker delta**, may likewise be carried out in the distributional sense using (A.14) again. This time however, it is more convenient to use a derivative of the test-function instead of the test-function itself, yielding

$$\langle \tilde{\delta}_{tt_0}, \psi^{(n-1)} \rangle = \langle \tilde{H}(t-t_0), \psi^{(n-1)} \rangle + \langle \tilde{H}(t_0-t), \psi^{(n-1)} \rangle - \langle \tilde{1}, \psi^{(n-1)} \rangle \quad (A.18)$$

since any derivative of such a test-function can also serve as a test-function, because a test-function is infinitely differentiable, as explained in §**3.3**. The choice of $n$ will become apparent at the resolution of the bracket (A.18). Moreover, the anti-derivative of a function as used in (A.18), is generally given by

$$\tilde{f}(t) = \int f(t) dt, \quad (A.19)$$

while neglecting the integration constant at this juncture, since additional integration steps as required by (A.18) are yet to be carried out. In particular, the anti-derivative of a step-function yields the ramp-function,

$$\tilde{H}(s(t-t_0)) = \int H(s(t-t_0)) dt = (t - |s|t_0) H(s(t-t_0)), \quad s \in \{-1, 0, 1\}. \quad (A.20)$$

The case of $s = 0$ corresponds to the indefinite integral of unity, which is required for the 3rd term of (A.18). Now (A.20) may be specialized to each term of (A.18), in order to carry out IBP on each bracket. IBP is an heuristic integration technique which can lead to difficulties, depending on how the integrand variables are chosen. To expedite the resolution of (A.18), the ramp-function (A.20) is differentiated while the test-functions are integrated, which collectively simplify (A.18) to

$$\langle \tilde{\delta}_{tt_0}, \psi^{(n-1)} \rangle = -\int_{-\infty}^{\infty} H(t-t_0) \psi^{(n-2)}(t) dt - \int_{-\infty}^{\infty} H(t_0-t) \psi^{(n-2)}(t) dt + \int_{-\infty}^{\infty} \psi^{(n-2)}(t) dt$$

(A.21)

after neglecting IBP by-products that involve ramp-functions and integrals of the test-function, all of which vanish at the integration bounds. Simplifying (A.21) yields

$$\langle \tilde{\delta}_{tt_0}, \psi^{(n-1)} \rangle = -\int_{t_0}^{\infty} \psi^{(n-2)}(t) dt - \int_{-\infty}^{t_0} \psi^{(n-2)}(t) dt + \int_{-\infty}^{\infty} \psi^{(n-2)}(t) dt. \quad (A.22)$$

The value of $n$ was left arbitrary in (A.18), but can now be set in (A.22) to 3 to further expedite the solution of (A.22), in order to reproduce the RHS of (A.17), whose outcome is already known. Subsequently carrying out the integrals leads to the conclusion that, **the integral of the Kronecker delta evaluates to zero in the distributional sense, like its derivative (A.17)**:

$$\langle \tilde{\delta}_{tt_0}, \psi'' \rangle = \psi(t_0) - \psi(t_0) + [\psi(t)]_{-\infty}^{\infty} = 0. \quad (A.23)$$

# APPENDIX B

The compactness to be attained for an expression depends on how many more new variables and/or functions are permitted to be used, beyond those of Okamoto's nomenclature. For instance, using the additional new parameters introduced in this report,

$$\lambda_\alpha = e^{j\pi\alpha}, \ \alpha \in \mathbb{Z}^+, \tag{7.12}$$

$$\phi_m = n\varphi + \psi_m, \ m \in \{1,2\}, n \in \mathbb{Z}^+ \tag{7.13}$$

then the complex generalized scalar equations (3.2.35) can be re-expressed as

$$E_{r\xi}(\rho,\varphi) = \frac{aA\beta}{2ju}\left(\frac{uJ_n(u)}{wK_n(w)}\right)^{r-1} \frac{(1-s+s\delta_{\xi z})\frac{J_{n-1}^{2-r}(u\rho/a)}{K_{n-1}^{1-r}(w\rho/a)} + \frac{(1+s-s\delta_{\xi z})\lambda_r}{\delta_{\xi\rho}-\delta_{\xi\varphi}-\delta_{\xi z}}\frac{J_{n+1}^{2-r}(u\rho/a)}{K_{n+1}^{1-r}(w\rho/a)}}{\delta_{\xi\rho}\sec\phi_m - \delta_{\xi\varphi}\csc\phi_m + jna^2\beta\frac{\lambda_r u^{2r-4}}{\rho w^{2r-2}}\delta_{\xi z}\sec\phi_m}$$

$$H_{r\xi}(\rho,\varphi) = \frac{aA\omega_0\varepsilon_r}{2ju}\left(\frac{uJ_n(u)}{wK_n(w)}\right)^{r-1} \frac{(1-s_r+s_r\delta_{\xi z})\frac{J_{n-1}^{2-r}(u\rho/a)}{K_{n-1}^{1-r}(w\rho/a)} - \frac{(1+s_r-s_r\delta_{\xi z})\lambda_r}{\delta_{\xi\rho}-\delta_{\xi\varphi}+\delta_{\xi z}}\frac{J_{n+1}^{2-r}(u\rho/a)}{K_{n+1}^{1-r}(w\rho/a)}}{\delta_{\xi\rho}\csc\phi_m + \delta_{\xi\varphi}\sec\phi_m - jna^2\beta\frac{\lambda_r u^{2r-4}}{\rho s_r w^{2r-2}}\delta_{\xi z}\csc\phi_m}$$

(B.1)

Some expressions, such as the EM-field vectors, can be rendered more compact at no sacrifice to clarity, by using the equivalence for the Heaviside step-function $f_r$ (3.3.33),

$$f_r(\rho) = \frac{1}{\Gamma(f_r(\rho))} \tag{B.2}$$

in terms of Euler's gamma function [27], **which is unity when its argument is unity, and is *effectively* infinite, when its argument is zero**, thus reproducing the properties of the step-function $f_r(\rho)$, when used in (B.2). Applying this relation to (3.4.1) for instance,

$$\vec{E}(\zeta) = \sum_{r=1}^{2}\sum_{\xi=\rho,\varphi,z} \xi \frac{\left[(1-s|\xi\times\mathbf{z}|)\frac{J_{n-1}^{2-r}(u\rho/a)}{K_{n-1}^{1-r}(w\rho/a)} + \frac{\lambda_r(1+s|\xi\times\mathbf{z}|)}{\xi\cdot\boldsymbol{\rho}-\xi\cdot\boldsymbol{\varphi}-\xi\cdot\mathbf{z}}\frac{J_{n+1}^{2-r}(u\rho/a)}{K_{n+1}^{1-r}(w\rho/a)}\right]\cos\phi_m}{\frac{2ju}{aA\beta}\left(\frac{uJ_n(u)}{wK_n(w)}\right)^{1-r}\left[\xi\cdot\boldsymbol{\rho}-\xi\cdot\boldsymbol{\varphi}\cot\phi_m + jna^2\beta\frac{\lambda_r u^{2r-4}}{\rho w^{2r-2}}\xi\cdot\mathbf{z}\right]\Gamma(f_r(\rho))}$$

$$\vec{H}(\zeta) = \sum_{r=1}^{2}\sum_{\xi=\rho,\varphi,z} \xi \frac{\left[(1-s_r|\xi\times\mathbf{z}|)\frac{J_{n-1}^{2-r}(u\rho/a)}{K_{n-1}^{1-r}(w\rho/a)} - \frac{\lambda_r(1+s_r|\xi\times\mathbf{z}|)}{\xi\cdot\boldsymbol{\rho}-\xi\cdot\boldsymbol{\varphi}+\xi\cdot\mathbf{z}}\frac{J_{n+1}^{2-r}(u\rho/a)}{K_{n+1}^{1-r}(w\rho/a)}\right]\varepsilon_r\sin\phi_m}{\frac{2ju}{aA\omega_0}\left(\frac{uJ_n(u)}{wK_n(w)}\right)^{1-r}\left[\xi\cdot\boldsymbol{\rho}+\xi\cdot\boldsymbol{\varphi}\tan\phi_m - jna^2\beta\frac{\lambda_r u^{2r-4}}{\rho s_r w^{2r-2}}\xi\cdot\mathbf{z}\right]\Gamma(f_r(\rho))}$$

(B.3)

which are more horizontally compact than (3.4.1), without a reduction in font-size. However, the gamma function is somewhat esoteric, as it is a non-elementary function, and is usually considered a "special function". In one definition, it is given by the integral

$$\Gamma(z) = \int_0^\infty t^{z-1} e^{-t} dt, \quad \text{Re}(z) > 0, \quad z \in \mathbb{C} \tag{B.4}$$

and is a *generalization* of the factorial function (*n*!) to complex arguments *z*, which includes the entire real number line as a subset. It is not widely used in electromagnetism, unless it occurs in the solutions of particular problems. This difficulty is somewhat exacerbated by its argument in (B.2), which itself is a generalized function.

In **§6**, generalized vector expressions were developed under the weakly-guided fiber approximation. In this section, more horizontally compact alternative expressions than those in **§6** are presented. After a re-arrangement, the generalized **complex** vectors of the EM-field can be found using (6.37) in the alternative forms of

$$\vec{E}(\zeta; n, s) = \sum_{r=1}^{2} \sum_{\xi=\rho,\varphi,z} \xi \frac{|\xi \times \mathbf{z}|\beta \frac{J_{n+s}^{2-r}(u\rho/a)}{K_{n+s}^{1-r}(w\rho/a)} + \xi \cdot \mathbf{z} \frac{u}{a} \frac{J_n^{2-r}(u\rho/a)}{K_n^{1-r}(w\rho/a)}}{\frac{us}{jaA} \frac{J_{n+s}^{1-r}(u)}{K_{n+s}^{1-r}(w)} (\xi \cdot \boldsymbol{\rho} + \xi \cdot \boldsymbol{\varphi} s \cot \phi_m + js\xi \cdot \mathbf{z})} f_r(\rho) \cos \phi_m$$

$$\vec{H}(\zeta; n, s) = \sum_{r=1}^{2} \sum_{\xi=\rho,\varphi,z} \xi \frac{|\xi \times \mathbf{z}|\omega_0 \varepsilon_r \frac{J_{n+s}^{2-r}(u\rho/a)}{K_{n+s}^{1-r}(w\rho/a)} + \xi \cdot \mathbf{z} \frac{u\beta}{a\omega_0\mu_0} \frac{J_n^{2-r}(u\rho/a)}{K_n^{1-r}(w\rho/a)}}{\frac{ju}{aA} \frac{J_{n+s}^{1-r}(u)}{K_{n+s}^{1-r}(w)} (\xi \cdot \boldsymbol{\rho} - \xi \cdot \boldsymbol{\varphi} s \tan \phi_m + js\xi \cdot \mathbf{z})} f_r(\rho) \sin \phi_m$$

(B.5)

whereas the generalized **bi-complex** vecsors of the same EM-field are found from (6.47) to be in the alternative forms

$$\tilde{\mathbf{E}}(\zeta; n, s) = \sum_{r=1}^{2} \sum_{\xi=\rho,\varphi,z} \xi \frac{|\xi \times \mathbf{z}|\beta \frac{J_{n+s}^{2-r}(u\rho/a)}{K_{n+s}^{1-r}(w\rho/a)} + \xi \cdot \mathbf{z} \frac{u}{a} \frac{J_n^{2-r}(u\rho/a)}{K_n^{1-r}(w\rho/a)}}{\frac{us}{jaA} \frac{J_{n+s}^{1-r}(u)}{K_{n+s}^{1-r}(w)} (\xi \cdot \boldsymbol{\rho} - is\xi \cdot \boldsymbol{\varphi} + js\xi \cdot \mathbf{z})} f_r(\rho) e^{-i\phi_m}$$

$$\tilde{\mathbf{H}}(\zeta; n, s) = \sum_{r=1}^{2} \sum_{\xi=\rho,\varphi,z} \xi \frac{|\xi \times \mathbf{z}|\omega_0 \varepsilon_r \frac{J_{n+s}^{2-r}(u\rho/a)}{K_{n+s}^{1-r}(w\rho/a)} + \xi \cdot \mathbf{z} \frac{u\beta}{a\omega_0\mu_0} \frac{J_n^{2-r}(u\rho/a)}{K_n^{1-r}(w\rho/a)}}{\frac{ju}{aA} \frac{J_{n+s}^{1-r}(u)}{K_{n+s}^{1-r}(w)} (\xi \cdot \boldsymbol{\rho} - is\xi \cdot \boldsymbol{\varphi} + js\xi \cdot \mathbf{z})} f_r(\rho) i e^{-i\phi_m}$$

(B.6)

Of course, if it is permissible to use the gamma function equivalence (B.2), even more *horizontally* compact versions of the above equations are possible.

# Nomenclature

## 1. Roman functions and variables

- $a$ : Radius of the core of the fiber, or the inner radius of its cladding
- $A$ : An EM-field amplitude variable
- $b$ : Outer radius of the cladding of the fiber
- c : Speed of light in vacuum
- $\mathbb{C}$ : The set of complex numbers
- $E$ : A component of the electric field in the **complex formulation** of the EM-field
- $\tilde{E}$ : A component phasor of the electric field in the **bi-complex formulation**
- $\vec{\bar{E}}$ : The electric field vector in the **complex formulation**, normalized to the power of the EM-field in that formulation
- $\tilde{\mathbf{E}}$ : The electric field vector, which is also a phasor, and thus termed a "**vecsor**" in this report
- $\bar{\tilde{\mathbf{E}}}$ : The electric field vecsor, normalized to the power of the EM-field
- $f$ : An orthogonal function defined by the Heaviside step-function
- $H$ : A scalar component of the magnetic field in the **complex formulation** of the EM-field
- $\tilde{H}$ : A phasor component of the magnetic field in the **bi-complex formulation**
- $\vec{\mathbf{H}}$ : The magnetic field vector in the **complex formulation**
- $\vec{\bar{\mathbf{H}}}$ : The magnetic field vector in the **complex formulation**, normalized to the power of the EM-field in that formulation
- $\tilde{\mathbf{H}}$ : The magnetic field vecsor, in either formulation, normalized to the power of the EM-field in that formulation
- $\bar{\tilde{\mathbf{H}}}$ : The magnetic field vecsor, normalized to the power of the EM-field
- H : The Heaviside unit-step function (**not-italicized**)
- $h$ : A variable used in the expression of modal power
- i : An imaginary number used to generate a bi-complex quantity
- $I$ : A generic integral expression
- $j$ : Another, distinct imaginary number used in Okamoto's nomenclature
- $J$ : A Bessel function of the first kind
- $K$ : A modified Bessel function of the second kind
- $k_0$ : The free-space wave-number
- $l$ : A summation index used in the expression of modal power
- $m$ : The polarization modal index
- $n$ : Another polarization modal index, used in orthogonality relations only
- $n$ : Azimuthal eigenvalue of the EM-field propagated in the fiber, **not used in the same expressions as those using $n$ as a polarization modal index**
- $n_1$ : The refractive index of the core of the fiber
- $n_2$ : The refractive index of the cladding of the fiber
- $n_{\text{eff}}$ : The effective index or eigenvalue of a mode of the fiber
- $n_r$ : A generalized, *r*egional refractive index that stands for either $n_1$ or $n_2$ depending on $r$
- $p$ : Another summation index used in the expression of modal power
- $P$ : Power of a mode propagating in the fiber
- $q$ : Another summation index used in the expression of modal power
- $r$ : A regional variable used to distinguish between the core ($r=1$) and the cladding ($r=2$)
- $\mathbb{R}$ : The set of real numbers
- $s$ : A parameter, expressed in terms of Bessel functions and their derivatives.
- $s_1$ : The variable $s$ multiplied by the squared-ratio $(n_{\text{eff}}/n_1)^2$
- $s_2$ : The variable $s$ multiplied by the squared-ratio $(n_{\text{eff}}/n_2)^2$

| | | |
|---|---|---|
| $s_r$ | : | Generalized modal variable that stands for either $s_1$ or $s_2$, depending on $r$ |
| $\vec{\mathbf{S}}$ | : | The Poynting vector |
| $\vec{S}$ | : | The time-averaged Poynting vector (italicized) |
| $S_1$ | : | The surface area of the core of the fiber |
| $T$ | : | A subscript used to denote a transverse field quantity |
| **U** | : | An EM-field vectorial quantity in the **bi-complex formulation**, that stands for either the E-field vector, or the H-field vector |
| $U$ | : | An EM-field scalar quantity in the **bi-complex formulation**, that stands for either an E-field scalar component, or an H-field scalar component |
| $\tilde{U}$ | : | An EM-field phasor quantity in the **bi-complex formulation**, that stands for either an E-field scalar component, or an H-field scalar component |
| $u$ | : | A non-dimensional transverse wave-number |
| $\vec{\mathbf{v}}$ | : | The real electromagnetic field vector, which is either that of the E-field of the M-field |
| **V** | : | An EM-field vector quantity in the **complex formulation**, that stands for either the E-field vector, or the H-field vector |
| $V$ | : | An EM-field scalar quantity in the **complex formulation**, that stands for either the E-field scalar component, or an H-field scalar component |
| $v$ | : | The *v*-number of an optical fiber, used in the report to distinguish between a single-mode and a multi-mode fiber |
| $w$ | : | A non-dimensional transverse wave-number, or: a bi-complex number used in §**2** only |
| $x, x_1, x_2$ | : | Real numbers used in §**2** and **7** only |
| $y, y_1, y_2$ | : | Real numbers used in §**2** and **7** only |
| $x, y$ | : | The 2 polarization states of a hybrid mode, used as such beyond §**2** |
| $z$ | : | A coordinate and the direction of EM-field propagation in cylindrical coordinates |
| $z_1, z_2$ | : | Complex numbers used in §**2** and **7** only |
| $\mathbb{Z}$ | : | The set of integers |

## 2. Greek functions and variables

| | | |
|---|---|---|
| $\alpha$ | : | The subscript, or the independent variable, of the $\lambda$-parameter |
| $\beta$ | : | The propagation constant of the EM-field propagated in the optical fiber |
| $\Gamma$ | : | Euler's gamma function |
| $\delta$ | : | Either the Dirac-delta function, or the Kronecker delta symbol, depending on context |
| $\varepsilon_r$ | : | The *r*egional (**and not the relative**) permittivity of a fiber's constituent material |
| $\varepsilon_1$ | : | The permittivity of the material of the fiber's core |
| $\varepsilon_2$ | : | The permittivity of the material of the fiber's cladding |
| $\zeta$ | : | Short form for the polar coordinate couple $(\rho, \varphi)$ |
| $\eta$ | : | A variable used in the generating function $\Lambda$, and stands for either 0, $s$ or $s_r$ |
| $\kappa$ | : | The ratio of the radius of the fiber's core to that of its cladding |
| $\lambda$ | : | A parameter used in the generating function $\Lambda$, **not to be confused with wavelength** |
| $\Lambda$ | : | The generating function used to produce the EM-field components |
| $\mu_0$ | : | The magnetic permeability constant of vacuum |
| $\mu$ | : | A complex number variable used in §**2** to denote either i or j. **Italicized** |
| $\xi$ | : | A variable that stands for any member of the cylindrical coordinate triplet $(\rho, \varphi, z)$ |
| $\rho$ | : | The radial coordinate in cylindrical coordinates |
| $\varphi$ | : | The azimuthal coordinate in cylindrical coordinates |
| $\psi_m$ | : | The *m*odal phase factor that determines the polarization state of a given mode |
| $\omega_0$ | : | The angular frequency of the EM-field |
| $\phi$ | : | A composite angle comprised of the algebraic sum of $\varphi$ and $\psi$ |